\documentclass[reqno,12pt]{article}

\usepackage[dvipsnames]{xcolor}
\usepackage{tikzit}
\usepackage{tikz}
\usepackage{tikz-3dplot}
\usepackage{tkz-euclide}
\usepackage{caption}
\usepackage{subcaption}
\usepackage[sorting=none,style=numeric-comp,backend=bibtex]{biblatex}
\bibliography{foo}

\tikzstyle{new style 0}=[fill=white, draw={rgb,255: red,206; green,0; blue,0}, shape=circle]

\tikzstyle{new edge style 0}=[draw=black, -]
\tikzstyle{new edge style 1}=[-, draw={rgb,255: red,206; green,0; blue,0}]

\usepackage{dsdshorthand}
\usepackage{tikz}
\usepackage{mdframed}
\usepackage{ae} 
\usepackage[T1]{fontenc}
\usepackage[ansinew]{inputenc}
\usepackage{amsmath}
\usepackage{amssymb}
\usepackage{graphicx}
\usepackage{caption}
\usepackage{subcaption}

\usetikzlibrary{decorations.pathmorphing}
\usetikzlibrary{decorations.markings}
\usetikzlibrary{decorations.pathreplacing,calligraphy}

\tikzset{snake it/.style={decorate, decoration=snake}}

\usepackage{todonotes}

\usepackage{amsmath}
\usetikzlibrary{arrows.meta}
\usepackage{color}
\definecolor{darkblue}{cmyk}{0.9,0.9,0,0}
\usepackage[colorlinks=true,linkcolor=darkblue,citecolor=darkblue,urlcolor=darkblue]{hyperref}
\usepackage{epsfig}

\usepackage{pgfplots}
\usepackage{float}
\usepackage{tikz,tikz-3dplot}
\usetikzlibrary{arrows}
\usepackage{cleveref}
\usepackage{makeidx}
\usepackage{mathtools}

\makeindex

\usepackage[english]{babel}
\pgfplotsset{compat=1.16}

\begin{document}

\thispagestyle{empty}

\renewcommand{\thefootnote}{\fnsymbol{footnote}}
\setcounter{page}{1}
\setcounter{footnote}{0}
\setcounter{figure}{0}

\vspace{-0.4in}

\begin{center}
  $$$$
  {\Large\textbf{\mathversion{bold}
        Conformal multi-Regge theory
      }\par}
  \vspace{1.0cm}

  \textrm{Miguel S. Costa$^\text{\tiny 1}$, Vasco Gon\c{c}alves$^\text{\tiny 1}$, Aaditya Salgarkar$^\text{\tiny 1}$, Jo\~ao Vilas Boas$^\text{\tiny 1, 2}$}
  \\ \vspace{1.2cm}
  \footnotesize{\textit{
  $^\text{\tiny 1}$Centro de F\'\i sica do Porto e Departamento de F\'\i sica e Astronomia, Faculdade de Ci\^encias da Universidade do Porto, Porto 4169-007, Portugal   \\
  $^\text{\tiny 2}$Fields and Strings Laboratory, Institute of Physics
  \'{E}cole Polytechnique F\'{e}d\'{e}ral de Lausanne (EPFL)
  Route de la Sorge, CH-1015 Lausanne, Switzerland \\
  }
  \vspace{4mm}
  }
\end{center}

\par\vspace{1.5cm}

\vspace{2mm}
\begin{abstract}
  We propose and explore the Regge limit for correlation functions of five local primary operators in conformal field theories.
  After reviewing some features of Regge theory for flat-space scattering amplitudes,	we analyze the analytic structure of
  conformal blocks both in position and Mellin space in the Regge limit and propose an extension of conformal Regge theory for five-point functions.
  As a byproduct of our analysis we also introduce a new basis of three-point correlation functions for operators with spin and the associated Euclidean conformal blocks.
\end{abstract}

\newpage

\setcounter{page}{1}
\renewcommand{\thefootnote}{\arabic{footnote}}
\setcounter{footnote}{0}

\tableofcontents

\section{Introduction}
The last decade has seen a lot of progress in the program of conformal bootstrap, an approach to classify conformal field theories (CFTs) by studying their correlation functions \cite{Rattazzi:2008pe}.
The main idea is to use the general principles of conformal field theories: conformal symmetry, unitarity and crossing symmetry, to constrain the observables \cite{Poland:2018epd}.
The paradigmatic example of success of this approach is the three-dimensional Ising model for which one can achieve outstanding numerical precision for the spectrum of this theory,
vastly outperforming the results from Monte-Carlo simulations~\cite{El-Showk:2012cjh,Kos:2016ysd}. Alternatively, one can also use the bootstrap philosophy to constrain observables analytically.
An example of that is the lightcone bootstrap \cite{Alday:2015ewa,Simmons-Duffin:2016wlq,Komargodski:2012ek}.

The vast majority of results of conformal bootstrap rely on the study of correlation functions of four primary operators. While the full set of these contain all the dynamical data of a theory, it is true that as the spin of these operators is increased the task of finding these data becomes more and more challenging. This is the reasoning why most conformal bootstrap works focus on correlation functions of scalar operators. In recent years, on the other hand, it has become more and more appreciated the fact that consistency conditions at the the level of scalar higher-point functions can be the appropriate setting to deal with this problem. Indeed, higher-point correlators give us access to more data than their lower-point counterparts and in particular can probe many spinning data. Due to the central role of conformal blocks in the conformal bootstrap, these have been considered for higher-point functions in
\cite{Rosenhaus:2018zqn,Goncalves:2019znr,Buric:2020dyz,Buric:2021kgy,Buric:2021ttm,Buric:2021ywo,Poland:2021xjs}.
Although their structure is generically intricate, it simplifies drastically in the lightcone limit where bootstrap studies have been performed in \cite{Antunes:2021kmm,Bercini:2020msp}.
Higher-point correlation functions have also been considered in multiple contexts, for instance in holographic  theories~\cite{Goncalves:2019znr,Alday:2022lkk,Goncalves:2023oyx} and more recently in numerical bootstrap \cite{Poland:2023vpn}.

An important tool in the analytical conformal bootstrap is the Regge limit \cite{Cornalba:2006xm,Cornalba:2007zb,Costa:2012cb}. The Regge limit of four-point correlation functions in CFTs is the conformal analogue of the limit of high centre-of-mass energy  at fixed impact parameter of scattering amplitudes in quantum field theory. Through AdS/CFT, it is thus relevant to study high-energy scattering in the bulk. In terms of cross-ratios, Regge limit resembles the behavior of Euclidean OPE. However, in Regge limit this happens after a branch-cut of the conformal block decomposition is crossed. Conformal Regge theory provides a resummation of the OPE in terms of families of operators called {\em Regge trajectories}~\cite{Costa:2012cb}. In doing so, one needed to assume a well-defined analytic continuation of OPE data to complex spin, which was later established by the inversion formula \cite{Caron-Huot:2017vep}.
This also puts the analytical conformal bootstrap using the lightcone limit on a firm footing, by showing that the large spin expansion is not asymptotic, but convergent. Regge limit has also been studied in the context of holography \cite{Cornalba:2006xm,Cornalba:2007zb,Antunes:2020pof}.
Recently, these ideas have been tested on several physical models, with great success \cite{Simmons-Duffin:2016wlq,Caron-Huot:2022eqs}. Regge limit and Regge behavior of correlation functions have also played an important role in imposing causality constraints~\cite{Afkhami-Jeddi:2016ntf,Costa:2017twz,Kulaxizi:2017ixa,Li:2017lmh,Caron-Huot:2021enk,Agarwal:2023xwl}.

This success in CFTs and the natural interest for multi-particle high-energy scattering calls for a deeper analysis of the Regge limit.
In this paper we start the discussion of the generalization of the Regge limit to higher-point functions, mostly focusing on the case of five-point functions. In the process we will also briefly review flat-space literature about Regge theory for higher-point amplitudes that has not been object of attention for a long time.

The outline of the paper is as follows.
In \cref{sec:FlatSpaceScattering}, we review the literature on the multi-Regge limit for S-matrix.
In \cref{sec:KinematicsFivePointCorrelators}, we discuss the setup of five-point correlation functions in conformal field theories.
We review the Euclidean OPE limit and the lightcone limit and contrast it with our proposal for Regge limit.
In \cref{sec:ReggeTheoryCFT}, we discuss analytic properties of the correlator as it is continued to Regge limit. Here we also consider the corresponding Mellin amplitude and Mellin partial wave and show that they produce the expected Regge behavior in position space.
In subsection \ref{subsec:SWtransform}, we consider the analytic continuation of the Mellin amplitude in three quantum numbers by means of Sommerfled-Watson transforms and resum the contribution of two leading Regge trajectories.
Finally, we conclude with some open directions in \cref{sec:conclusions}.

\section{Scattering in flat-space and Regge theory}
\label{sec:FlatSpaceScattering}
In this section, we review Regge theory for scattering amplitudes in flat space.
We begin by reviewing the key ingredients in the case of $2 \to 2$ scattering process in four-dimensional Lorentzian spacetime.
Then, we review the generalization to the case of higher-point scattering amplitudes.
It will serve as the main inspiration for the conformal Regge theory that we will consider later.

Let us restrict to the more familiar case of $2 \to 2$ scattering, where we define  the Mandelstam invariants as
\begin{equation}
  \label{eq:mandelstamst}
  s= - ( k_1+k_3)^2\,,\qquad t=-(k_1+k_2)^2\,,\qquad u=-(k_1+k_4)^2,
\end{equation}
with  $k_i$ the external incoming momenta.
The Regge limit corresponds to the high-energy limit of an amplitude, where $s\to \infty$ with fixed $t$.
Regge theory, on the other hand, comes in play to address the experimental observation that $s$-channel
processes exhibit a small $t$ peak whenever there are exchanges of particles or resonances in the $t$-channel.
One would like to understand this behavior by considering a partial wave decomposition of the amplitude.
Consider the scattering amplitude of four spinless particles with equal mass $m$ in the $t$-channel decomposition
\begin{align}
  \label{eq:4ptpartialwavedef}
  A(s,t)=\sum_{J=0}^{\infty} (2J+1) a_{J}(t) P_{J}(z)\,,
\end{align}
where $z=\cos \theta=1+ 2s/(t-4m^2)$ and
$P_J(z)$ is a Legendre polynomial of first kind of degree $J$. $\theta$ denotes the $t$-channel scattering angle and $J$ is the angular momentum of the exchanged particles.
This series converges in the $t$-channel physical region, namely the region $t>4m^2$ and $s<0$, and therefore is not reliable to study the large $s$ limit.
Note that the large $s$ limit of a spin $J$ partial wave behaves as $ s^{J} $.
The infinite sum over $ J $ gets more and more contributions from the higher $ J $ partial waves, in this limit.
Regge theory provides a rewriting of~(\ref{eq:4ptpartialwavedef}) in a form that can be analytically continued to this large $s$ region.
This is done by complexifying angular momentum, extending Regge's idea~\cite{Regge:1959mz}, and allows to resum the contribution of a family of
particles that correctly predict the observed large s behavior.

Naturally, one would like to extend Regge theory to multi-particle exchange amplitudes. The analytic structure of these amplitudes is less well understood than the four-point analogue.
The increasing number of Mandelstam invariants turns this into a more technically-challenging goal, but there have been important contributions in the 70's that we now briefly review
for the case of five-point amplitudes (see \cite{Goddard:1971fq, White:1972sc,White:1972rq,White:1973ola,White:1976qm,Brower:1974yv,Weis:1972tbu,Abarbanel:1972ayr,Stapp:1982mq,PhysRev.176.2003, NotesOnMultiRegge} for more details).

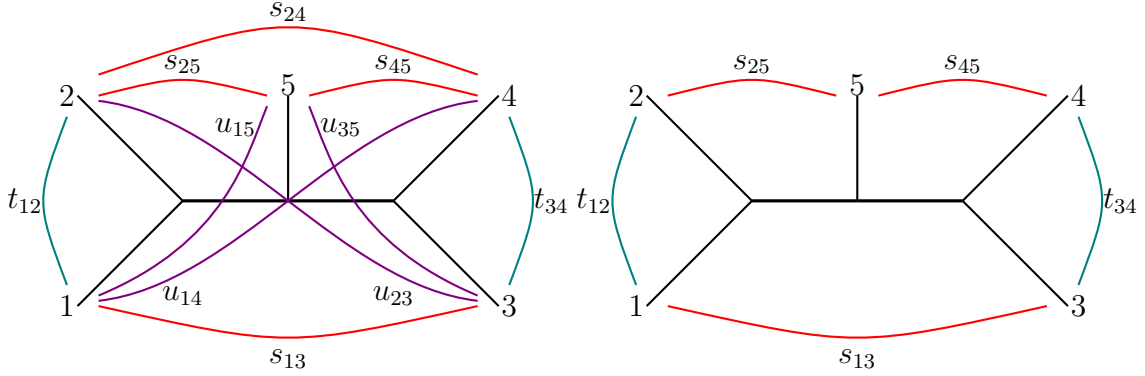
\begin{figure}[t]
  \centering
  \begin{subfigure}[b]{0.45\textwidth}
    \centering
    \begin{tikzpicture}[scale=1.4]

      \draw[thick] (-1,1)--(0,0);
      \draw[thick] (-1,-1)--(0,0);
      \draw[very thick] (0,0)--(1,0);
      \draw[thick] (1,1) -- (1,0);
      \draw[very thick] (1,0)--(2,0);
      \draw[thick] (2,0)--(3,1);
      \draw[thick] (2,0)--(3,-1);

      \node at (-1.1, 1) {2};
      \node at (-1.1, -1) {1};
      \node at (3.1, 1.0) {4};
      \node at (3.1, -1.0) {3};
      \node at (1, 1.1) {5};

      \draw[teal, thick] (-1.1,0.8)..controls (-1.4,0)..(-1.1,-0.8);
      \draw[teal, thick] (3.1,0.8)..controls (3.4,0)..(3.1,-0.8);
      \draw[red, thick] (-0.8,-1.0)..controls (1.,-1.4)..(2.8,-1);
      \draw[red, thick] (-0.8,1)..controls (0.,1.2)..(0.8,1.);
      \draw[red, thick] (1.2,1)..controls (2.,1.2)..(2.8,1.);
      \draw[red, thick] (-0.8,1.2)..controls (1.,1.8)..(2.8,1.2);
      \draw[violet,thick] (-0.8,-0.9)..controls (0.4,-0.4) and (0.6,0.4)..(0.8,0.9);
      \draw[violet,thick] (2.8,-0.9)..controls (1.6,-0.4) and (1.4,0.4)..(1.2,0.9);
      \draw[violet,thick] (2.8,-0.95)..controls (1.5,-0.8) and (0.5,0.8)..(-0.8,0.95);
      \draw[violet,thick] (2.8,0.95)..controls (1.5,0.8) and (0.5,-0.8)..(-0.8,-0.95);

      \node at (-1.5,0) {$t_{12}$};
      \node at (3.5,0) {$t_{34}$};
      \node at (1.,-1.5) {$s_{13}$};
      \node at (0, 1.3) {$s_{25}$};
      \node at (2,1.3) {$s_{45}$};
      \node at (1,1.8) {$s_{24}$};
      \node at (0,-0.9) {$u_{14}$};
      \node at (2,-0.9) {$u_{23}$};
      \node at (0.5,0.7) {$u_{15}$};
      \node at (1.5,0.7) {$u_{35}$};

    \end{tikzpicture}
  \end{subfigure}
  \begin{subfigure}[b]{0.45\textwidth}
    \centering
    \begin{tikzpicture}[scale=1.4]

      \draw[thick] (-1,1)--(0,0);
      \draw[thick] (-1,-1)--(0,0);
      \draw[very thick] (0,0)--(1,0);
      \draw[thick] (1,1) -- (1,0);
      \draw[very thick] (1,0)--(2,0);
      \draw[thick] (2,0)--(3,1);
      \draw[thick] (2,0)--(3,-1);

      \node at (-1.1, 1) {2};
      \node at (-1.1, -1) {1};
      \node at (3.1, 1.0) {4};
      \node at (3.1, -1.0) {3};
      \node at (1, 1.1) {5};

      \draw[teal, thick] (-1.1,0.8)..controls (-1.4,0)..(-1.1,-0.8);
      \draw[teal, thick] (3.1,0.8)..controls (3.4,0)..(3.1,-0.8);
      \draw[red, thick] (-0.8,-1.0)..controls (1.,-1.4)..(2.8,-1);
      \draw[red, thick] (-0.8,1)..controls (0.,1.2)..(0.8,1.);
      \draw[red, thick] (1.2,1)..controls (2.,1.2)..(2.8,1.);

      \node at (-1.5,0) {$t_{12}$};
      \node at (3.5,0) {$t_{34}$};
      \node at (1.,-1.5) {$s_{13}$};
      \node at (0, 1.3) {$s_{25}$};
      \node at (2,1.3) {$s_{45}$};
    \end{tikzpicture}
  \end{subfigure}
  \caption{The ten two-body Mandelstam invariants of a five-point scattering amplitude (left) and our choice of five independent ones (right).}
  \label{fig:mandelstaminv5pt}
\end{figure}

As represented in figure \ref{fig:mandelstaminv5pt} (left), one can define  ten two-body Mandelstam invariants for a five-point function,
in an analogous way to the $2 \to 2$ scattering definition (\ref{eq:mandelstamst}), i.e.
\begin{equation}
  s_{ij}=-(k_i+k_j)^2\,,
\end{equation}
where $k_i$ are again external incoming momenta. Similarly, we define $t_{ij}$- and $u_{ij}$-type invariants.
Since not all the invariants are independent, we shall choose
the following five as independent, $s_{13},s_{25},s_{45},t_{12},t_{34}$, as shown in figure \ref{fig:mandelstaminv5pt} (right).
We will be focusing on the double Regge limit where $s_{25}, s_{45}\to \infty$, and necessarily $s_{13}\to \infty$, while $t_{12}$ and $t_{34}$ are held fixed.
It is also possible to define a single Regge limit by considering either $s_{25}\to \infty$ with $s_{13}/s_{25}$ also fixed, or $s_{45}\to \infty$ with $s_{13}/s_{45}$ fixed.\footnote{Another interesting limit to consider is the helicity-pole limit where $s_{25}\to \infty$ with $s_{13}/s_{25}\to \infty$ while $t_{12}, t_{34}$ and $s_{45}$ are fixed (or the one where  the roles of $s_{25}$ and $s_{45}$ are swapped). This limit is experimentally accessible in inclusive cross-sections~\cite{PhysRevLett.26.675}. It is also possible to consider the limit $s_{13}\to \infty$ with all the other Mandelstam invariants kept fixed.}

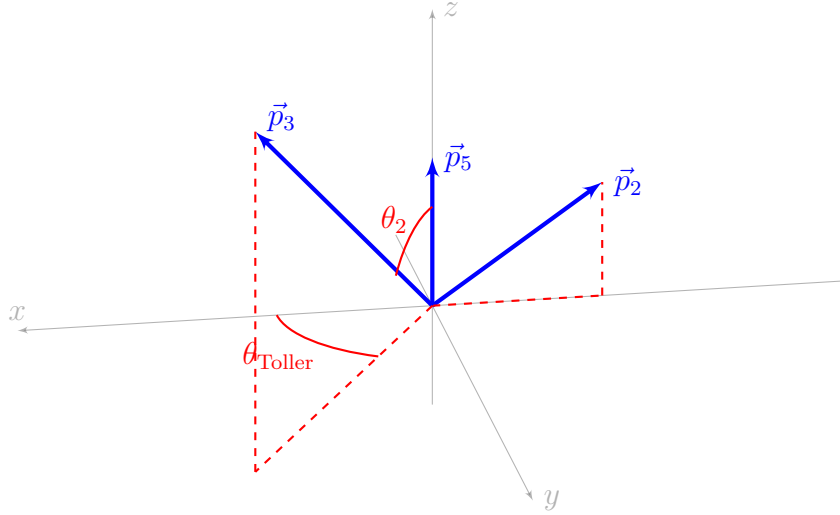
\begin{figure}[t!]
  \centering
  \tdplotsetmaincoords{70}{170}
  \begin{tikzpicture}
    [scale=1.4,tdplot_main_coords, axis/.style={->,-latex',gray, opacity=0.6},
      vector guide/.style={dashed,red,thick},angle/.style={red,thick}]
    \draw[axis] (0,0,-1)--(0,0,3) node[anchor=west]{$z$};
    \draw[axis] (0,-2,0)--(0,5.5,0) node[anchor=west]{$y$};
    \draw[axis] (-4,0,0)--(4,0,0) node[anchor=south]{$x$};

    \draw[blue,ultra thick,->,-latex'] (0,0,0)--(0,0,1.5) node[anchor=west, blue] {$\vec{p}_{5}$};
    \tdplotsetcoord{P2}{2}{55}{180};
    \draw[blue,ultra thick,->,-latex'](0,0,0)--(P2) node[anchor=west, blue] {$\vec{p}_{2}$};
    \draw[vector guide] (0,0,0)--(P2xy);
    \draw[vector guide] (P2xy)--(P2);

    \tdplotsetcoord{P3}{6}{55}{60};
    \draw[blue,->,ultra thick,-latex'](0,0,0)--(P3) node[anchor=west,yshift=+2mm, blue] {$\vec{p}_{3}$};
    \draw[vector guide] (0,0,0)--(P3xy);
    \draw[vector guide] (P3xy)--(P3);

    \tdplotdrawarc[angle]{(0,0,0)}{1.5}{0}{60}{anchor=east,xshift=+2mm,yshift=-2mm}{$\theta_{\text{Toller}}$};
    \tdplotsetthetaplanecoords{55}
    \tdplotdrawarc[tdplot_rotated_coords,angle]{(0,0,0)}{1}{0}{55}{anchor=south east, xshift=+1mm,yshift=-1.5mm}{$\theta_{2}$};
  \end{tikzpicture}
  \caption{Scattering process shown in the resting frame of exchanged momentum $q_2$. This defines the angles $\theta_2$ and $\theta_{\text{Toller}}$. $ \theta_1 $ is defined analogously in the rest frame of exchanged momentum $q_1$.}
  \label{fig:anglesframeq2}
\end{figure}

Let us consider the partial-wave decomposition of an amplitude $ A =  A(s_{25}, s_{45},\eta; t_{12}, t_{34}) $ of five identical massive particles in
the  $t_{12}, t_{34}$ channels,
\begin{align}
  \label{eq:partialwavet12t34}
  A
  =\sum_{m=-\infty}^{\infty}\sum_{J_{1,2}=|m|}^{\infty} \prod_{i=1}^{2} (2 J_i+1) a_{J_1,J_2,m}(t_{12},t_{34}) z^{m} d_{0 m}^{J_1}(\cos\theta_1)d_{m 0}^{J_2}(\cos\theta_2)
  \,,
\end{align}
where  $\eta\equiv s_{13}/(s_{25}s_{45})$ and $z\equiv e^{i \theta_{\text{Toller}}}$, as defined below. Here $m$ denotes the allowed helicities of exchanged particles.
We also use Wigner-$d$ functions which can be written in terms of Jacobi polynomials $\mathcal{P}_{J}^{\alpha,\beta}$ as
\begin{align}
  d_{m' m}^{J}(\cos\theta)=\left(\frac{\left(J+m\right)!\left(J-m\right)!}{\left(J+m'\right)!\left(J-m'\right)!}\right)^{\frac{1}{2}}\left(\sin \frac{\theta}{2}\right)^{m-m'}\left(\cos \frac{\theta}{2}\right)^{m+m'}\mathcal{P}_{J-m}^{m-m',m+m'}(\cos \theta)\,,
\end{align}
with
\begin{align}
  \mathcal{P}_{J}^{\alpha,\beta}(z)=\sum_ {n=0}^{J} \frac{(-1)^n}{n!} \frac{J!}{(J-n)!} \frac{\Gamma(n+\alpha+1)\Gamma(J-n+\beta+1)}{\Gamma(J+\alpha+\beta+2)} z^{n+\alpha} \left(1-z\right)^{J-n+\beta}
  .\end{align}
The scattering angles $\theta_1, \theta_2$ and $\theta_{\text{Toller}}$ have a clear physical meaning - see e.g.~\cite{Brower:1974yv}.
Consider the scattering process shown in the figure~\ref{fig:mandelstaminv5pt} with the exchanged momenta $q_1^2=t_{12}$ and $q_2^2=t_{34}$.
Firstly, we lump together particles $3,4$ and treat them as a single particle of momentum $q_2$.
The angle $\theta_1$ is defined as the scattering angle of the process $12\to 5 (34)$ which happens in a single plane in the center of mass frame.
Secondly, we consider the rest frame of the exchanged momentum $q_2$.
We denote the three momentum of particle-$i$ by $\vec{p}_i$.
As depicted in figure~\ref{fig:anglesframeq2}, we can choose a coordinate system where $\vec{p}_5$ is aligned with the $z$-axis and $\vec{p}_2$ lies somewhere in $xz$-plane.
We define $\theta_2$ as the zenith-angle of $ \vec{p}_3 $, whereas $\theta_{\text{Toller}}$ is the azimuth angle. Alternatively, the Toller angle can   be thought of as the angle
between the two scattering planes corresponding to the $q_1$ and $q_2$ rest frames, respectively.

The scattering angles are related to the Mandelstam invariants in a nontrivial way - see e.g. ~\cite{Weis:1972tbu,White:1973ola},
\begin{align}
  \label{eq:sijtoangles}
    & s_{25}= 2m^2+\frac{1}{2}\left(t_{34}-m^2-t_{12}\right)+\frac{1}{2}\left(\frac{(t_{12}-4m^2)\lambda(t_{12},t_{34},m^2)}{t_{12}}\right)^{1/2} \cos\theta_1\,,          \nonumber                  \\
    & s_{45}= 2m^2+\frac{1}{2}\left(t_{12}-m^2-t_{34}\right)+\frac{1}{2}\left(\frac{(t_{34}-4m^2)\lambda(t_{12},t_{34},m^2)}{t_{34}}\right)^{1/2} \cos\theta_2\,,           \nonumber                 \\
    & s_{13}=2m^2+\frac{1}{4}\left(m^2-t_{12}-t_{34}\right)+\frac{1}{4}\left(\frac{(t_{12}-4m^2)\lambda(t_{12},t_{34},m^2)}{t_{12}}\right)^{1/2} \cos\theta_1                                         \\
  + & \frac{1}{4}\left(\frac{(t_{34}-4m^2)\lambda(t_{12},t_{34},m^2)}{t_{34}}\right)^{1/2} \cos\theta_2+\frac{1}{4}\left(\frac{(t_{12}-4m^2)(t_{34}-4m^2)}{t_{12} t_{34}}\right)^{1/2}\times\nonumber \\
    & \left(m^2-t_{12}-t_{34}\right)\cos \theta_1 \cos \theta_2-\frac{1}{2}\left((t_{12}-4m^2)(t_{34}-4m^2)\right)^{1/2}\sin \theta_1\sin \theta_2 \cos\theta_{\text{Toller}}\,,\nonumber
\end{align}
where we use $\lambda(a,b,c)=a^2+b^2+c^2-2ab-2bc-2ca$. Note that here $m$ corresponds to the mass of the exchanged particles and should not be confused with helicity. We believe the context makes clear which one we refer to.
We emphasize that only $s_{13}$ depends on $\theta_{\text{Toller}}$.
Moreover, in the double Regge limit,
\begin{align}
  \label{eq:etaandToller}
  \eta\sim \frac{2(t_{12} t_{34})^{1/2}\cos \theta_{\text{Toller}}+m^2-t_{12}-t_{34}}{\lambda(t_{12},t_{34},m^2)}\,,
\end{align}
independently of $\theta_1$ and $\theta_2$.
This map suffers from some kinematical singularities in terms of the variables $t_{12}, t_{34}$.
These will be extracted from the possible types of singularities that we study below and we focus only on dynamical singularities.

To explore the double Regge region from the partial wave decomposition~(\ref{eq:partialwavet12t34}), we need a well-defined analytic continuation of the amplitude.
In contrast to the $2\to2$ scattering, for multiparticle scattering besides considering complex angular momentum, one also needs to account for helicity  dependence.
As stressed in~\cite{Goddard:1971fq, White:1972rq, White:1972sc, White:1973ola},  the analytic continuation of the amplitude to complex helicity values is also required.
The proper procedure for analytic continuation and its uniqueness deserve some comments.
Let us first review some concepts in the four-point case that will straightforwardly generalize to the five-point case we wish to analyze in more detail.
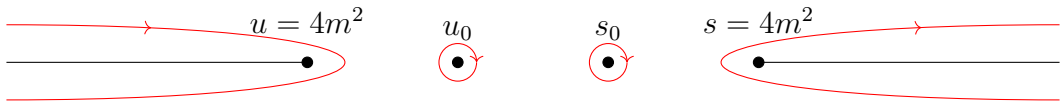
\begin{figure}[t!]
  \centering
  \begin{tikzpicture}
    \tikzset{middlearrow/.style={
          decoration={markings,
              mark= at position 0.8 with {\arrow{#1}} ,
            },
          postaction={decorate}
        }
    }
    \draw[] (-8,0)--(-4,0);
    \filldraw[black] (-4,0) circle (2 pt)node[anchor=south, yshift=2.5mm]{${ u=4m^2}$};
    \draw[] (2,0)--(6,0);
    \filldraw[black] (2,0) circle (2 pt) node[anchor=south, yshift=2.5mm]{${ s=4m^2}$};
    \draw[] (5.5,2.)--(6.,2.);
    \draw[] (5.5,2.5)--(5.5,2.);
    \node at (5.75,2.25) {$s$};

    \path [draw=red, middlearrow={<}]
    (-8.0,-0.5)..controls (-2.,-0.5) and (-2.,0.5) ..(-8.0,0.5);
    \path [draw=red, middlearrow={>}]
    (6.0,-0.5)..controls (0.,-0.5) and (0.,0.5) .. (6.0,0.5);

    \filldraw[black] (-2.,0) circle (2 pt) node[anchor=south,yshift=1.5mm]{$u_{0}$};
    \filldraw[black] (0,0) circle (2 pt)node[anchor=south,yshift=1.5mm]{$s_{0}$};

    \draw[red,<-] (-1.75,0) arc (0:360:0.25);
    \draw[red,<-] (0.25,0) arc (0:360:0.25);

  \end{tikzpicture}
  \caption{Singularities of $A(s,t)$ in the $s$ complex-plane at fixed $t$.}
  \label{fig:cutssplane}
\end{figure}

We assume that a $2\to2$ scattering amplitude has only singularities with dynamical origin. Namely, we only consider poles associated with bound states and branch-cuts starting at physical thresholds for particle production.
\footnote{In particular we ignore the possible existence of anomalous thresholds. However, as long as these lie on the real axis and the analytic structure resembles figure~\ref{fig:cutssplane} with a different branch point for some fixed $t$, the discussion that follows remains valid.\label{footanom4pt}}

 In figure~\ref{fig:cutssplane} we depict these singularities at fixed $t$ in the complex $s$ plane. We assume that the following dispersion relation at fixed $t$
\begin{align}
  A(s,t) & =\frac{1}{2\pi i} \left(
  \int_{0}^{+\infty} \!\!\! ds' \, \frac{\text{Disc}_{s}(s',t,u')}{s'-s} + \int_{0}^{+\infty} \!\!\! du' \,\frac{\text{Disc}_{u}(s',t,u')}{u'-u} \right) =A_{R}(s,t)+A_{L}(u,t)\,.
\end{align}
Here, we have extended the notion of discontinuities $\text{Disc}_{s}$ and $\text{Disc}_{u}$ to include the contributions of bound-state poles.\footnote{We  assumed that no subtractions were needed in order to neglect contributions from arcs at infinity from the Cauchy integral that gives rise to the dispersion relation. If this is not the case, one should proceed considering instead a subtracted amplitude.}
We have also defined $A_{R}$ and $A_{L}$ with respect to each of the terms.
As it is clear from the definition, each of the terms has only cut contributions along one of the directions in the $s$ complex plane.
This fact is crucial in performing the analytic continuation of the amplitude with good large $J$ behavior which happens to be unique as guaranteed by Carlson's theorem.
It is also useful to define the {\em signatured amplitude}~\footnote{The reader might be familiar with an equivalent decomposition of the full amplitude in terms of even and odd angular momentum contributions.
These are composed of signatured amplitudes.
Indeed we have $A^{\text{even}}=A^{+}(s,t)+A^{+}(-s,t)$ and $A^{\text{odd}}=A^{-}(s,t)-A^{-}(-s,t)$, where we use $u\sim-s$ in Regge limit.}
\begin{align}
  \label{eq:signaturedef}
  A^{\delta}(s,t)=\frac{1}{2}\big(A_{R}(s,t)+\delta\, A_{L}(s,t)\big) ,
\end{align}
where $\delta=\pm 1$.
Note that we replace $u$ by $s$ in $A_{L}$ ensuring that there are only cuts in a single direction.
The full amplitude can be easily reconstructed from the signatured amplitudes as
\begin{align}
  A(s,t)=\sum_{\delta=\pm 1}\left(A^{\delta}(s,t)+\delta\, A^{\delta}(u,t)\right) .
\end{align}
In what follows, we assume that the signatured amplitudes have the same analytic structure as the full amplitude.
This assumption greatly simplifies the discussion of dynamical singularities of partial-wave amplitudes.
We are entitled to consider the partial wave expansion of the signatured amplitude
\begin{align}
  \label{eq:partialwavesignature}
  A^{\delta}(s,t)=\sum_{J=0}^{\infty} (2J+1) a_{J}^{\delta}(t) P_{J}(z)\,.
\end{align}
Using the orthogonality of Legendre polynomials $P_J$ and (\ref{eq:signaturedef}), we can write
\begin{align}
  \label{eq:partialwavesignatureHelicity}
  a_{J}^{\delta}(t)=\frac{1}{4\pi i}\int_{z_0}^{+\infty} dz'\Big( \text{Disc}_{s}A_{R}(s',t)+\delta\,\text{Disc}_{s}A_{L}(s',t)\Big) Q_J(z')\,,
\end{align}
where $z_0$ is the branch point of the discontinuity and
$Q_J$ is the Legendre polynomial of the second kind  given by
\begin{align}
  Q_J(z) = \int_{-1}^{1} d\zeta \,\frac{P_J(\zeta)}{z-\zeta} \,.
\end{align}

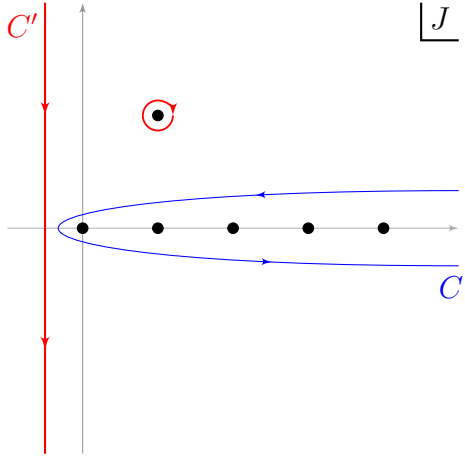
\begin{figure}[t]
  \centering
  \begin{tikzpicture}
    \tikzset{middlearrow/.style={
          decoration={markings,
              mark= at position 0.77 with {\arrow{#1}} , mark=at position 0.25 with {\arrow{#1}},
            },
          postaction={decorate}
        }
    }
    \draw[->,gray, opacity=0.6,-latex'] (-1.0, 0.0)--(5.0,0.0);
    \draw[->,gray, opacity=0.8,-latex'] (0.0, -3.0)--(0.0,3.0);

    \filldraw[black] (0.0,0.0) circle (2pt);
    \filldraw[black] (1.0,0.0) circle (2pt);
    \filldraw[black] (2.0,0.0) circle (2pt);
    \filldraw[black] (3.0,0.0) circle (2pt);
    \filldraw[black] (4.0,0.0) circle (2pt);

    \draw[thick, black] (4.5,3.0)--(4.5,2.5)--(5.0,2.5) node[midway,anchor=south]{$J$};
    \path [draw=red,thick, middlearrow={latex'}] (-0.5, 3.0)--(-0.5,-3.0);
    \node[red] at (-0.8,2.7) {$C'$};
    \path [draw=blue, middlearrow={latex'}] (5.0, 0.5)..controls (-2.1,0.5) and (-2.1,-0.5)..(5.0,-0.5) node[blue,below,xshift=-1mm]{$C$};

    \filldraw[black] (1.0,1.5) circle (2pt);
    \draw[line width=0.7, red,<-,latex'-] (1.2, 1.5) arc (0:360:0.2);
  \end{tikzpicture}
  \caption{Contour integrals for the Sommerfeld-Watson transform for the four particle scattering in the $J$-complex plane. As one deforms the contour from $C$ to $C'$ one has to consider the contribution from dynamical singularities which here we assume to be a Regge pole.
  }
  \label{fig:Jplane4pt}
\end{figure}
Using the symmetry $P_J(z)=(-1)^{J}P_J(-z)$, we perform a Sommerfeld-Watson transform of~(\ref{eq:partialwavesignature})
\begin{align}
  A^{\delta}(s,t)=\int_{C} \frac{dJ}{2\pi i} \frac{\pi}{\sin\pi J} \, (2J+1)a^{\delta}(J,t)P_{J}(-z) \,,
\end{align}
where $C$ is the closed contour depicted in figure~\ref{fig:Jplane4pt}. Due to the good large $J$ behavior of the partial-wave $P_{J}$, one can continuously deform the contour from $C$ to $C'$, as
shown in the same figure.
We should account for all the possible singularities that   may be encountered during this deformation.
In this paper, we always assume that these are pole type singularities \footnote{Other type of singularities like Regge cuts and nonsense-wrong-signature-fixed poles also exist. Moreover, singularities can also appear in a multiplicative manner but we neglect this scenario here for simplicity. The interested reader can find a discussion on those in~\cite{Weis:1972tbu} and references thereafter.}
\begin{align}
  a^{\delta}(J,t)\simeq \frac{\beta(t)}{J-\alpha(t)} \,,
\end{align}
where $\alpha(t)$ is a Regge trajectory and $\beta(t)$ is regular in $t$.
Regularity follows from the assumption that $A^{\delta}(s,t)$ has the same analytic structure of the full amplitude $A(s,t)$.
We also use the fact that Steinmann relations~\cite{SteinmannThesis} impose the latter to have no double discontinuity in $s$ and $t$.
At large $s$, we keep the rightmost pole as it gives the leading contribution and write
\begin{align}
  A^{\delta}(s,t)\sim \frac{1}{2\pi i}(-s)^{\alpha(t)} \Gamma\big(-\alpha(t)\big)\beta(t)\,,
\end{align}
where we   absorbed nonsingular factors into the definition of $\beta(t)$.

For the five-particle case  we consider a similar analytic structure of the amplitude in the $s_{25}, s_{45}$ and $s_{13}$ complex planes as in the four-particle case.~\footnote{Generically one expects anomalous thresholds to exist in multipoint amplitudes. Here, however, we consider the simpler case where they don't appear. The same is done in the literature we are briefly reviewing (see for instance, \cite{Goddard:1971fq, White:1972rq, White:1972sc, White:1973ola} and section 1.4 of~\cite{Brower:1974yv} where there is brief discussion about anomalous thresholds) and the counting of necessary signatured amplitudes follows from this assumption. It would be interesting to understand how this counting is (or not) affected by the existence of anomalous thresholds and how the partial-wave coefficients can be written as analytic functions of spin and helicity in that case. Moreover, it would be relevant to understand if the existence of anomalous thresholds indeed alters the asymptotic behaviour of the amplitude in the multi-Regge limit we describe here. Note, however, that similarly to the four-point case we commented in footnote \ref{footanom4pt}, if the anomalous thresholds lie on the real line and the analytic structure still resembles that of figure~\ref{fig:cutssplane}, we do expect the counting and the discussion of signatured amplitudes we review here to remain valid.}
We would like to decompose the full amplitude in terms of signatured amplitudes with only right-hand or left-hand cuts for each $s$-type invariant.
We immediately see that we have to consider $2^3=8$ possible signatures.
Indeed, one writes
\begin{align}
  \label{eq:the8signatures}
   & A(s_{ij}, t_{ij})=  \sum_{\delta_{ij}=\pm 1} \left\{
  \left(A^{\delta_{25}\delta_{45}\delta_{13}}(s_{25},s_{45},\eta, t_{12},t_{34})+\delta_{25} A^{\delta_{25}\delta_{45}\delta_{13}} (-s_{25},s_{45},\eta, t_{12},t_{34}) + \right.\right. \\
   & \left.\left.  \delta_{45} A^{\delta_{25}\delta_{45}\delta_{13}} (s_{25},-s_{45},\eta, t_{12},t_{34}) +
  \delta_{25}\delta_{45} A^{\delta_{25}\delta_{45}\delta_{13}} (-s_{25},-s_{45},\eta, t_{12},t_{34})\right)+ \delta_{13} (\eta\to-\eta)
  \right\}\,,
  \nonumber
\end{align}
where we make a slight abuse of notation by writing $u_{ij}\sim-s_{ij}$ as dictated by the double-Regge limit we are exploring.
Indeed, note that each of the signatured amplitudes $A^{\delta_{25}\delta_{45}\delta_{13}}$ is a function with only right-hand cuts in each of the invariants $s_{25},s_{45}$ and $s_{13}$.
While $\delta_{25}, \delta_{45}$ are the already familiar signatures associated with angular momenta in the $t_{12}$ and $t_{34}$ channels,  $\delta_{13}$ is a new signature related to the helicity at the
central vertex.

\begin{figure}[t]
  \centering
  \begin{tikzpicture}
    \draw[->,-latex',opacity=0.3] (0,0) -- (-3,0);
    \draw[->,-latex',opacity=0.3] (0,0) -- (0,3);
    \draw[->,-latex',opacity=0.3] (0,0) -- (0,-3);
    \draw[->,-latex',opacity=0.3] (0,0) -- (3,0);
    \draw[] (-1,3) -- (-1,2.5) -- (-1.5,2.5);
    \node[left] at (-1,2.8) {$ z $};
    \node[below right] at (2,0) {$ 1 $};
    \node[below right] at (4,-0.2) {$ a_> $};
    \node[below right] at (1,-0.2) {$ a_< $};
    \draw[line width=0.7,black ,-latex'] (2,0) to[out=90,in=0] (0,2) to[out=180,in=90] (-2,0) to[out=-90,in=180] (0,-2) to[out=0,in=-90] (2,0);
    \draw[thick,red,snake it] (0,0) -- (2,0);
    \draw[thick,blue,snake it] (2,0) -- (6,0);
  \end{tikzpicture}
  \caption{ Contour deformation in   $ z = e^{i \theta_{\mathtt{Toller}}} $ for doing the Froissart-Gribov continuation.
    The orthogonality relation holds on the black contours.
    We show the two different branch cuts corresponding to $ a_{\gtrless} $ discussed in (\ref{eq:aForsignaturedHelicity}).
  }
  \label{fig:tollerContour}
\end{figure}
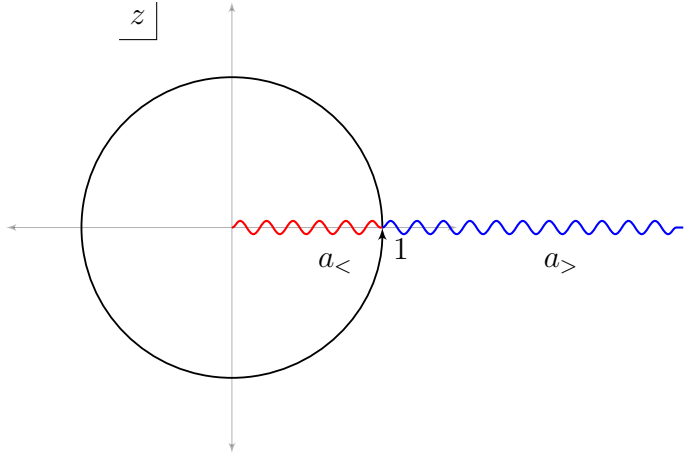

\begin{figure}[t]
  \centering
  \begin{subfigure}{0.5\textwidth}
    \centering
    \begin{tikzpicture}
      \tikzset{middlearrow/.style={
            decoration={markings,
                mark= at position 0.5 with {\arrow{#1}} ,
              },
            postaction={decorate}
          }
      }
      \draw[->,gray, -latex',opacity=0.6] (-1.0, 0.0)--(5.0,0.0);
      \draw[->,gray, -latex',opacity=0.8] (0.0, -3.0)--(0.0,3.0);

      \filldraw[black] (1.0,0.0) circle (2pt) node[below,yshift=-0.8mm]{${\scriptstyle m}$};
      \filldraw[black] (2.0,0.0) circle (2pt)node[below]{${\scriptstyle m+1}$};
      \filldraw[black] (3.0,0.0) circle (2pt)node[below]{${\scriptstyle m+2}$};
      \filldraw[black] (4.0,0.0) circle (2pt)node[below]{${\scriptstyle m+3}$};

      \draw[thick, black] (4.5,3.0)--(4.5,2.5)--(5.0,2.5) node[midway,anchor=south,yshift=-1mm]{$J_i$};

      \path [draw=red,thick, middlearrow={latex'}] (0.5,3.0)--(0.5,1.7);
      \path [draw=red,thick, middlearrow={latex'}] (0.5,1.3)--(0.5,-3.0);
      \path[draw=red,thick] (0.5,1.7)..controls (1.4,1.7) and (1.4,1.3).. (0.5,1.3);
      \filldraw[black] (1.0,1.5) circle (2pt);

    \end{tikzpicture}
  \end{subfigure}%
  \begin{subfigure}{0.5\textwidth}
    \centering
    \begin{tikzpicture}
      \tikzset{middlearrow/.style={
            decoration={markings,
                mark= at position 0.5 with {\arrow{#1}} ,
              },
            postaction={decorate}
          }
      }
      \draw[->,gray,-latex', opacity=0.6] (-4.0, 0.0)--(4.5,0.0);
      \draw[->,gray, -latex',opacity=0.8] (0.0, -3.0)--(0.0,3.0);

      \filldraw[black] (0.0,0.0) circle (2pt);
      \filldraw[black] (1.0,0.0) circle (2pt);
      \filldraw[black] (2.0,0.0) circle (2pt);
      \filldraw[black] (3.0,0.0) circle (2pt);
      \filldraw[black] (4.0,0.0) circle (2pt);

      \draw[thick, black] (4.,3.0)--(4.,2.5)--(4.5,2.5) node[midway,anchor=south]{$m$};

      \path [draw=red,thick, middlearrow={latex'}] (-0.3,3.0)--(-0.3,1.9);
      \path [draw=red,thick, middlearrow={latex'}] (-0.3,1.1)--(-0.3,-3.0);
      \path[draw=red,thick] (-0.3,1.9)..controls (2.4,2) and (2.4,1).. (-0.3,1.1);

      \filldraw (1.2,1.5) circle(2pt);
      \filldraw (0.2,1.5) circle(2pt);
      \filldraw (-0.8,1.5) circle(2pt);
      \filldraw (-1.8,1.5) circle(2pt);
      \filldraw (-2.8,1.5) circle(2pt);
      \filldraw (-3.8,1.5) circle(2pt);
      \draw [decorate, very thick,
        decoration = {brace, mirror}] (0,-0.2) --  (4,-0.2);
      \node at (2.0,-0.6) {{\small Poles of} $ {\scriptstyle\Gamma(-m)}$};
      \draw [decorate, very thick,
        decoration = {brace,mirror}] (1.2,+1.7) --  (-4,1.7) node[midway,above]{\begin{tabular}{c} {\small Poles of}\\${\scriptstyle\Gamma(-J_i+m)}$\end{tabular}};
    \end{tikzpicture}
  \end{subfigure}
  \caption{Contour of integration in $J_i$ and $m$-complex planes when the respective variable is integrated first.
    Here, we only account for dynamical singularities given by Regge poles and ignore the existence of Regge cuts and fixed poles.
    Note that there are no dynamical singularities in the $m$-complex plane.}
  \label{fig:jandmplanescontours}
\end{figure}
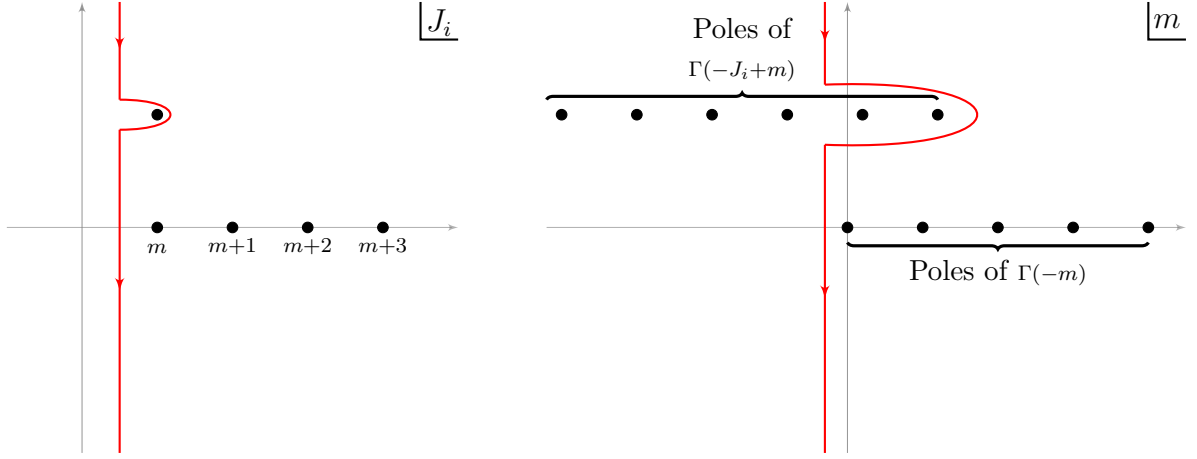

Following~\cite{Goddard:1971fq}, we first analyze the analytic continuation of helicity $m$ to complex values.
Inspired by the form of the partial-wave decomposition~(\ref{eq:partialwavet12t34}), one expects the following dispersion relation to hold in the
$z$-complex plane,
\begin{equation}
  A(s_{ij},\eta, t_{ij})=\frac{1}{2\pi i} \left(\int_{-\infty}^{-1-\epsilon}+\int_{-1+\epsilon}^{1-\epsilon}+\int_{1+\epsilon}^{+\infty}\right)\frac{\text{Disc}_{z}A(s_{ij},\eta', t_{ij})}{z'-z}\,dz'\,.
\end{equation}
To have a well-defined analytic continuation, we need to consider amplitudes with cuts either only on the right or only on the left half plane in the respective Mandelstam variable.
Thus, we consider signatured amplitudes as introduced in~(\ref{eq:the8signatures}).
We can write
\begin{align}
  A^{\delta_{13}}(z)=\sum_{m=-\infty}^{+\infty} a_m^{\delta_{13}} z^{m}\,,
\end{align}
where we suppress the dependence on labels or parameters that are irrelevant for this discussion.
Using the fact that signatured amplitudes are functions with only right-hand cuts,~\footnote{Note that taking $z\to-z$ is related with $\eta\to -\eta$ as one can see from~(\ref{eq:etaandToller}).} we have
\begin{align}
  a_m^{\delta_{13}}=\left(\frac{1}{2\pi i}\right)^2 \left(\int_{0}^{1-\epsilon}+\int_{1+\epsilon}^{+\infty}\right)\int_{|z|=1}\frac{\text{Disc}_{z}A^{\delta_{13}}(z')}{z'-z} \, z^{-m-1}dz' dz\,.
\end{align}
For $z'>1$ and $m<0$ the $z$-integral gives 0, while for $m\ge0$, it gives $z'^{-m-1}$. On the other hand, if $0<z'<1$ and $m\ge 0$, the residues at the two poles cancel and the integral yields $ 0 $, whereas for $m<0$ we find $-z'^{-m-1}$.
We can then define, as shown in figure \ref{fig:tollerContour},
\begin{align}
  \label{eq:aForsignaturedHelicity}
   & a_{>}^{\delta_{13}}(m)=\frac{1}{2\pi i} \int_{1+\epsilon}^{\infty}(z')^{-m-1}\text{Disc}_{z}A^{\delta_{13}}(z') dz'\,, \\
   & a_{<}^{\delta_{13}}(m)=\frac{1}{2\pi i} \int_{0}^{1-\epsilon}(z')^{-m-1}\text{Disc}_{z}A^{\delta_{13}}(z')dz'\,,
\end{align}
where it is clear that $a_{>}^{\delta_{13}}$ has a good asymptotic behavior in the right half-plane in the complex $m$ variable, whereas $a_{>}^{\delta_{13}}$ does so on the left-half plane.
We can thus perform a Sommerfeld-Watson transform in $m$ and write
\begin{align}
  A^{\delta_{13}}(z) & =\sum_{m=0}^{\infty}a_{>}^{\delta_{13}}(m) z^{m}-\sum_{m=-\infty}^{-1}a_{<}^{\delta_{13}}(m) z^{m}\nonumber                                                               \\
                     & =\frac{1}{2\pi i}\int_{C_R}dm\frac{\pi a_{>}^{\delta_{13}}(m)(-z)^m}{\sin \pi m}-\frac{1}{2\pi i}\int_{C_L}dm\frac{\pi a_{<}^{\delta_{13}}(m)(-z)^m}{\sin \pi m}\nonumber \\
                     & =\frac{1}{2\pi i}\int_{C}dm\frac{\pi \left(a_{>}^{\delta_{13}}(m)+a_{<}^{\delta_{13}}(m)\right)}{\sin \pi m}(-z)^m\,,
\end{align}
where the contours $C_R, C_L$ and $C$ are shown in figure~\ref{fig:ccontoursmplane}. Recovering the previously suppressed dependence and parameters, we have
\begin{figure}[t]
  \centering
  \begin{tikzpicture}
    \tikzset{middlearrow/.style={
          decoration={markings,
              mark= at position 0.77 with {\arrow{#1}} , mark=at position 0.25 with {\arrow{#1}},
            },
          postaction={decorate}
        }
    }
    \draw[-latex',gray, opacity=0.6] (-5.0, 0.0)--(5.0,0.0);
    \draw[-latex',gray, opacity=0.8] (0.0, -3.0)--(0.0,3.0);

    \filldraw[black] (0.0,0.0) circle (2pt);
    \filldraw[black] (1.0,0.0) circle (2pt);
    \filldraw[black] (2.0,0.0) circle (2pt);
    \filldraw[black] (3.0,0.0) circle (2pt);
    \filldraw[black] (4.0,0.0) circle (2pt);
    \filldraw[black] (-1.0,0.0) circle (2pt);
    \filldraw[black] (-2.0,0.0) circle (2pt);
    \filldraw[black] (-3.0,0.0) circle (2pt);
    \filldraw[black] (-4.0,0.0) circle (2pt);

    \draw[thick, black] (4.3,3.0)--(4.3,2.5)--(5.0,2.5) node[midway,anchor=south]{$m$};
    \path [draw=red,thick, middlearrow={latex'}] (-0.5, 3.0)--(-0.5,-3.0);
    \node[red,left] at (-0.8,-2.7) {$C=-\frac{1}{2}+i \mathbb{R}$};
    \path [draw=blue,thick, middlearrow={latex'}] (5.0, 0.5)..controls (-2.1,0.5) and (-2.1,-0.5)..(5.0,-0.5) node[blue,below,xshift=-2mm]{$C_{R}$};
    \path [draw=blue,thick, middlearrow={latex'}] (-5.0, 0.5)..controls (0.8,0.5) and (0.8,-0.5)..(-5.0,-0.5) node[blue,below,xshift=+3mm]{$C_{L}$};
  \end{tikzpicture}
  \caption{Contour integrals for the Sommerfeld-Watson transform in the $m$-complex plane.}
  \label{fig:ccontoursmplane}
\end{figure}
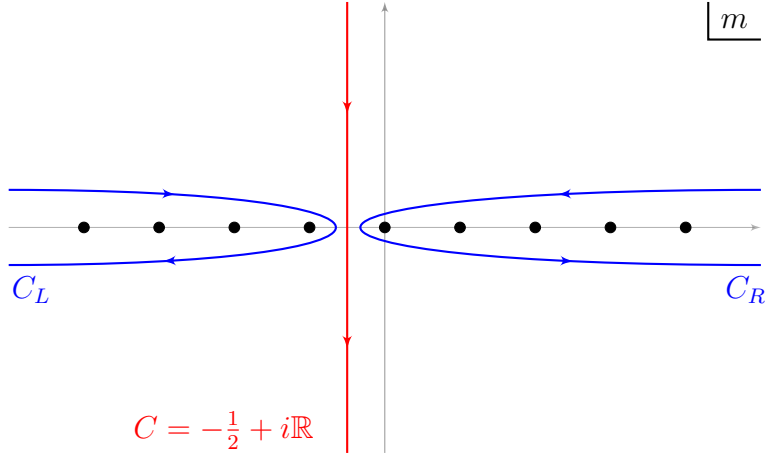
\begin{align}
  a_{\gtrless}^{\delta_{13}}(m) & = \sum_{J_1,J_2=m}^{\infty} \left(\prod_{i=1}^{2} (2 J_i+1)\right) a_{\gtrless,J_1,J_2,m}^{\delta_{25}\delta_{45}\delta_{13}}(t_{12},t_{34}) d_{0 m}^{J_1}(\cos\theta_1)d_{m 0}^{J_2}(\cos\theta_2)\,,
\end{align}
which only makes sense if we also analytically continue in the two angular momenta,
\begin{align}
  a_{\gtrless}^{\delta_{13}}(m)=\left(\prod_{i=1}^{2}\int_{C_i} \frac{dJ_i}{2\pi i} \frac{\pi(2 J_i+1)}{\sin \pi(J_i-m)}\right) a_{\gtrless}^{\delta_{25}\delta_{45}\delta_{13}}(J_1,J_2, m,t_{12},t_{34}) d_{0 m}^{J_1}(-z_1)d_{m 0}^{J_2}(-z_2)\,,
\end{align}
with contours $C_i$ as shown in figure~\ref{fig:jandmplanescontours} (left) and where $z_i=\cos\theta_i$.
This is a reasonable but non-trivial claim.
In fact, ~\cite{White:1972rq,White:1972sc} was only able to check a well-defined analytic continuation for a single angular momentum and helicity, but not the three simultaneously.
To the best of our knowledge, there is no derivation of the latter.
In the following, we assume that this defines a satisfactory analytic continuation of the signatured amplitude in terms of the scattering angles and of $t_{12}$ and $t_{34}$.
However, we would like to rewrite it in terms of the Mandelstam invariants alone.
This can be done by using the map   (\ref{eq:sijtoangles}).
To find the dependence on $s_{25}$ and $s_{45}$, we mimic the analysis of the four-particle case.
On the other hand, the $\eta$ dependence requires one more comment.
We assume that $A^{\delta_{13}}$ is an even function of the Toller angle and, in particular, a function of $\cos\theta_{\text{Toller}}$ (and thus invariant under $z\to 1/z$)\footnote{See~\cite{Weis:1972tbu} for whenever this is not the case.}.
This requirement follows from the realization that $\eta$ is an even function of $\theta_{\text{Toller}}$ and therefore only even functions of $\theta_{\text{Toller}}$ can be rewritten in terms of $\eta$.
This ends up imposing $a_{>}^{\delta_{13}}(m)=-a_{<}^{\delta_{13}}(-m)$ and justifies dropping the subscripts when we write
\begin{align}
  A^{\delta_{13}}(\eta)=\frac{1}{2\pi i}\int_{C}dm\,\frac{\pi a^{\delta_{13}}(m)}{\sin \pi m}(-\eta)^m\,.
\end{align}
Note that, as we write $z$ in terms of $\eta$, we redefine what we mean by $a^{\delta_{13}}$.~\footnote{In particular, as commented before, there are kinematical singularities in the map that we shall ignore when we discuss dynamical singularities in $a^{\delta_{13}}(m)$.}
We can summarize the discussion on analytic continuations of five-particle amplitudes by writing
\begin{align}
  \label{eq:5ptreggeamplitude}
   & A^{\delta_{25}\delta_{45}\delta_{13}}(s_{25}, s_{45},\eta, t_{12}, t_{34})=\left(\frac{1}{2\pi i}\right)^{3}\int_C dm \left(  \prod_{i=1}^2   \int_{C_i} dJ_i  (2J_i+1)\right) \nonumber                             \\
   & \qquad\qquad \frac{\pi^3 d_{0 m}^{J_1}(-\cos\theta_1)d_{m 0}^{J_2}(-\cos\theta_2)(-\eta)^{m}}{\sin \pi m\sin \pi (J_1-m)\sin \pi(J_2-m)} \,a^{\delta_{25}\delta_{45}\delta_{13}}(J_1,J_2, m,t_{12},t_{34}) \nonumber \\
   & \qquad\qquad=\left(\frac{1}{2\pi i}\right)^{3}\int_C dm   \Gamma(-m)    \left(  \prod_{i=1}^2   \int_{C_i} dJ_i  (2J_i+1)\Gamma(-J_i+m) \right)                                                                      \\
   & \qquad\qquad (-s_{25})^{J_1-m}(-s_{45})^{J_2-m}(-s_{13})^{m}a^{\delta_{25}\delta_{45}\delta_{13}}(J_1,J_2, m,t_{12},t_{34})\,,\nonumber
\end{align}
where we used $\eta=s_{13}/(s_{25}s_{45})$ and
in the second equality $a^{\delta_{25}\delta_{45}\delta_{13}}(J_1,J_2, m,t_{12},t_{34})$ was redefined.

Under the assumption that the analytic continuation of the signatured amplitude has a good asymptotic behavior in $J_1, J_2$ and $m$ such that we can ignore arcs at infinity, we focus on possible singularities that one might encounter as we move the contours to the left.
In figure~\ref{fig:jandmplanescontours}, we draw both $m$ and $J_i$ complex planes when the respective variable is integrated first.
In particular, we show the possible singularities.
As before, we restrict our analysis to Regge-pole-type of singularities and we refer interested readers to~\cite{Weis:1972tbu,Brower:1974yv} for more details on other type of singularities.
One expects the singularities in $m$ to the left of contour and that determine the asymptotic behavior of the amplitude to be completely determined by the dynamical singularities in angular momenta. The reason for that is bi-folded. First, note that the
amplitude has the asymptotic behavior
\begin{align}
  (-s_{25})^{J_1-m}(-s_{45})^{J_2-m}(-s_{13})^{m}\,.
\end{align}
Generically, this expression has a nonzero double discontinuity in the partially-overlapping channel invariants, namely $s_{25}$ and $s_{45}$.
However, this is forbidden by Steinmann relations~\cite{SteinmannThesis}.
Therefore, it must be that either $J_1-m$ or $J_2-m$ is a non-negative integer after the capture of poles.
It then follows that, in this limit, helicity singularities are fully determined by angular momentum ones as
\begin{align}
  m=\alpha-N\,,
\end{align}
where $\alpha$ is the location of a dynamical singularity in $J_1$ or $J_2$ and $N$ is a non-negative integer.
In the above argument, we naturally assume that the asymptotic behavior is attained within a physical region for the amplitude.
It is conceivable, however, that such asymptotics do  not correspond to a physical behavior and thus the argument would require an extension of validity of Steinmann relations for those configurations.
The second reason concerns the special nature of the helicity quantum number.
The physical interpretation of dynamical singularities are associated with the existence of particles.
As helicity is not a good Lorentz invariant and does not classify particles, as mass and spin do, we do not expect dynamical singularities in $m$~\cite{Brower:1974yv, Abarbanel:1972ayr}.
Besides, these assumptions seem to work well with specific models~\cite{Brower:1974yv,Schwarz:1973yz} as we will see below.

We now focus on our particular case of interest, the contribution of two Regge poles $\alpha_1(t)$ and $\alpha_2(t)$ in the double Regge limit with
\begin{align}
  a^{\delta_{25}\delta_{45}\delta_{13}}(J_1,J_2, m,t_{12},t_{34})\approx \frac{\beta(m, t_{12},t_{34})}{\big(J_1-\alpha_1(t_{12})\big)\big(J_2-\alpha_2(t_{34})\big)}\,.
\end{align}
In the Regge limit we move the $C_1$ and $C_2$ contours to the left in~(\ref{eq:5ptreggeamplitude}) and capture the poles in complex angular momentum. The leading contributions come from the rightmost poles. We find
\begin{align}
  \label{eq:5ptreggetheory}
   & A^{\delta_{25}\delta_{45}\delta_{13}}(s_{25}, s_{45},\eta, t_{12}, t_{34}) \sim\frac{1}{2\pi i}\int_C dm(2 \alpha_1+1)(2 \alpha_2+1)\Gamma(-m) \Gamma(-\alpha_1+m)\Gamma(-\alpha_2+m)\nonumber \\
   & \qquad\qquad\qquad\qquad\qquad\qquad \times(-s_{25})^{\alpha_1-m}(-s_{45})^{\alpha_2-m}(-s_{13})^{m}\beta(m,t_{12},t_{34})\nonumber                                                            \\
   & \sim(-s_{25})^{\alpha_1}(-s_{45})^{\alpha_2} \left((-\eta)^{\alpha_1}\sum_{i}\Gamma(-\alpha_1+i)\Gamma(\alpha_1-\alpha_2-i)\beta(\alpha_1-i,t_{12},t_{34})\frac{\eta^{-i}}{i!}\right.          \\
   & \qquad\qquad\qquad\qquad\quad
  \left.+(-\eta)^{\alpha_2}\sum_{i}\Gamma(-\alpha_2+i)\Gamma(\alpha_2-\alpha_1-i)\beta(\alpha_2-i,t_{12},t_{34})\frac{\eta^{-i}}{i!}\right).\nonumber
\end{align}
From the first to second line we closed the $C$ contour to the left, capturing all the $\alpha_i$-dependent poles, and absorbed overall constants into $\beta$.
In particular, if we consider the limit $\eta=s_{13}/(s_{25}s_{45})\to \infty$, we can just keep the leading contribution
\begin{align}
  A^{\delta_{25}\delta_{45}\delta_{13}}(s_{25}, s_{45},\eta, t_{12}, t_{34}) & \sim (-s_{13})^{\alpha_1}(-s_{45})^{\alpha_2-\alpha_1}\Gamma(-\alpha_1)\Gamma(\alpha_1-\alpha_2)\beta(\alpha_1,t_{12},t_{34})\nonumber \\
                                                                             & +(-s_{13})^{\alpha_2}(-s_{25})^{\alpha_1-\alpha_2}\Gamma(-\alpha_2)\Gamma(\alpha_2-\alpha_1)\beta(\alpha_2,t_{12},t_{34})\,,
\end{align}
which clearly does  not have double discontinuities in $s_{25}$ and $s_{45}$, as follows from our construction.
Note that the apparent singularities in $\alpha_1=\alpha_2$ are just spurious, as they cancel each other.

There are many subtleties and unproven statements in deriving the Regge theory result~(\ref{eq:5ptreggetheory}), but the final form seems very reasonable in physical terms.
We can analyze these claims in   specific models.
We consider a dual resonance model of a five-particle amplitude in the so-called Bardakci-Ruegg representation~\cite{Bardakci:1969ue}
\begin{align}
  B_5 =  \int \frac{dx_1}{x_1}\, \frac{dx_2}{x_2}\, &
  x_1^{-\alpha(t_{12})} \left( 1-x_1 \right)^{-1-\alpha\left( s_{25} \right)}
  x_2^{-\alpha(t_{34})} \left( 1-x_2 \right)^{-1-\alpha\left( s_{45} \right)}
  \nonumber                                                                                                                                                                        \\
                                                    & \times \left( 1-x_1 x_2 \right)^{-\alpha\left( s_{13} \right) + \alpha\left( s_{25} \right) + \alpha\left( s_{45} \right) },
  \label{eq:BRrepresentation}
\end{align}
where the  integral ranges from $0$ to $1$ in $x_1$ and $x_2$. We defined   $\alpha(x)  = \alpha_0 + x$ with  $\alpha_0$  the intercept of the Regge trajectory.
As stated above, a single Regge limit happens when $ s_{25}$ (or  $s_{45}$),  $s_{13} \rightarrow \infty $ with their ratio fixed. In this limit, it can be shown  \cite{Brower:1974yv} that the region $ x_1 \approx 0 $ dominates in the integral~(\ref{eq:BRrepresentation}).
For the values $ 0<s_{25}/s_{13}<1 $, it can be shown that
\begin{align}
  B_5 =
  \left(
  -s_{13}
  \right)^{\alpha(t_{12})}
  \sum_{n=0}^{\infty} p_n \left( -\frac{s_{25}}{s_{13}} \right)^{n}
  +
  \left(
  -s_{13}
  \right)^{\alpha(t_{34})}
  \left(
  -s_{25}
  \right)^{\alpha(t_{12})-\alpha(t_{34})}
  \sum_{n=0}^{\infty} q_n \left( -\frac{s_{25}}{s_{13}} \right)^{n}
  ,\label{eq:singleReggeBR}
\end{align}
where
\begin{align}
  p_n (t_{12},t_{34}, s_{45}) & =
  \frac{\Gamma\big( n - \alpha(t_{12})\big)\, \Gamma( -n+t_{12}-t_{34})\, \Gamma\big( n - \alpha( s_{45} ) \big)}{\Gamma\big( t_{12}-t_{34}-\alpha( s_{45} ) \big) \, n!}\,,
  \\
  q_n (t_{12},t_{34}, s_{45}) & =
  \frac{\Gamma\big( n - \alpha(t_{34}) \big) \,\Gamma( -n+t_{34}-t_{12})\, \Gamma\big( n +t_{12}-t_{34}- \alpha( s_{45}) \big)}{\Gamma\big( t_{12}-t_{34}-\alpha( s_{45} ) \big) \, n!}\,.
  \label{eq:pqExpressions}
\end{align}
Note that there are no simultaneous singularities in the overlapping Mandelstam invariants.
This follows from the explicit expressions of $ p_n $ and $ q_n $.
The first term has power-law behavior in $ s_{13} $ and poles in $s_{45} $, while having no singularities in $ s_{25} $.
The second term, on the other hand, has power-law behavior in both $ s_{25} $ and $ s_{13} $ times a function without any singularities in $s_{45}$.
This is an instance of the Steinmann relations, which hold for the full amplitude.
The double Regge limit corresponds to taking a further limit $ s_{45} \rightarrow \infty $ with the ratio $\eta= s_{13}/\left( s_{25} s_{45} \right) $ fixed.
It leads to \cite{Brower:1974yv}
\begin{align}
  B_5 = \left( -s_{25} \right)^{\alpha(t_{12})} \left( -s_{45} \right)^{\alpha(t_{34})}
  \int_{-i \infty}^{i \infty}  \frac{dm}{2\pi i}  \, \Gamma\big( m- \alpha(t_{12}) \big)\Gamma\big( m- \alpha(t_{34}) \big)
  \Gamma (-m) \left( -\eta\right)^{m}\,,
  \label{eq:doubleReggeBR}
\end{align}
which is of the same form as~(\ref{eq:5ptreggetheory}).

With the knowledge of the multi-Regge limit in S matrix theory, we are now in a position to study the multi-Regge limit in conformal field theories.

\section{Kinematics of five-point conformal correlators}
\label{sec:KinematicsFivePointCorrelators}

Correlation functions of local primary operators in any conformal field theory can be written in terms of a simple prefactor, that absorbs the weight of external operators, and a  non-trivial function that depends on conformal invariant variables, usually called cross ratios, that contains all the dynamics of the correlator.
In this paper, we will be mostly focused in correlators involving five operators.
These depend on five different cross ratios through\footnote{This is the same number as independent Mandelstam invariants in flat space scattering amplitudes as reviewed in the previous section. The connection between correlation functions in conformal field theories and scattering amplitudes is more clear in Mellin space, as we shall see in the next section. }
\begin{align}
  \langle \mathcal{O}(x_1) \mathcal{O}(x_2) \mathcal{O}(x_3) \mathcal{O}(x_4) \mathcal{O}(x_5)  \rangle  = \frac{\left(\frac{x_{23}^2}{x_{13}^2}\right)^{\frac{\Delta_{12}}{2}} \left(\frac{x_{14}^2}{x_{13}^2}\right)^{\frac{\Delta_{34}}{2}}   }{(x_{12}^2)^{\frac{\Delta_1+\Delta_2}{2}}(x_{34}^2)^{\frac{\Delta_3+\Delta_4}{2}}} \left(\frac{x_{13}^2}{x_{15}^2x_{35}^2}\right)^{\frac{\Delta_5}{2}}\mathcal{G}(u_1\dots u_5)\,,
  \label{eq:fiveptcorrelatordef}
\end{align}
where $x_{ij}^2=(x_i-x_j)^2$, we used the shorthand notation $\Delta_{ij}\equiv \Delta_i-\Delta_j$ and the cross ratios are defined as
\begin{align}
  u_1= \frac{x_{12}^{2} x_{35}^{2}}{x_{13}^{2} x_{25}^{2}}\,,  \qquad\qquad u_{i+1} = u_{i}|_{x_{i} \rightarrow x_{i+1}}\,.
  \label{eq:crossRatiosuiFive}
\end{align}
It is worth emphasizing that this is just a particular choice of cross ratios which is obviously not unique.
For instance, $\tilde{u}_3\equiv u_3u_2$ would be as valid a choice as $u_3$.
The choice (\ref{eq:crossRatiosuiFive}) has the nice feature that the cross ratios can be defined by transforming the $x_i$ cyclically, {\em i.e.} $x_{i}\rightarrow x_{i+1}$.
This is particularly interesting when studying observables that are cyclically symmetric~\cite{Bercini:2020msp,Bercini:2021jti,Antunes:2021kmm}.

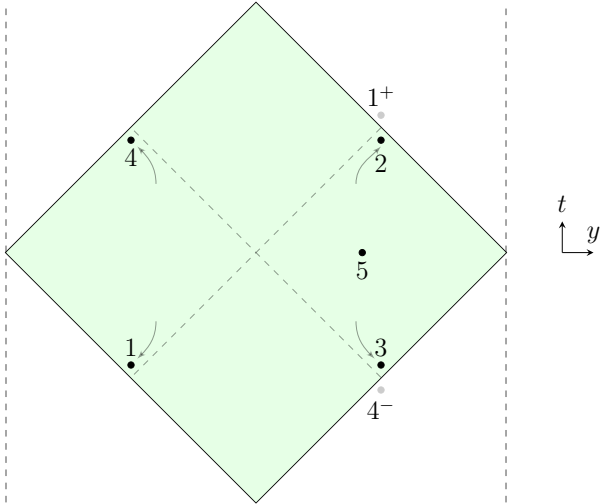
\begin{figure}[t!]

  \centering
  \begin{subfigure}[]{0.5\textwidth}
    \centering
    \resizebox{\columnwidth}{!}{
      \begin{tikzpicture}
        \draw[thick] (4,0) -- (0,4)  -- (-4,0) -- (0,-4)-- cycle;

        \fill[green!10,draw=none]  (4,0) -- (0,4)  -- (-4,0) -- (0,-4)-- cycle;
        \draw[dashed,opacity=0.4] (2,2) -- (-2,-2);
        \draw[dashed,opacity=0.4] (2,-2) -- (-2,2);

        \node[below] at (1.7,0) {$5$};
        \draw[fill,black](1.7,0) circle (0.05cm);

        \node[below] at (2,1.7) {$2$};
        \draw[fill,black](2,1.8) circle (0.05cm);
        \node[above] at (2,-1.8) {$3$};
        \draw[fill,black](2,-1.8) circle (0.05cm);
        \node[below] at (2,-2.2) {$4^{-}$};
        \draw[fill,gray,opacity=0.4](2,-2.2) circle (0.05cm);
        \node[above] at (-2,-1.8) {$1$};
        \draw[fill,black](-2,-1.8) circle (0.05cm);
        \node[above] at (2,2.2) {$1^{+}$};
        \draw[fill,gray,opacity=0.4](2,2.2) circle (0.05cm);
        \node[below] at (-2,1.8) {$4$};
        \draw[fill,black](-2,1.8) circle (0.05cm);

        \draw[->,opacity=0.4,-latex'] (1.6,1.1) to[out=90,in=225] (2,1.7);
        \draw[->,opacity=0.4,-latex'] (-1.6,1.1) to[out=90,in=-45] (-1.9,1.7);
        \draw[->,opacity=0.4,-latex'] (1.6,-1.1) to[out=270,in=135] (1.9,-1.7);
        \draw[->,opacity=0.4,-latex'] (-1.6,-1.1) to[out=270,in=45] (-1.9,-1.7);

        \draw[opacity=0.4,thick,dashed] (4,-4) -- (4,4);
        \draw[opacity=0.4,thick,dashed] (-4,-4) -- (-4,4);
        \draw[-stealth] (4.9,0) -- (4.9,0.5) node[anchor=south]{$ t $ };
        \draw[-stealth] (4.9,0) -- (5.4,0) node[anchor=south]{$ y $ };
      \end{tikzpicture}
    }
  \end{subfigure}
  \caption{
    We show our proposal for the Regge limit of the five-point correlator.
  }
  \label{fig:ReggeLimitVsOthers}
\end{figure}
In general, $\mathcal{G}(u_i)$ is an intricate function of the cross ratios with a complex analytic structure. One interesting question is, {\em what are the allowed singularities of a correlation function of five local operators and what is their physical meaning? }
This is a hard question that we will not try to answer here in full generality (see \cite{Maldacena:2015iua} for progress in this direction).
Instead, we shall focus on a particular singularity that is associated with the limit described in figure \ref{fig:ReggeLimitVsOthers} and that is similar to the Regge limit of scattering amplitudes reviewed in the previous section.
There are two other more common (and simpler) singularities, the Euclidean and lightcone OPE limits which will be relevant for the Regge limit analysis.
Indeed, it is possible to extract some information about these singularities from the conformal block decomposition of five points
\begin{align}
  \mathcal{G}(u_i)=  \sum_{k_1k_2,\ell}P_{k_1k_2 }^{\ell}G_{k_1k_2}^{\ell}(u_1,\dots u_5)\,,
  \label{eq:confblockDecom}
\end{align}
where $G_{k_1k_2}^{\ell}(u_1,\dots u_5)$ are conformal blocks in the channel $(12)$ and $(34)$, $P_{k_1k_2}^{\ell}$ are products of three-point coefficients (to be described in more detail in the following subsection) and the sum is over all primary operators.

In the following subsections, we will review and explore the Euclidean and lightcone singularities and introduce the Regge limit for five-point correlation functions.

\subsection{Euclidean limit}\label{subsec:Euclideanlimitsubsection}
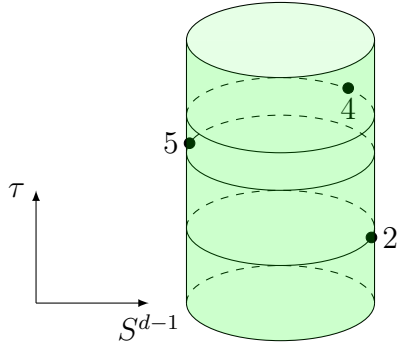
\begin{figure}[tbp]
  \centering

  \begin{tikzpicture}
    \draw (0,0) ellipse (1.25 and 0.5);
    \draw (-1.25,0) -- (-1.25,-3.5);
    \draw (1.25,-3.5) -- (1.25,0);
    \filldraw[black ] (1.25 * 0.72, 0.5 *0.72-1 ) circle (2pt);

    \node[below] at(1.25 * 0.72, 0.5 *0.72 -1 ){$4$};
    \filldraw[black ] (1.25 * -0.967251, 0.5 *0.253823 -1.5 ) circle (2pt);
    \node[left] at (-1.25 * 0.967251, 0.5 *0.253823 -1.5 ){$5$};

    \filldraw[black ] (1.25 * 0.967251, -0.5 *0.253823 -2.5 ) circle (2pt);
    \node[right] at (1.25 * 0.967251, -0.5 *0.253823 -2.5 ){$2$};
    \draw[->,-latex] (-3.25,-3.5) -- (-3.25,-2);
    \draw[->,-latex] (-3.25,-3.5) -- ( -1.75 ,-3.5);

    \node[left] at (-3.25,-2) {$ \tau $};
    \node[below] at (-1.75, -3.5) {$ S^{d-1} $};

    \draw (-1.25,-1.5) arc (180:360:1.25 and 0.5);
    \draw [dashed] (-1.25,-1.5) arc (180:360:1.25 and -0.5);
    \draw (-1.25,-3.5) arc (180:360:1.25 and 0.5);
    \draw [dashed] (-1.25,-3.5) arc (180:360:1.25 and -0.5);

    \draw (-1.25,-1.0) arc (180:360:1.25 and 0.5);
    \draw [dashed] (-1.25,-1.0) arc (180:360:1.25 and -0.5);

    \draw (-1.25,-2.5) arc (180:360:1.25 and 0.5);
    \draw [dashed] (-1.25,-2.5) arc (180:360:1.25 and -0.5);
    \fill [green , opacity = 0.2] (-1.25,0) -- (-1.25,-3.5) arc (180:360:1.25 and 0.5) -- (1.25,0) arc (0:180:1.25 and -0.5);
    \fill [green , opacity = 0.1] (-1.25,0) arc (180:0:1.25 and 0.5) -- (-1.25,0) arc (180:0:1.25 and -0.5);

  \end{tikzpicture}
  \caption{Position of points on the Euclidean cylinder. Two points $ 1 $ and $ 3 $, are at $ \tau =  -\infty $ and $ \tau = \infty $.}
  \label{fig:cylinder5point}
\end{figure}
The simplest limit in a CFT is when two operators are brought close to each other.
In this setup, the operator product expansion (OPE) is convergent and can be used safely.
The OPE is perhaps one of the most important properties of a CFT.
This feature tells that the product of two operators at distinct points can be replaced by a linear combination of operators
\begin{align}
  \mathcal{O}(x_1)\mathcal{O}(x_2) \approx \sum_{k} \frac{C_{12k}}{(x_{12}^2)^{\frac{\Delta_1+\Delta_2-(\Delta_k-J_k)}{2}}}
  F_k \big(x_{12}, D_{z},\partial_{x_1}\big)\mathcal{O}_k(x_1,z)\,,
  \label{eq:OPEequation}
\end{align}
where the sum runs over all primary operators, $C_{12k}$ are the OPE coefficients and $F_{k}$ is a differential operator that takes into account the contribution of descendants.
The auxiliary null variable $z$ is used to encode the open indices of a symmetric and traceless spin $J$ operator as
\begin{align}
  \mathcal{O}(x,z) \equiv z^{\mu_1}\dots z^{\mu_J} \mathcal{O}^{\mu_1\dots \mu_J}(x)\,,
\end{align}
while
\begin{align}
  D_{z^{\mu}} =\left( \frac{d}{2}-1+z\cdot\frac{\partial}{\partial z} \right)\frac{\partial}{\partial z^\mu}-\frac{1}{2} z^ \mu \frac{\partial^2}{\partial z \cdot \partial z}
  \label{eq:TodorovOPE}
\end{align}
is used to recover the information about the indices.
The exact form of $F_k$ can be determined from consistency of two- and three-point correlation functions of local operators.
It follows from a simple computation that, at the leading order and in the limit $x_2\rightarrow x_1$, the function $F_k$ is given by
\begin{align}
  F_k (x_{12}, D_{z},\partial_{x_1}) = \frac{(x_{12}\cdot D_{z})^{J_k}}{J_k!\left(\frac{d}{2}-1\right)_{J_k}} +\dots\,,
  \label{eq:OPEEuclideanLEading}
\end{align}
where $\dots$ represent  subleading terms.
One feature of this simple result is that it is evident that the limit is dominated by operators with lowest dimension $ \Delta_k $.
In particular, this determines the dominant contribution of a five-point conformal block  in the limits $x_{2}\rightarrow x_1$ and $x_4\rightarrow x_3$
\begin{align}
  \sum_{\ell} P_{k_1k_2}^{\ell}G_{k_1k_2}^{\ell}(u_1,\dots u_5) \approx \tfrac{C_{12k_1}C_{34k_2} (x_{12}\cdot D_{z})^{J_1}(x_{34}\cdot D_{z'})^{J_2}}{(x_{12}^2)^{\frac{J_{1}-\Delta_{k_1}}{2}} (x_{34}^2)^{\frac{J_{2}-\Delta_{k_2}}{2}}} \langle \mathcal{O}_{k_1}(x_1,z)\mathcal{O}_{k_2}(x_3,z') \mathcal{O}(x_5) \rangle\,.
  \label{eq:leadingtermOPEBoundaryCond}
\end{align}
Note that the double limit in the pair of points $(12)$ and $(34)$ was taken to reduce the correlator to a three-point function which is fixed by symmetry as
\begin{align}
   & \langle \mathcal{O}_{k_1}(x_1,z_1)\mathcal{O}_{k_2}(x_2,z_2) \mathcal{O}(x_3) \rangle  = \sum_{\ell=0}^{\textrm{min}(J_1,J_2)}\frac{C_{123}^{\ell} V_{123}^{J_1-\ell} V_{213}^{J_2-\ell} H_{12}^{\ell}}{(x_{12}^2)^{\frac{h_1+h_2-h_3}{2}} (x_{13}^2)^{\frac{h_1+h_3-h_2}{2}} (x_{23}^2)^{\frac{h_2+h_3-h_1}{2}}   } \,,
  \label{eq:threepontfunction}
\end{align}
where $h_i\equiv \Delta_i+J_i$ and
\begin{align}
   & H_{12} = (z_1\cdot x_{12})(z_2\cdot x_{12})- \frac{x_{12}^2  (z_1\cdot z_2)}{2}\,, \ \ \ \ \ V_{123} = \frac{(z_1\cdot x_{12} )x_{13}^2- (z_1\cdot x_{13})x_{12}^2}{x_{23}^2}\,.
\end{align}
It follows  from (\ref{eq:leadingtermOPEBoundaryCond}) that the constants $P_{k_1k_2}^{\ell}$ are given by
\begin{align}
  P_{k_1k_2}^{\ell} & =C_{12k_1}C_{34k_2}C_{k_1k_25}^{\ell} \,.
  \label{eq:OPE_coeffs}
\end{align}

Conformal blocks are complicated functions which are not known in closed form for general dimensions.
However, it is possible to compute them as an expansion around some limits.
One method to obtain them takes advantage of the fact that they are eigenfunctions of the conformal Casimir differential equation
\begin{align}
  \left(\mathcal{D}_{12} -c_{\Delta_{k_1},J_1}\right) G_{k_1k_2}^{\ell}=0\,,
\end{align}
with
\begin{align}
  c_{\Delta,J} =\Delta  (\Delta -d)+J (d+J-2)\,, \qquad
  \mathcal{D}_{12} = 2u_1^2\partial_{u_1}^2+\dots\,,
\end{align}
where $\dots$ represent other subleading terms.
We omitted an analogous equation in the $(34)$ channel that can be obtained using symmetry.

The cross ratios (\ref{eq:crossRatiosuiFive}) are not appropriate for all situations.
For instance, in the limit considered above where $x_2\rightarrow x_1$ and $x_4\rightarrow x_3$, one has
\begin{align}
  u_1,u_3\rightarrow 0\,, \ \ \ \  \ \ \ u_i\rightarrow 1 \ \ \ (i=2,4,5)\,,
\end{align}
which is insensitive to the angle at which the operators approach each other.
For this limit, it is preferable to use instead another set of cross ratios\footnote{We have decided to use slightly different angles as compared with \cite{Goncalves:2019znr} to make it appear more symmetric in the variables $u_i$. }\cite{Goncalves:2019znr}
\begin{align}
  \xi_1  =\frac{1-u_5}{2 \sqrt{u_1}}                                        \,,\qquad
  \xi_2  =\frac{1-u_4}{2 \sqrt{u_3} }                               \,,     \qquad
  \xi_3  =\frac{ u_2-1}{2 \sqrt{u_1} \sqrt{u_3}}\,,
  \label{eq:anglesdef}
\end{align}
which remain finite.
These are related to the angles just mentioned above.
The leading behavior, in the Euclidean OPE limit, of the five-point conformal block can be written in terms of these new cross ratios as
\begin{align}
   & G_{k_1k_2}^{\ell} = u_1^{\frac{\Delta_{k_1}}{2}}u_3^{\frac{\Delta_{k_2}}{2}}\mathcal{H}_{\ell}(\xi_i)      \,,
\end{align}
with
\begin{align}
  \mathcal{H}_{\ell}(\xi_i)=\prod_{i=1}^2\frac{1}{J_i!\left(\frac{d}{2}-1\right)_{J_i}} \frac{(x_{12}\cdot D_{z})^{J_1}(x_{34}\cdot D_{z'})^{J_2}}{(x_{12}^2)^{\frac{J_{k_1}}{2}} (x_{34}^2)^{\frac{J_{k_2}}{2}}} \frac{V_{135}^{J_1-\ell} V_{315}^{J_2-\ell} H_{13}^{\ell}}{(x_{13}^2)^{\frac{J_1+J_3}{2}} (x_{15}^2)^{\frac{J_1-J_2}{2}} (x_{35}^2)^{\frac{J_2-J_1}{2}}   }\,.
  \label{eq:nonfactorizedH}
\end{align}
A brute force implementation of the action of the operators $D_{z}$ and $D_{z'}$ on the previous expression for the function $\mathcal{H}_{\ell}$ will lead to a rather complicated sum \cite{Goncalves:2019znr} that we do not show since it will not be important in the discussion. A simple analysis reveals that the leading term of $\mathcal{H}_\ell$ in the limit $\xi_{1,2}\rightarrow \lambda \xi_{1,2},\, \xi_3\rightarrow \xi_3\lambda^2$ for large
$\lambda$, which corresponds to considering lightcone limits\footnote{In this limit we can discard the second term in the differential operator $D_z$ which in turn makes its action easier to implement. This just corresponds to throwing away the contribution of terms associated with traces.  } $x_{12}^2,x_{34}^2\rightarrow 0$, is of the form
\begin{align}
  \mathcal{H}_\ell \approx \xi_{1}^{J_1-\ell}\xi_{2}^{J_2-\ell} \xi_3^{\ell} +\dots\,, \label{eq:leadingtermHfunction}
\end{align}
where the $\dots$ represent subleading terms.  Alternatively we can use the Casimir differential equation, in the Euclidean limit, to obtain subleading terms in (\ref{eq:leadingtermHfunction})
\begin{align}
   & \displaystyle {\!\!\!\!\big[(1-\xi_1^2)\partial_{\xi_1}^2+(1-\xi_3^2)\partial_{\xi_3}^2-(d-1)(\xi_1\partial_{\xi_1}+\xi_3\partial_{\xi_3})-2(\xi_1\xi_3+\xi_2)\partial_{\xi_1}\partial_{\xi_3}+C_{J_1}\big]\mathcal{H}_{\ell}=0}\,,
  \label{eq:AngulardifferentialEquation}                                                                                                                                                                                                 \\
   & \displaystyle {\!\!\!\!\big[(1-\xi_2^2)\partial_{\xi_2}^2+(1-\xi_3^2)\partial_{\xi_3}^2-(d-1)(\xi_2\partial_{\xi_2}+\xi_3\partial_{\xi_3})-2(\xi_2\xi_3+\xi_1)\partial_{\xi_2}\partial_{\xi_3}+C_{J_2}\big]\mathcal{H}_{\ell}=0}\,,
  \nonumber
\end{align}
with $C_J=J(J+2h-2)$.
It is essential in extracting the dots in (\ref{eq:leadingtermHfunction}) from the Casimir equation to assume that $\mathcal{H}_{\ell}$ is polynomial in the variables $\xi_i$.
However, this follows from the definition (\ref{eq:nonfactorizedH}).

It turns out that, after changing the cross ratio $\xi_3$ to $\zeta$ defined by\footnote{These cross ratios were introduced in the context of conformal field theories in \cite{Buric:2021kgy}. }
\begin{align}
  \xi_3= -\xi_1\xi_2 +\zeta \sqrt{(1-\xi_1^2)(1-\xi_2^2)}\,,
\end{align}
the Casimir differential equation becomes much simpler
\begin{align}
  \bigg[J_1 \left(d+J_1-2\right) +\frac{(d-2) \zeta \partial_{\zeta}+(\zeta^2-1) \partial_{\zeta}^2 }{\xi _1^2-1}+(1-d) \xi _1\partial_{\xi _1} +(1-\xi _1^2) \partial_{\xi _1}^2 \bigg] \mathcal{H}\,,=0\label{eq:differentialequationHNewangle}
\end{align}
with an analogous equation for $J_2$. This form of the differential equation allows to look for solutions with a factorized form
\begin{align}
  \tilde{\mathcal{H}} = f_1(\xi_1) f_2(\xi_2) g(\zeta) \,,
  \label{eq:factorizePolynomial}
\end{align}
where we have used tilde to emphasize that the solution is factorized and possibly different from (\ref{eq:nonfactorizedH}).
The function $g(\zeta)$ satisfies a differential equation that can be read from (\ref{eq:differentialequationHNewangle})
\begin{align}
  \big[(\zeta^2-1)\partial_{\zeta}^2 +(d-2)\zeta \partial_{\zeta}+\ell' (\ell'+d-3) \big] g_{\ell'} = 0\,,
  \label{eq:ellDiffequation}
\end{align}
where the separation constant $\ell' (\ell'+d-3) $ was chosen for convenience.
One solution to this differential equation that is polynomial in $\zeta$ is given by
\begin{align}
  g_{\ell'}  = \,_2F_1\left(-\ell',\ell'+d-3,\frac{d-2}{2},\frac{1-\zeta}{2}\right)
   & = \,\frac{\ell'!\Gamma(2h-3)}{\Gamma(2h+\ell'-3)} \,C_{\ell'}^{\frac{d-3}{2}}  (\zeta)\,.
\end{align}
This is clearly a polynomial of degree $\ell'$.
It is also simple to check that
\begin{align}
  f_1(\xi_1) = (1-\xi_1^2)^\frac{\ell'}{2} C_{J_1-\ell'}^{\frac{d-2}{2}+\ell'}(\xi_1)\,,
\end{align}
is a solution to the differential equation arising from (\ref{eq:differentialequationHNewangle}). The solution $f_2$ can be obtained analogously.
It can also be checked that this new solution  $ \tilde{\mathcal{H}}_{\ell'}$ is consistent with the non-factorized $\mathcal{H}_\ell$ in (\ref{eq:nonfactorizedH}). Let us see how in more detail.

Both $\mathcal{H}_{\ell}$ and $ \tilde{\mathcal{H}}_{\ell'}$ satisfy the same differential equation, however they are not the same function. Nevertheless
it is possible to express  $\mathcal{H}_{\ell}$ in terms of $\tilde{\mathcal{H}}$ and vice-versa, that is
\begin{align}
   & \tilde{\mathcal{H}}_{\ell'} = \sum_{\ell=0}^{\ell'}  C_{\ell \ell'} \mathcal{H}_{\ell} \,, \qquad
  \label{eq:equationchangebasis}
\end{align}
The coefficients $C_{\ell\ell'}$ can be thought as a change of basis of three-point functions. To determine them it is useful to take the limit $\xi_{1,2}\rightarrow \lambda \xi_{1,2}$ and $\xi_3 =\xi_3 \lambda^2$, with $\lambda$ large. In this limit the functions $\mathcal{H}_{\ell}$ and $\tilde{\mathcal{H}}_{\ell} $ behave as
\begin{align}
  \mathcal{H}_\ell \approx \xi_1^{J_1-\ell}\xi_2^{J_2-\ell} \xi_3^{\ell} +\dots \,,\qquad
  \tilde{\mathcal{H}}_{\ell'} \approx \tilde{c} \,\xi_1^{J_1}\xi_2^{J_2} g_{\ell'} (\zeta)+\dots\,, \\
  \tilde{c}=\frac{\Gamma (h+J_1-1) \Gamma (h+J_2-1) 2^{J_1+J_2-2 \ell'  }}{\Gamma (h+\ell'  -1)^2 \Gamma (J_1-\ell'  +1) \Gamma (J_2-\ell'  +1)},
\end{align}
where $\zeta\rightarrow (\xi_1\xi_2+\xi_3)/(\xi_1\xi_2)$ and the $\dots$ represent subleading terms.
Using the previous equation and (\ref{eq:equationchangebasis}) we can find the coefficients.
Let us start by  $C_{\ell\ell'}$,
\begin{align}
  \xi_1^{J_1}\xi_2^{J_2} \sum_{\ell=0}^{\ell'} C_{\ell \ell'}  \left(\frac{\xi_3}{\xi_1\xi_2}\right)^\ell  = \xi_1^{J_1}\xi_2^{J_2}  g_{\ell'} (\zeta) = \xi_1^{J_1}\xi_2^{J_2}  \sum_{k=0}^{\ell'} \frac{\left(-\ell'\right)_k\left(\ell'+d-3\right)_k}{k!\left(\frac{d-2}{2}\right)_k} \left(\frac{1-\zeta}{2}\right)^k \,,
\end{align}
where $\xi_3/(\xi_1\xi_2)=\zeta-1$. The coefficients $C_{\ell\ell'}$ can be obtained straightaway leading to
\begin{align}
  C_{\ell\ell'}= \tilde{c}(-1)^{\ell} \frac{\left(-\ell'\right)_\ell\left(\ell'+d-3\right)_\ell}{\ell!\left(\frac{d-2}{2}\right)_\ell 2^{\ell}} \,.
\end{align}
To find the inverse relation we make use of the identity
\begin{align}
  \sum_{\ell'=0}^\ell \frac{(c)_{\ell} {{\ell}\choose{\ell'}} (b+2\ell')(-1)^{\ell'}}{(b+1+\ell')_{\ell} (b+\ell')} \,_2F_1\left(-\ell',b+\ell',c,\frac{x}{2}\right) = \left(\frac{x}{2}\right)^{\ell}\,,
\end{align}
for any variable $x$ and constants $b$ and $c$. Using this  equation  the  inverse matrix  $\tilde{C}_{\ell'\ell}$ follows immediately
\begin{align}
  \tilde{C}_{\ell'\ell} = \frac{1}{\tilde{c}}\frac{(-1)^{\ell' } (d+2 \ell' -3) \binom{\ell }{\ell' } \left(\frac{d-2}{2}\right)_{\ell }}{(d+\ell' -3) (d+\ell' -2)_{\ell }} \,.
\end{align}
This concludes the change of basis from (\ref{eq:threepontfunction}) to the one that leads to  (\ref{eq:factorizePolynomial}), which we call factorized basis.
In this basis, the three-point function can be written as~\footnote{Note that, for integer $\ell$, the hypergeometric reduces to a polynomial,
  $$_2F_1\left(-\ell,\ell+d-3,\frac{d-2}{2},\frac{H_{12}}{2V_{123}V_{213}}\right)=\sum_{\ell'=0}^{\ell} \frac{\left(-\ell\right)_{\ell'}\left(\ell+d-3\right)_{\ell'}}{\ell'!\left(\frac{d-2}{2}\right)_{\ell'} 2^{\ell'}}\left(\frac{H_{12}}{2V_{123}V_{213}}\right)^{\ell'}=\sum_{\ell'=0}^{\ell} C_{\ell'\ell}\left(\frac{H_{12}}{2V_{123}V_{213}}\right)^{\ell'}\,.$$}
\begin{align}
  \langle \mathcal{O}_{k_1}(x_1,z_1)\dots\mathcal{O}(x_3) \rangle=\frac{V_{123}^{J_1}V_{213}^{J_2}\sum_{\ell=0}^{\textrm{min}(J_1,J_2)}\tilde{C}_{\ell}\ \,_2F_1\left(-\ell,\ell+d-3,\frac{d-2}{2},\frac{H_{12}}{2V_{123}V_{213}}\right) }{(x_{12}^2)^{\frac{h_1+h_2-h_3}{2}} (x_{13}^2)^{\frac{h_1+h_3-h_2}{2}} (x_{23}^2)^{\frac{h_2+h_3-h_1}{2}}}\,,
\end{align}
where $\tilde{C}_{\ell}$ are the OPE coefficients in the new basis.
Let us remark that this is still  polynomial in the structures $V$ and $H$, as it should. The factorized basis for the leading behavior of the block in the Euclidean OPE limit is a new result. It would be interesting to construct conformal blocks in a radial expansion\cite{Hogervorst:2013sma,Goncalves:2019znr,Poland:2023vpn} using this new basis

\subsection{Lightcone limit}\label{sec:lightcone}

The distance between two operators, in Lorentzian kinematics, can be small when one of them approaches the lightcone of the other.
This is in contrast with what has been analyzed in the previous subsection where the operators were actually close in the Euclidean sense.
The OPE and more generally correlation functions are naturally organized, in this limit, in terms of distances between the almost null related operators.
For example, the leading term in $F_k$ of (\ref{eq:OPEequation}), in the limit $x_{12}^2\rightarrow 0$, is given by
\begin{align}
  F_{k} = (x_{12}\cdot \partial_{z_1})^{J_k} \int_{0}^{1} [dt]  \, e^{tx_{12}\cdot \partial_{x_1} }\,,\label{eq:lightconeLeadingSpin}
\end{align}
where
\begin{align}
  [dt] \equiv \frac{\Gamma(\Delta_k+J_k)}{\Gamma^2(\frac{\Delta_k+J_k}{2})} \,\big(t(1-t)\big)^{\frac{\Delta_k+J_k}{2}-1}dt
\end{align}
for spin $J_k$ operators.
For exchanged scalar operators, it is also easy to write down the formula for $F_k$, including all subleading corrections,
\begin{align}
  F_{k} = \sum_{n=0}^{\infty} \frac{(-x_{12}^2)^{n} \left(\frac{\Delta-a}{2}\right)_n\left(\frac{\Delta+a}{2}\right)_n}{2^{2n} (\Delta)_{2n} (\tfrac{2\Delta-d}{2})_n\, n!} \,_1F_1\left(\frac{2n+\Delta+a}{2},2n+\Delta,x_{21}\cdot \partial_{x_1}\right) (\partial_{x_1}^2)^{n} \,,
  \label{eq:lightconeScalarAll}
\end{align}
with $a=\Delta_{12}$ and $\,_1F_1(a,b,x)=\int_{0}^{1}dt \, \tfrac{t^{a-1}(1-t)^{b-a-1}\Gamma(b)}{\Gamma(a)\Gamma(b-a)} \,e^{t x}$.
In turn, these two formulae can be used to derive the five-point conformal blocks in the lightcone limit by just applying the OPE formula to a five-point correlator. For the leading term of spinning lightcone conformal blocks we have
\begin{align}
  G_{k_1k_2,J_1,J_2}^{\ell} =u_1^{\frac{\Delta_{J_1}-J_1}{2}} u_3^{\frac{\Delta_{J_2}-J_2}{2}} (1-u_2)^{\ell}u_5^{\frac{\Delta_\phi}{2}}  \int_{0}^{1}[dt_1][dt_2] \, \mathcal{I}
  \label{eq:lightconeblocksspins}
\end{align}
with
\begin{align}
  \mathcal{I}=\tfrac{\big(1-t_1(1-u_2)u_4-u_2u_4\big)^{J_2-\ell}\big(1-t_2(1-u_2)u_5-u_2u_5\big)^{J_1-\ell}}{ \big(1-(1-u_4)t_2\big)^{\frac{h_2-\tau_1-2\ell+\Delta_\phi}{2}}\big(1-(1-u_5)t_1\big)^{\frac{h_1-\tau_2-2\ell+\Delta_\phi}{2}} \big(1-(1-t_1)(1-t_2)(1-u_2)\big)^{\frac{h_1+h_2-\Delta_\phi}{2}}}\,.
\end{align}
For the scalar blocks in the  lightcone  we can write %
\begin{align}
  G_{k_1k_200}^{0} = \sum_{n_1,n_2=0}^{\infty} u_1^{\frac{\Delta_{k_1}+2n_1}{2}} u_3^{\frac{\Delta_{k_2}+2n_2}{2}} u_{2}^{\frac{\Delta_{21}}{2}}u_4^{\frac{2n_1+\Delta_{34}-\Delta_5+\Delta_{k_1}}{2}}u_5^{\frac{2n_2+\Delta_{21}+\Delta_{k_2} }{2}}\int_{0}^{1}dt_1dt_2  \, \tilde{\mathcal{I}}_{n_1,n_2}\,,
  \label{eq:scalarblockslightcone}
\end{align}
where the formula for $\tilde{\mathcal{I}}_{n_1,n_2}$ is shown in appendix \ref{app:Blocks}. The cross ratios $u_i$ are appropriate to describe the lightcone limit $x_{12}^2, x_{34}^2\rightarrow 0$, as only two of them go
to zero while the others remain fixed.

One feature that is evident from the formulae above is that this limit is dominated by operators that have lowest twist, defined by $\Delta-J$. Hints of this property are already present in (\ref{eq:OPEequation}) and (\ref{eq:OPEEuclideanLEading}).

Another interesting attribute of the lightcone block is that it allows to probe Lorentzian regimes, this in sharp contrast with the Euclidean expansion (\ref{eq:factorizePolynomial}) that is only valid when the point $x_2$ is in the vicinity of $x_1$. In particular, the integral formulation of both (\ref{eq:lightconeLeadingSpin}) and (\ref{eq:lightconeScalarAll}) is specially suitable to study  monodromies of the block.

\subsection{Regge limit}
\label{sec:Regge_limit}

The limits described in the previous section shared a common feature as they could be taken in a kinematics where all points are still spacelike separated from each other. This is a significant restriction on the positions of   operators and the physics that one is probing with a given correlation function. The goal of this subsection is to introduce and describe   another limit, the Regge limit,   as depicted in figure \ref{fig:ReggeLimitVsOthers}. The main novelty is that some points are timelike related, while others are still spacelike separated, more concretely the pairs of points $(1,4), (2,3),(3,5), (2,5)$ are timelike, while the other pairs remain spacelike. The configuration represented in figure    \ref{fig:ReggeLimitVsOthers} can be parametrized by the following variables
\begin{align}
  x_1 & = -r \left( \sinh  \delta_1  , \cosh   \delta_1 ,\textbf{0}_{d-2}\right), \label{eq:Reggelimitcoordinates}                                       &
  x_2 & = r \left( \sinh   \delta_2 , \cosh  \delta_2  ,\textbf{0}_{d-2}\right),                                                                           \\
  x_3 & =  \left( -\sinh   \delta_2 , \cosh  \delta_2  ,\textbf{0}_{d-2}\right),                                                                         &
  x_4 & =  \left( \sinh   \delta_1 , -\cosh  \delta_1   ,\textbf{0}_{d-2}\right),   \qquad	x_5  =  \left( 0,h_1,h_2,\textbf{0}_{d-3} \right) .  \nonumber
\end{align}
where $\delta_{i}$ are being taken to infinity and $r$ and $h_i$ can assume generic values. Here we also use a $d$-dimensional vector of zeros denoted by $\textbf{0}_{d}$. This configuration can also be written in terms of the cross ratios $u_i$ as
\begin{align}
   & u_1 = \frac{4 r^2 \left(x_5^2+1-2 h_1 \cosh\delta _2\right)}{\left(1+r^2+2 r \cosh\delta\right) \left(x_5^2+r^2-2 h_1 r \cosh \delta _1\right)}\,,
  \qquad u_2= \left(\frac{1+r^2-2 r \cosh \delta}{1+r^2+2 r \cosh\delta}\right)^2,
  \nonumber                                                                                                                                             \\
   & u_3= \frac{4 \left(x_5^2+r^2-2 h_1 r \cosh\delta _1\right)}{ \left(1+r^2+2 r \cosh\delta\right)\left(x_5^2+1-2 h_1 \cosh\delta_2\right)}\,,
  \\
   & u_4= \frac{1}{\sqrt{u_2}}\frac{x_5^2+1+2 h_1 \cosh \delta_2 }{x_5^2+1-2 h_1 \cosh\delta_2} \,,
  \qquad\qquad\qquad\ \
  u_5= \frac{1}{\sqrt{u_2}}\frac{x_5^2+r^2+ 2 h_1 r \cosh\delta _1}{x_5^2+r^2-2 h_1 r \cosh\delta _1} \,,
  \nonumber
\end{align}
where  $\delta=\delta_{1}+\delta_2$ and $x_5^2=h_1^2+h_2^2$. It is simple to see that both $u_1$ and $u_3$ approach zero as the $\delta_i$ are sent to infinity and that the remaining $u_i$ go to $1$
(note that $u_2$ approaches $1$ faster then the other two cross ratios). This limit, in terms of cross ratios, is the same as the Euclidean OPE limit discussed in section \ref{subsec:Euclideanlimitsubsection}.  The main distinction between these two limits resides in the different causal ordering of the operators. The similarity to the Euclidean OPE limit should come as  no surprise to the reader that is familiar with Regge limit for four points.
In reality there is a simple reason for this to be the case as one can also interpret this configuration as an OPE limit between $1^+$ and $2$, as well as $3$ and $4^-$, where  $1^+$ and $4^-$ are defined
respectively as the image of the points $1$ and $4$ on the next and previous Poincar\'e patch on the Lorentzian cylinder. This is shown in  figure \ref{fig:ReggeLimitVsOthers}.

The fifth point is kind of a spectator in this limit.
Nonetheless, it is important as it allows to introduce other parameters to differentiate the gaps $\delta_1$ and $\delta_2$.
This is essentially the same as we already see in the Regge limit of five-point scattering amplitudes.

Note that in this section we made a choice of analytic continuation but there are other possible ways to attain Regge kinematics. Indeed, with some care, one can even move the fifth point in other directions and even boost it and find similar OPE behavior after lightcones are crossed. The latter can be used as a guiding principle when we look for Regge kinematics. In Appendix~\ref{sec:otherregge}, we present some additional kinematics and path continuations that might be useful in understanding single-Reggeon exchanges or the Regge limit six-point functions in CFTs .

As mentioned before, the different causal relations between the points have important consequences. The analysis of the correlator in this setting is more elaborate and for this reason we devote the next section to it.

\subsection{Conformal partial waves}
The conformal block decomposition (\ref{eq:confblockDecom}) is not the most appropriate option to analyze the Regge limit of correlation functions.
A better alternative is to do the so-called conformal partial wave decomposition
\begin{align}
  \mathcal{G}(u_i) = \sum_{J_i=0}^{\infty}\sum_{\ell=0}^{\textrm{min}(J_1,J_2)}\int_{-\infty}^{\infty}\frac{d\nu_1}{2 \pi i }\frac{d\nu_2}{2 \pi i } \,b_{J_1J_2}^{\ell}(\nu_1,\nu_2)\, F_{\nu_1,\nu_2,J_1,J_2,\ell}(u_i)\,,
\end{align}
where the conformal partial wave coefficient  $b_{J_1J_2}^{\ell}(\nu_1,\nu_2)$ contains all the dynamical information of the correlation function, {\em {i.e.}} dimensions and OPE coefficients. The function $F_{\nu_1,\nu_2,J_1,J_2,\ell}(u_i)$ is the conformal partial wave defined by the integral
\begin{align}
  \label{eq:confPartialWaveFivePoints}
  F_{\nu_1,\nu_2,J_1,J_2,\ell}(u_i) = & \tfrac{(x_{12}^2x_{34}^2)^{\Delta_{\phi}}(x_{15}^2x_{35}^2)^{\frac{\Delta_{\phi}}{2}} }{(x_{13}^2)^{\frac{\Delta_{\phi}}{2}}} \int d^dx_6 \, d^dx_7\, \langle \mathcal{O}_{\frac{d}{2}-i\nu_1}(x_6,D_{z_1})\mathcal{O}_{\frac{d}{2}-i\nu_2}(x_7,D_{z_2})  \mathcal{O}(x_5) \rangle^{(\ell)}  \nonumber \\
                                      & \times \langle \mathcal{O}(x_1)\mathcal{O}(x_2) \mathcal{O}_{\frac{d}{2}+i\nu_1}(x_6,z_1) \rangle \langle \mathcal{O}(x_3)\mathcal{O}(x_4) \mathcal{O}_{\frac{d}{2} +i\nu_2}(x_7,z_2) \rangle\,,
\end{align}
where the $\langle \rangle^{(\ell)}$ should be understood as the term proportional to  $C_{123}^{\ell}$ in (\ref{eq:threepontfunction}) (in other words, it is just the space dependence of the three-point function) and $D_{z}$ is the differential operator defined in (\ref{eq:TodorovOPE}).  It is simple to see that both integrals in $x_6$ and $x_7$ are conformal and that $F_{\nu_i,J_i,\ell}$ should satisfy the conformal Casimir equation in the channels $(12)$ and $(34)$ with eigenvalue $C_{\frac{d}{2}+i\nu_1,J}$ and $C_{\frac{d}{2}+i\nu_2,J_2}$, respectively. In particular, this implies that the conformal partial wave can be written as a linear combination of
conformal blocks which solve the same equation
\begin{align}
  \!\!F_{\nu_1,\nu_2,J_i,\ell}  = \sum_{\tilde{\ell}}\sum_{\alpha_1,\alpha_2 =\pm}
  A_{\alpha_1,\alpha_2}^{\ell\tilde{\ell}} G_{\frac{d}{2}+i \alpha_1 \nu_1,\frac{d}{2}+i\alpha_2 \nu_2,J_i }^{\tilde{\ell}} (u_i) \,,
  \label{eq:conformalpartialwave}
\end{align}
where we used the symmetry of the eigenvalue $C_{\frac{d}{2}+i\nu_i,J_i} = C_{\frac{d}{2}-i\nu_i,J_i}$. The sum
over $\tilde{\ell}$ appears because the Casimir equation  is not able to fix it, and so in principle we can have a sum over this number.
The coefficients $A^{\ell\tilde{\ell}}$ were determined in \cite{Antunes:2021kmm} and are expressed in terms of several sums. It would be interesting to see if the coefficients in the new basis introduced in \ref{subsec:Euclideanlimitsubsection} are simpler and, more importantly for this work, analytic in spin.
The conformal partial waves have the advantage that are Euclidean single valued\footnote{Conformal partial waves are single valued for integer $J$. It should be possible to add a term to them to make them single valued for positive real $J$ as was done in \cite{Caron-Huot:2020nem} for four points. We hope to return to this point in the future.  }.
Recall that the correlator also enjoys this property in contrast with a single conformal block.

\section{Regge theory}
\label{sec:ReggeTheoryCFT}
\subsection{Wick rotation or how to go Lorentzian}\label{sec:EucToLorent}
The Regge limit of a correlation function is an intrinsically Lorentzian limit that explores a specific causal configuration of the operators.
On the other hand, CFTs have been better understood in Euclidean space.
It is thus important to understand how to analytically continue from Euclidean to Lorentzian space and what can we say about convergence and other properties of the Lorentzian correlator from CFT axioms. These questions have only very recently been discussed in firmer grounds in~\cite{Kravchuk:2020scc,Kravchuk:2021kwe}, extending the works of L\"{u}scher and Mack~\cite{Luscher:1974ez, Mack:1976pa}.
However, there the analysis focuses only on correlation functions of $n\leq 4$ points and no systematic study for higher-point functions exists to date.~\footnote{This seems to be technically challenging (see discussion of Appendix B of~\cite{Kravchuk:2021kwe}) but we hope that our results may also increase the motivation of community to tackle these questions on higher-point functions.}

We want to consider Lorentzian invariant correlation functions of local operators that commute at spacelike separated points,
\begin{align}
  \mathcal{W}\left(x_1, x_2,\dots,x_n\right)=\langle \mathcal{O}(x_1)\mathcal{O}(x_2)\dots \mathcal{O}(x_n)\rangle\,.
\end{align}
These are called Wightman functions (or distributions). In particular, note that up to spacelike separated points, different orders of local operators give rise to different Wightman functions. We stress that these are not the standard time-ordered correlation functions one encounters in QFT textbooks. In fact, one can decompose time-ordered correlation functions in terms of Wightman functions\footnote{We  assume the existence of a Hilbert space with a unique vacuum $\Omega$ under the unitary action of the Poincar\'e group. We can however talk about Wightman distributions without making any such assumption since Wightman's reconstrution theorem guarantees that we would find a Hilbert space once we assumed spectral and positivity properties of the distributions - see ~\cite{Kravchuk:2021kwe, Streater:1989vi}.}
\begin{align}
   & \langle \Omega | T\{\mathcal{O}(x_1) \mathcal{O}(x_2) \dots \mathcal{O}(x_n)\} | \Omega \rangle=                                                                                         \\
   & =\langle \Omega| \mathcal{O}(t_1,\mathbf{x_1}) \mathcal{O}(t_2, \mathbf{x_2}) \dots \mathcal{O}(t_n,\mathbf{x_n})| \Omega\rangle \theta(t_1>t_2>\dots>t_n)+ \text{permutations}\nonumber \\
   & =\mathcal{W}(x_1,x_2,\dots, x_n)\theta(t_1>t_2>\dots>t_n)+ \text{permutations}\nonumber\,.
\end{align}
One Wightman axiom states that Wightman functions are indeed tempered distributions even at coincident points.
This means that when integrated against test functions belonging to Schwartz class $f(x_i)\in \mathcal{S}$, the following integral is finite
\begin{equation}
  \int d^d x_1\dots d^d x_1 \mathcal{W}(x_1,\dots,x_n) f(x_1)\dots f(x_n)<\infty\,.
\end{equation}

Our goal is to reach a Wightman correlation function with a given order starting from a translational- and rotational-invariant Euclidean one. The basic idea is that there should be some holomorphic function $G(x_1,\dots,x_n)$ that reduces to a Lorentzian correlator in a given limit and to a Euclidean one in another. Let us then consider a real-analytic (away from coincident points) Euclidean correlator,
with operators at $x_i=(\tau_i, \textbf{x}_i)$,
\begin{equation}
  \label{eq:euclcorr0}
  \langle  \mathcal{O}(\tau_1,\textbf{x}_1)\mathcal{O}(\tau_2,\textbf{x}_2)\dots\mathcal{O}(\tau_n,\textbf{x}_n)\rangle^{E}\,,
\end{equation}
where Euclidean times $\tau_i$ are ordered $\tau_1>\tau_2>\dots>\tau_n$. Recall that this ordering is necessary. If we assume the existence of a Hilbert space and a Hamiltonian that is bounded from below, we get that our Euclidean correlator can be rewritten as
\begin{equation}
  \label{eq:euclcorr1}
  \langle \Omega| \mathcal{O}(0,\textbf{x}_1)e^{-H(\tau_1-\tau_2)}\mathcal{O}(0,\textbf{x}_2)e^{-H(\tau_2-\tau_3)}\dots\mathcal{O}(\tau_n,\textbf{x}_n) |\Omega\rangle^{E}\,,
\end{equation}
where we use the Heisenberg representation of the field operators $\mathcal{O}$. To avoid high-energy states being exponentially enhanced,
we immediately recognize that the Euclidean correlator needs to be ``time-ordered".

To move towards a Lorentzian configuration, we want to consider an analytic continuation of the Euclidean correlator. This is achieved  by taking $\tau_i\to \epsilon_i+i t_i$. Heuristically, adding the imaginary parts does not harm the convergence,  as long as we keep  $\epsilon_1>\dots>\epsilon_n$. This analytic continuation defines our function $G(x_1,\dots,x_n)$ that is holomorphic in $\tau_i= \epsilon_i+i t_i$ and real-analytic in $\textbf{x}_i$. We can then find a Lorentzian correlator by sending $\epsilon_i\to0$ while keeping the order of limits,
\begin{align}
  \label{eq:wightmancorrdef}
  \langle \Omega| \mathcal{O}(t_1,\textbf{x}_1)\dots\mathcal{O}(t_n,\textbf{x}_n) |\Omega\rangle\equiv \lim_{\underset{\epsilon_1>\dots>\epsilon_n}{\epsilon_i\to0}}	\langle \Omega| \mathcal{O}(\epsilon_1+i t_1,\textbf{x}_1)\dots\mathcal{O}(\epsilon_n+i t_n,\textbf{x}_n) |\Omega\rangle^{E}\,.
\end{align}
This formally defines our Wightman function $\mathcal{W}(x_1,\dots,x_n)$. Note that to achieve different orderings we should start from an Euclidean correlator in a different ordering.
Holomorphicity may however be lost as we take $\epsilon_i \to 0$. We expect nonetheless the correlator to converge at least in a distributional sense. For CFT Wightman functions, the authors in~\cite{Kravchuk:2021kwe} found power law bounds and used Vladimirov's theorem to assure that indeed this limit converges at least in the distributional sense (even at coincident points) for $n\leq 4$-point functions in Minkowski space.\footnote{All the remaining Wightman axioms were also proved from standard axioms of translational- and rotational- invariant Euclidean correlators.}

We want to consider the Regge limit of CFT five-point functions of identical scalars.
In this context, we are interested in correlation functions where the operator ordering is consistent with time ordering. Using the causal relations of figure~\ref{fig:ReggeLimitVsOthers}, we take
\begin{align}
  \langle\phi(x_4)\phi(x_1)\phi(x_2)\phi(x_5)\phi(x_3)\rangle\,,
\end{align}
where permutations between spacelike separated operators are equivalent. As we approach the Regge kinematics, starting from a configuration where all operators are spacelike separated (essentially equivalent to a Euclidean configuration), we find branch-cut singularities whenever an operator crosses the lightcone of another. The way we deal with branch-cuts depends on the $i \epsilon$ prescription we adopted to reach this ordering of the Wightman function. In particular, as we move from fully spacelike separated points to the Regge kinematics we have $\{x_{14}^2, x_{23}^2,x_{25}^2,x_{35}^2\}\to \{|x_{14}^2|, |x_{23}^2|,|x_{25}^2|,|x_{35}^2|\}\times \exp(\pi i)$ which implies that the cross-ratios $u_2, u_4$ and $u_5$ go around 0 with the first going anticlockwise and the last two in clockwise direction. At the branch-cuts, OPEs $\phi_1\times \phi_2$ and $\phi_3\times \phi_4$, in which we block decompose our correlation function, are no longer convergent. We should then worry about boundedness in Regge limit. For a four-point function in the Regge limit and with operator ordering consistent with time ordering one can prove its boundedness. The general proof uses Rindler positivity~\cite{Casini:2010bf,Caron-Huot:2017vep, Maldacena:2015waa,Kologlu:2019bco} and bounds the latter Wightman function with another correlator of different ordering where the OPE does converge.
This proof does not work however with five-point functions. Nonetheless, we expect to be possible to find these type of bounds between different ordered Wightman functions or different channel decompositions but we will not make these considerations any more precise here.
Conformal Regge theory, on the other hand, provides a method to resum divergent OPEs and exhibit the dominant Reggeon-exchange contributions. This resumation invokes an analytic continuation of OPE data in spin for which, in the case of four-point functions, the justification follows from the Lorentzian inversion formula~\cite{Caron-Huot:2017vep,Simmons-Duffin:2017nub}. For higher-point functions, there are additional representation labels associated with the possible three-point structures between spinning operators.

In what follows we focus on double Reggeon-exchanges but similar analysis can be performed at the level of the single Reggeon exchanges, that we briefly discuss in Appendix~\ref{sec:otherregge}. The proper $i\epsilon$ prescription for these  cases follows straightforwardly from the corresponding kinematics since we want to consider the operator orderings consistent with time ordering.

\subsection{Mellin amplitudes}\label{sec:MellinMellin}
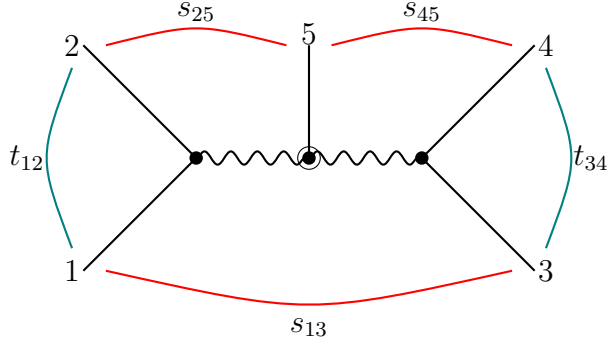
\begin{figure}[t!]
  \centering

  \begin{tikzpicture}[scale=1.5]
    \draw[thick] (-1,1)--(0,0);
    \draw[thick] (-1,-1)--(0,0);
    \draw[thick,snake it] (0,0)--(1,0);
    \draw[thick] (1,1) -- (1,0);
    \draw[thick,snake it] (1,0)--(2,0);
    \draw[thick] (2,0)--(3,1);
    \draw[thick] (2,0)--(3,-1);

    \draw[thick,fill] (1,0) circle (0.05cm);
    \draw[] (1,0) circle (0.1cm);
    \draw[thick,fill] (0,0) circle (0.05cm);
    \draw[thick,fill] (2,0) circle (0.05cm);
    \node at (-1.1, 1) {2};
    \node at (-1.1, -1) {1};
    \node at (3.1, 1.0) {4};
    \node at (3.1, -1.0) {3};
    \node at (1, 1.1) {5};

    \draw[teal, thick] (-1.1,0.8)..controls (-1.4,0)..(-1.1,-0.8);
    \draw[teal, thick] (3.1,0.8)..controls (3.4,0)..(3.1,-0.8);
    \draw[red, thick] (-0.8,-1.0)..controls (1.,-1.4)..(2.8,-1);
    \draw[red, thick] (-0.8,1)..controls (0.,1.2)..(0.8,1.);
    \draw[red, thick] (1.2,1)..controls (2.,1.2)..(2.8,1.);

    \node at (-1.5,0) {$t_{12}$};
    \node at (3.5,0) {$t_{34}$};
    \node at (1.,-1.5) {$s_{13}$};
    \node at (0, 1.3) {$s_{25}$};
    \node at (2,1.3) {$s_{45}$};
  \end{tikzpicture}
  \caption{ Regge kinematics for scattering amplitudes can be defined as $s_{13},s_{25}^2,s_{45}^2\rightarrow \frac{1}{x^2},x\rightarrow 0$ while keeping $t_{12}$ and $t_{34}$ fixed. As can be seen in Mellin space the dominant contribution to the kinematics described in figure \ref{fig:ReggeLimitVsOthers} is the same.
  }
  \label{fig:ReggeKineScatt}
\end{figure}

The similarities of Mellin and flat space scattering amplitudes make the former a suitable tool  to build intuition. The goal of this section is to analyze the
Regge limit for Mellin amplitudes \cite{Costa:2012cb}.
We shall see that  the Regge limit for five operators, as defined in the previous section, is dominated by the same kinematics of flat space scattering amplitudes reviewed in
section \ref{sec:FlatSpaceScattering}.
In the following, we will review the definition of Mellin amplitudes, some of its properties and then analyze the Regge limit  in this language.
The definition of a Mellin amplitude, $\mathcal{M}(\delta_{ij})$,   is given by\footnote{Here we are assuming that there exists a Mellin amplitude. This might only be true if one performs some subtractions of the correlator, such as the contribution of the identity. }
\begin{align}
  \langle \mathcal{O}(x_1)\dots \mathcal{O}(x_n) \rangle  = \int [d\delta_{ij}] \,\mathcal{M}(\delta_{ij})\,\prod_{1\leq i <j \leq n} \frac{\Gamma(\delta_{ij})}{(x_{ij}^2)^{\delta_{ij}}}\,,
  \label{eq:scalarMellinamplitude}
\end{align}
where we decided to extract a  standard prefactor containing $\Gamma$ functions and  the integration variables $\delta_{ij}$ run parallel to the imaginary axis.
Since the Mellin variables are restricted by the condition $\sum_{j}\delta_{ij}=0$, with   $\delta_{ii}=-\Delta_i $,
we shall use the following set of independent Mellin variables
\begin{align}
  t_{12} & = 2\Delta_{\phi}-2\delta_{12}\,,
         & t_{34}                           & = 2\Delta_{\phi}-2\delta_{34}\,,
  \\
  s_{13} & = \Delta_{\phi} +2\delta_{13}\,,
         & s_{25}                           & =-2\delta_{25}\,,
  \qquad\qquad\qquad
  s_{45} = -2\delta_{45}\,,
  \nonumber
\end{align}
which is the same number as conformal cross ratios - see figure~\ref{fig:ReggeKineScatt}.
One advantage of Mellin amplitudes is that it is easy to analytically continue from the Euclidean configuration to Lorentzian, as the space-time dependence is simple \cite{Mack:2009mi}.  For example, the configuration
of figure \ref{fig:ReggeLimitVsOthers} can be obtained just by adding a phase to the integrand \cite{Costa:2012cb}
\begin{align}
  \label{eq:MellinAmpDefinition}
  G^{\circlearrowleft} (u_i) = & \int [dt_{ij}ds_{ij}] \,u_4^{\frac{s_{45}}{2}} u_1^{\frac{t_{12}}{2}} u_3^{\frac{t_{34}}{2}} u_2^{\frac{s_{13}+s_{45}-t_{12}}{2} } u_5^{\frac{t_{34}-s_{25}-t_{12}}{2}} \mathcal{M}(s_{ij},t_{ij})
  \,e^{ -i\pi\frac{2 \left(s_{13}+s_{25}+s_{45}\right)+\Delta_{\phi} }{2}   }\nonumber                                                                                                                                                                          \\
                               & \Gamma \!\left(-\frac{s_{25}}{2}\right) \Gamma \!\left(-\frac{s_{45}}{2}\right) \Gamma \!\left(\frac{s_{13}+s_{25}+s_{45}}{2} \right) \Gamma \!\left(\frac{s_{13}-\Delta_\phi }{2}  \right)                                    \\
                               & \Gamma \!\left(\frac{t_{12}-s_{13}-s_{45}}{2} \right) \Gamma \!\left(\frac{t_{34}-s_{13}-s_{25}}{2} \right) \Gamma \!\left( \frac{2\Delta_\phi-t_{12}}{2}\right) \Gamma \!\left(\frac{2\Delta_\phi-t_{34}}{2}\right) \nonumber \\
                               & \Gamma \!\left(\frac{\Delta_\phi +s_{25}+t_{12}-t_{34}}{2} \right) \Gamma \!\left(\frac{\Delta_\phi +s_{45}-t_{12}+t_{34}}{2} \right)
  \nonumber
\end{align}
where $G^{\circlearrowleft} $ is the correlator analytically continued to the Regge kinematics.
This particular phase seems to make the integrand divergent for large imaginary values of $s_{ij}$. However, the $\Gamma$ functions in the definition of the Mellin amplitude cancel this apparent divergence.  To see this in more detail we just have to use the identity
\begin{align}
   & \Gamma\left(a+i\frac{x_i}{2}\right)\Gamma\left(b-i\frac{x_i}{2}\right) \approx 2\pi e^{i\frac{\pi}{2}(a-b)} \left(\frac{x_i}
  {2}\right)^{a+b-1}e^{-\frac{\pi}{2}x_i} \,,
\end{align}
in a regime where $s_{13}$ goes faster to infinity than $s_{45}$ and $s_{25}$. In the Regge limit, as defined in section \ref{sec:Regge_limit}, we have that the cross ratios  %
$u_2\rightarrow 1+\sigma_1\sigma_2 \xi_3$,  $u_4 \rightarrow 1-\sigma_2\xi_2$, $u_5 \rightarrow 1-\sigma_1\xi_1 $, with $u_1=\sigma_1^2,u_3=\sigma_2^2$ going to zero while $\xi_i$ are left fixed.  This simplifies the dependence of the Mellin amplitude on the cross ratios
\begin{align}
  u_4^{\frac{s_{45}}{2}} u_1^{\frac{t_{12}}{2}} u_3^{\frac{t_{34}}{2}} u_2^{\frac{s_{13}+s_{45}-t_{12}}{2} } u_5^{\frac{t_{34}-s_{25}-t_{12}}{2}} \rightarrow u_1^{\frac{t_{12}}{2}}  u_3^{\frac{t_{34}}{2}} e^{\frac{i(y_{25}\sigma_1\xi_1+y_{45}\sigma_2\cosh \xi_2-y_{13}\sigma_1\sigma_2\xi_3)}{2}}\,,
\end{align}
where we made the change $s_{ij} = iy_{ij}$. Note that the exponent is not small provided $\sigma_{i}$ and $y_{ij}$ scale appropriately.
By putting every piece together we obtain that in the Regge limit
\begin{align}
   & G^{\circlearrowleft}(u_i) = \pi^4\int [dt_{ij}] \,\Gamma \!\left(\frac{2\Delta_{\phi}-t_{12}}{2}\right)\Gamma \!\left(\frac{2\Delta_{\phi}-t_{34}}{2}\right)\frac{\sigma_1^{t_{12}}  \sigma_2^{t_{34}}  }{2^{\frac{t_{12}+t_{34}+\Delta_{\phi}-16 }{2}} e^{\frac{i\pi (\Delta_{\phi} +3t_{12}-t_{34})}{4}}  } \label{eq:ReggeMellinLimit1} \\
   & \int [dy_{ij}] \, y_{13}^{\frac{t_{12}+t_{34}-4-\Delta_{\phi} }{2}}y_{25}^{\frac{t_{12}+\Delta_{\phi}-t_{34}-2}{2}}y_{45}^{\frac{t_{34}+\Delta_{\phi}-t_{12}-2}{2}}e^{\frac{i(y_{25}\sigma_1\xi_1+y_{45}\sigma_2\cosh \xi_2-y_{13}\sigma_1\sigma_2\xi_3)}{2}} \mathcal{M}(t_{ij},y_{ij})\,,
  \nonumber
\end{align}
where we have defined
$u_1=\sigma_1^2$, $u_3=\sigma_2^2$
and we should take the leading behavior in $\mathcal{M}(t_{ij},y_{ij})$ when $y_{ij} \rightarrow \infty$ with $y_{13}/y_{25}y_{25}$ fixed. Thus, in the remaining part of the section we shall  analyze the Mellin amplitude in this limit. Let us just remark that the region of integration that dominates in the Regge limit is the same as for flat space scattering amplitudes.  %

One of the reasons to use Mellin amplitudes is their simple analytic structure. They are meromorphic functions of the Mellin variables $\delta_{ij}$ with just simple poles. This property follows, in a loose sense, from the structure of the OPE \cite{Goncalves:2014rfa}.  The exchange of primary operator with dimension $\Delta$ and spin $J$ (and its conformal family) implies that the Mellin amplitude has a infinite set of poles whose residues are given by a dynamical part (related to OPE data) and a kinematical one, {\em{i.e.}} determined by symmetry\footnote{This formula should be valid for any CFT. The fact that we decided to extract $\Gamma$ functions in the definition of Mellin amplitudes does not imply that we are assuming the existence of double trace like operators. If they do not exist in the spectrum then the Mellin amplitude should have zeros that cancel the poles of these $\Gamma$ functions.}
\begin{align}
  \mathcal{M}(\delta_{ij}) \approx  \mathcal{M}_{\Delta} \equiv  \frac{R_{m}(\delta_{ij})}{\delta_{LR}-(\Delta-J+2m)} \,,\,
  \qquad\qquad  m=0,1,\,\dots\,,
\end{align}
where
\begin{align}
  \delta_{LR} = \sum_{a=1}^{k}\sum_{i=k+1}^{n}\delta_{ai}\,,
  \label{eq:factorizationformula}
\end{align}
$m$ labels subleading twists and $R_m$ is related with lower-point Mellin amplitudes whose precise form has been studied in \cite{Goncalves:2014rfa}. This property is analogous to the factorization in flat space scattering amplitudes.

The residue itself, depending on the number of points, can have poles. To see this,  take as an example the Mellin amplitude of a five-point correlator and look, without loss of generality, to poles in $\delta_{12}$ (this corresponds to setting $k=2$ and $n=5$ in (\ref{eq:factorizationformula})). The residue $R_m(\delta_{ij})$, as mentioned before, depends on a kinematical part and on the four-point Mellin amplitude $\mathcal{M}_{{\cal O}345}$, where ${\cal O}$ is the operator being exchanged. A four-point Mellin amplitude can also have poles for the very same argument.

In this language, the exchange of operators of dimension $\Delta_1$ and $\Delta_2$ in the channels $(12)$ and $(34)$ is  respectively encoded  by the presence of poles in the Mellin amplitude
$\mathcal{M}_{5}(s_{ij},t_{ij})$ at $t_{12}=(\Delta_1-J_1+2m_1)$ and $t_{34} =  (\Delta_2-J_2+2m_2)$,
\begin{align}
  \mathcal{M}_{5}(s_{ij},t_{ij}) \approx \sum_{m_i}\frac{Q_{m_1,m_2}(s_{25},s_{45},s_{13})}{\big(t_{12}-(\tau_1+2m_1)\big)\big(t_{34}-(\tau_2+2m_2)\big)}+\dots\,,
  \label{eq:MellinamplitudepolesQ}
\end{align}
where the $\dots$ represent regular terms (or poles at other locations). Notice that the poles with $m_1=m_2=0$ are associated with the position space lightcone blocks \ref{eq:lightconeblocksspins} and $m_i>0$ correspond to corrections around the lightcone. The residue for these sequential poles is related to three-point functions involving the operators that are exchanged.

Now it remains to analyze the large $s_{ij}$ limit of the Mellin amplitude $\mathcal{M}(t_{ij},s_{ij})$.
As for the four-point case, the Casimir differential equations can be translated into Mellin space, where it transforms to a recurrence relation
that we defer to (\ref{eq:recurrencerelationsMellinCasi}) in appendix \ref{app:Blocks}.
For the $m_i=0$ sector,  the difference equation simplifies considerably. Moreover, for each pair of spins $(J_1,J_2)$, there are $1+\textrm{min}(J_1,J_2)$ polynomial solutions which can be labeled by an integer $n$
and have the leading large $s_{ij}$ behavior
\begin{align}
  Q_{m_1,m_2}(s_{25},s_{45},s_{13}) = c_{\ell,m_1,m_2}s_{25}^{J_1-\ell} s_{45}^{J_2-\ell}s_{13}^{\ell} +\dots \,,
  \label{eq:equationleadingsijMellin}
\end{align}
where $\dots$ represent lower degree terms in the Regge limit.
Note that the $\ell$ denotes a different basis of tensor structure compared to the position space.
We have Mellin transformed the lightcone blocks (\ref{eq:lightconeblocksspins}) and verified the behavior  (\ref{eq:equationleadingsijMellin}).

The recurrence relation (\ref{eq:recurrencerelationsMellinCasi}) can be used to derive relations between $c_{\ell,m_1,m_2}$ with different values of $m_i$
\begin{align}
   & 2 m_1 c_{\ell ,m_1,m_2} \big(d-2 (J_1+m_1+\tau _1)\big)-c_{\ell ,m_1-1,m_2} \left(\Delta _{\phi }+2 J_2-2 m_{12}-\tau _{12}-2 \ell \right)    \nonumber                 \\
   & \times\left(2 m_1+\tau _1-2 \Delta _{\phi }\right)+c_{\ell ,m_1-1,m_2-1} \left(2 m_1+\tau _1-2 \Delta _{\phi }\right) \left(2 m_2+\tau _2-2 \Delta _{\phi }\right)=0\,,
  \label{eq:RecurrencerelationMellin}
\end{align}
for the $(12)$ channel where $m_{ij}=m_i-m_j$. This particular limit is important in the Regge kinematics.
It gives  two recurrence relations for the coefficients $c_{\ell,m_1,m_2}$ that allow  to fix them all in terms of the seed $c_{\ell,0,0}$. As mentioned before, the label $m_i$ in the poles are related to corrections around the lightcone blocks.  Fortunately, we have worked out all these corrections for scalar operators in position space in (\ref{eq:blockscalarposallmi})  and it is a simple exercise to translate the result into Mellin space, written in (\ref{eq:ScalarblockMellin}). In particular, this solution is consistent with (\ref{eq:RecurrencerelationMellin}).

It is possible to construct another solution to the scalar Casimir equation, written in Mellin space, by studying conformal partial waves (or alternatively, exchanged Witten diagrams using the split representation \cite{Costa:2014kfa}). The idea behind this approach is simple, however the computation involves several steps and for this reason is given in the appendix \ref{app:ScaMellinpar}.
The five-point scalar partial wave can be defined by
\begin{align}
  \mathcal{M}_{\nu_1, \nu_2,0,0,0}(\delta_{ij})= & {\textstyle\frac{\pi^{2h}\left[\prod_{i=1}^{2}\Gamma\left(\frac{\Delta_{2i-1}+\Delta_{2i}-t_{2i-1\,2i}}{2}\right)\left(\prod_{\sigma=\pm}\Gamma\left(\frac{h+\sigma \left(\Delta_{2i-1}-\Delta_{2i}\right)+i \nu_i}{2}\right)\right)\right]^{-1}}{\Gamma \left(\Delta _5\right) \Gamma \left(\frac{\Delta _5-i \nu _1+i \nu _2}{2}\right) \Gamma \left(\frac{t_{12}-t_{34}+\Delta _5}{2}\right) \Gamma \left(\frac{2 h-\Delta _5-i \nu _1-i \nu _2}{2}\right) \Gamma\! \left(\frac{h-t_{12}+\Delta _5-i \nu _2}{2}\right)}}\nonumber \\
                                                 & {\textstyle\bigg[\Big(\prod_{\sigma=\pm}\Gamma\!\left(\frac{h-t_{12}+\sigma\Delta_5-i \nu_2}{2}\right)\Gamma\!\left(\frac{\Delta_5+\sigma i \nu_1+i \nu_2}{2}\right)\!\Big)\,\Gamma\!\left(\frac{h-t_{34}+i \nu_2}{2}\right)\Gamma\!\left(\frac{t_{12}-t_{34}+\Delta_5}{2}\right)}\nonumber                                                                                                                                                                                                                                          \\
                                                 & {\textstyle\left. _3F_2\left(_{\Delta_5\,,\frac{2-h+t_{12}+\Delta_5+i\nu_2}{2}}^{\frac{t_{12}-t_{34}+\Delta_5}{2}\,,\frac{\Delta_5-i \nu_1+i \nu_2}{2},\frac{\Delta_5+i \nu_1+i \nu_2}{2}};1\right)+\Gamma\!\left(\Delta_5\right)\Gamma\!\left(\frac{t_{12}-h+\Delta_5+i \nu_2}{2}\right)\right.}\label{eq:scalarpartialMellin}                                                                                                                                                                                                      \\
                                                 & {\textstyle \left(\prod_{\sigma=\pm}\prod_{i=1}^{2}\Gamma\!\left(\frac{h-t_{2i-1\,2i}+\sigma i\nu_i}{2}\right)\right)\,_3F_2\left(_{\frac{2+h-t_{12}-\Delta_5-i\nu_2}{2}\,,\frac{h-t_{12}+\Delta_5-i\nu_2}{2}}^{\frac{h-t_{12}-i\nu_1}{2}\,,\frac{h-t_{12}+i\nu_1}{2}\,,\frac{h-t_{34}-i\nu_2}{2}};1\right)\bigg]}\,,\nonumber
\end{align}
where we use the notation $\delta_{ij}=(\Delta_i+\Delta_j-t_{ij})/2$.
Obviously, the Mellin amplitude of the scalar conformal partial wave only depends on the variables $t_{12}$ and $t_{34}$ and it is symmetric under $\nu\rightarrow -\nu$. More importantly,  it gives a solution valid at finite $t_{ij}$ and reduces to the solution (\ref{eq:ScalarblockMellin}) when $t_{ij}$ are at the poles. This leads us to study the casimir equation away from the poles.
For this purpose let us write the Mellin amplitude as
\begin{align}
  \label{eq:ReggeizePartialWaveMellinSpace}
  \mathcal{M}_{J_1,J_2}(s_{ij},t_{ij}) = s_{25}^{J_1-\ell}s_{45}^{J_2-\ell}s_{13}^{\ell}\, f(t_{12},t_{34})\,,
\end{align}
and plug it in the the recurrence relation (\ref{eq:recurrencerelationsMellinCasi}).  In turn, this leads difference equation for $t_{12}$ and $t_{34}$ that reads
\begin{align}
   & f_{00} \left(t_{12}-\tau _1\right) \left(d-t_{12}-\tau _1-2J_1\right)+f_{-20} \left(2 \Delta_\phi -t_{12}\right) \left(t_{34}-t_{12}+\Delta_\phi +2 J_2-2 \ell \right)\nonumber \\
   & +f_{-2-2} \left(2 \Delta_\phi -t_{12}\right) \left(2 \Delta_\phi -t_{34}\right)=0\,,
\end{align}
where the subindices denote $f_{a_1a_2} \equiv f(t_{12}+a_1,t_{34}+a_{2})$. This difference equation can be further simplified by redefining $f(t_{12},t_{34})$
\begin{align}
  \tilde{f}_{00} \left(\tau_1-t_{12}\right) \left(d-2 J_1-\tau_1-t_{12}\right)+2 \tilde{f}_{-20} \left(t_{12}-t_{34}-\Delta_{\phi}-2 J_2+2 \ell \right)-4 \tilde{f}_{-2-2}=0
\end{align}
where $\tilde{f}$ is given by
\begin{align}
  \label{eq:HfunctionMellin}
  f(t_{12},t_{34}) =\frac{\tilde{f}(t_{12},t_{34})}{\Gamma(\frac{2\Delta_{\phi}-t_{12}}{2})\Gamma(\frac{2\Delta_{\phi}-t_{34}}{2})} .
\end{align}
Note that this prefactor is precisely the same as the one that comes from the Gamma functions in the definition of Mellin amplitudes (\ref{eq:scalarMellinamplitude}).
It is now simple to see that the equation for $J_1=J_2$ and generic $\ell$ can be obtained from the scalar difference equation by doing the following shifts
\begin{align}
  \Delta_{\phi}\rightarrow \Delta_{\phi}+2(J_1-\ell)	,\ \ \ \ d\rightarrow d-2J_1 .
\end{align}
This suggests that the Mellin partial wave for equal spin $J_1=J_2$ and generic $\ell$ can be obtained from (\ref{eq:scalarpartialMellin}) by doing these replacements. One way to check this statement is to build solutions with  the recursion relations in spin derived in \cite{Poland:2021xjs} (we have rederived parts of these relations in the appendix \ref{app:Blocks} using lightcone blocks) and verify that it agrees with the solution that we proposed above.%

These solutions for Mellin amplitudes can then be inserted in (\ref{eq:ReggeMellinLimit1}) to obtain the conformal block in the Regge limit, that is
\begin{align}
   & G_{J_1,J_2,\ell,\nu_1,\nu_2}^{\circlearrowleft} (\sigma_i,\rho_i)=  \sigma_{1}^{1-J_1} \sigma_{2}^{1-J_2} \bar{\mathcal{H}}_{\nu_1\nu_2} (\xi_1,\xi_2,\xi_3)  \,,
\end{align}
with
\begin{align}
   & \bar{\mathcal{H}}_{\nu_1\nu_2} (\xi_1,\xi_2,\xi_3)=\int dt_{12}dt_{34}\, \Gamma\! \left(\frac{2\Delta_{\phi}-t_{12}}{2}\right)\Gamma\! \left(\frac{2\Delta_{\phi}-t_{34}}{2}\right)\Gamma\! \left(\frac{2 \ell -\Delta_{ \phi} +t_{12}+t_{34}-2}{2} \right)\nonumber \\
   & \qquad\qquad\qquad
  \Gamma\!\left(\frac{2 J_1-2 \ell +\Delta_{ \phi} +t_{12}-t_{34}}{2} \right) \Gamma \!\left(\frac{2 J_2-2 \ell +\Delta_{ \phi}-t_{12}+t_{34}}{2} \right)\mathcal{M}_{\nu_1,\nu_2}(t_{12},t_{34})  \nonumber                                                              \\
   & \qquad\qquad\qquad
  \xi_1 ^{\frac{t_{34}-t_{12}-\Delta_\phi -2 J_1+2 \ell }{2} } \xi_2 ^{\frac{t_{12}-t_{34}-\Delta_\phi -2 J_2+2 \ell }{2} }\xi_3 ^{\frac{2-t_{12}-t_{34}+\Delta_\phi -2 \ell }{2}}\,,
  \label{eq:HReggeLimitMellin}
\end{align}
where $\mathcal{M}_{\nu_1,\nu_2}(t_{12},t_{34})$ is the conformal partial wave in Mellin space in the Regge limit. This expression highlights two properties of the Regge limit, firstly the limit is dominated by operators of
high spin and, secondly, it depends on three fixed cross ratios that can be thought of as angles, which is similar to  what happens in the Euclidean OPE limit as we mentioned before. In fact $\bar{\mathcal{H}}_{\nu_1\nu_2} $ solves the Casimir differential equation in the Euclidean region (\ref{eq:AngulardifferentialEquation}) but with a different eigenvalue $C$.  Let us point out that the integral (\ref{eq:HReggeLimitMellin}) can be done by picking up poles.

It follows from what was said above that  $\bar{\mathcal{H}}_{\nu_1\nu_2} (\xi_1,\xi_2,\xi_3)$ must have the same form as (\ref{eq:factorizePolynomial}), as it solves the same conformal Casimir equation.

\subsection{Comment on position space}
The analysis of the Regge limit in Mellin space of the previous section exposed the similarities to flat space scattering amplitudes but it does not emphasize enough the role of analytic continuation in the cross ratios in changing the behavior of the conformal block. This aspect is clearer in  position space, in particular, in the lightcone expressions introduced in subsection \ref{sec:lightcone}. The kinematics of the Regge limit (where some pair of points are timelike while others are spacelike) can be reached from the
Euclidean configuration after doing analytic continuations in $u_2,u_4$ and $u_5$ around $0$ as explained in section \ref{sec:EucToLorent}.

The analysis is simpler for the discontinuities around  $u_4, u_5=0$ in the lightcone blocks (\ref{eq:lightconeblocksspins}) and contains most of the physics we want to highlight in this subsection. These discontinuities come from the first two terms in the denominator of (\ref{eq:lightconeblocksspins}),  provided that $u_2> 0$.
The origin of branch point at, say $u_5=0$, comes from the region $t_1 \approx 1/(1-u_5)$ where the denominator $(1-(1-u_5)t_1)$ changes sign. To deal with this it is convenient to divide the integration region in two parts,
\begin{align}
  \int_{0}^{1} dt_1 \,\mathcal{I}\ \rightarrow \   \int_{0}^{\frac{1}{1-u_5}} dt_1 \,\mathcal{I}+(-1)^{(h_1-\tau_2-2\ell+\Delta_\phi )} \int_{\frac{1}{1-u_5}} ^{1} dt_1\,\mathcal{I}\,,
\end{align}
where the phase comes from the change of sign in the factor $(1-(1-u_5)t_1)$. The first term drops out when taking the discontinuity and so we obtain
\begin{align}
  \textrm{Disc}_{u_5=0} \int_{0}^{1} dt_1\, \mathcal{I}= \big( 1-(-1)^{(h_1-\tau_2-2\ell+\Delta_\phi )} \big) \frac{u_5}{u_5-1}\int_{0}^{1} d\tau_1 \,\mathcal{I}\,,
\end{align}
where we have changed variables to $t_1 = (u_5\tau_1 -1)/(u_5-1)$ in order to have the integration running from $0$ to $1$ again.
It is possible to repeat the same steps to take the discontinuity of $u_4$.

Recall that the cross ratios $u_4$, $u_5$ and $u_2$ approach $1$ with $\tfrac{(1-u_2)}{(1-u_4)(1-u_5)}=\tfrac{1 + \zeta}{2}$ fixed in the Regge limit.
The discontinuity in $u_4$ and $u_5$  of the lightcone block after  the Regge limit is given by
\begin{align}
  \lim_{\substack{   u_4,u_5\rightarrow 1 \\   \zeta \, \textrm{fixed}}}\textrm{Disc}_{u_5,u_4=0} \mathcal{G}
  =\frac{u_1^{\frac{1-J_1}{2}}u_3^{\frac{1-J_2}{2}  }
  (1+\zeta)^{\ell} }{\xi_2^{\Delta_2-1}\xi_1^{\Delta_1-1} 2^{\ell} }\sum_{m=0}^{\infty}
  \frac{	\dbinom{\frac{\Delta_{\phi}-\tau_1-\tau_2-2J_1-2J_2}{2}}{m}
    \left( -\frac{1+\zeta}{2} \right)^{m}
    F_1 F_2
  }{
    \Gamma\!\left(\frac{2J_1+\Delta_{\phi}-2\ell+\tau_{12}}{2}\right) \Gamma\!\left(\frac{2J_2+\Delta_{\phi}-2\ell+\tau_{21}}{2}\right)
  }\,,
  \label{eq:discInu4u5}
\end{align}
where we have used $\tau_{ij}=\tau_i-\tau_j$, the cross ratios (\ref{eq:anglesdef}) and  %
\begin{align}
  F_i & =
  \frac{
    \pi
    \Gamma^2(\frac{\tau_i+2J_i}{2})\, \Gamma(\tau_i+2J_i)\,\Gamma(\tau_i+2J_i+m-1)}{
    2^{J_i-\frac{1}{2}}
    \Gamma\!\left(\frac{\tau_1+\tau_2+2J_i+2m+2\ell-\Delta_{\phi}}{2}\right)
  }
  \, \pFq{2}{1}{ \ell-J_i,\tau_{i+1}+2J_{i+1}+m-1}{\tfrac{2J_{i+1}+2m+2\ell+\tau_1+\tau_2-\Delta_{\phi}}{2}}{\frac{\zeta+1}{2} }.\nonumber
\end{align}
The discontinuities in $u_4$ and $u_5$ are enough to reveal that the discontinuities of conformal block behave with $\sigma_1^{1-J_1}\sigma_2^{1-J_2}$ in the Regge limit, which compares with $\sigma_1^{\Delta_1}\sigma_2^{\Delta_2}$ of the Euclidean block\footnote{We also need to consider the monodromy of the lightcone block around the branch point at $u_2=0$.  It is possible to do  a Mellin transform of the lightcone block and apply the method of the previous subsection to derive all discontinuities. In the appendix, we provide several checks that the discontinuity of the block in $u_2$ has the same behavior. }.
It can also be shown from the previous formula that three sequential discontinuities,  $\textrm{Disc}_{u_2,u_4,u_5}$, evaluate to zero.
Recall that four-point conformal blocks have vanishing double discontinuity.
We believe that conformal blocks have this property away from the lightcone limit.

\subsection{Conformal Regge theory for five points}
\label{subsec:SWtransform}

Let us consider the representation of the five-point correlation function in terms of  conformal partial waves, and its implications for the Regge limit.
This basis  is complete and orthogonal.
Since we have more control over the analytic properties of the  partial waves in Mellin space, we consider the  expansion
\begin{align}
  \mathcal{M}(s_{ij},t_{ij} )= \sum_{J_1,J_2=0}^{\infty} \sum_{\ell=0}^{\textrm{min} \left( J_1, J_2 \right)}\int \frac{d \nu_1}{2\pi i}\,\frac{d \nu_2}{2\pi i}\, b_{J_1,J_2,\ell}(\nu_1,\nu_2)
  \mathcal{M}_{J_1,J_2,\ell}(s_{ij},t_{ij})
  \,.
  \label{eq:partialWave}
\end{align}
We suppress the dependence of the Mellin partial wave $ \mathcal{M}_{J_1,J_2,\ell} $ on the scaling dimensions, as the nontrivial analytic continuation occurs in other quantum numbers.
We have introduced poles in the variables $ \nu_1$ and  $\nu_2 $ with residues corresponding to the OPE coefficients, using
\begin{align}
  b_{J_1,J_2,\ell}\left( \nu_1,\nu_2 \right) & \approx
  \frac{
    P_{\nu_1,\nu_2,J_1,J_2}^{\ell}
  }{
    \big( \nu_1^2+ \left( \Delta_1 - h \right)^2 \big) \big( \nu_2^2+ \left( \Delta_2 - h \right)^2 \big)
  }\,,
  \label{eq:residueValue}
\end{align}
where $\Delta_i=\Delta_i(J_i)$ is the dimension of the $i$-th exchanged operator of spin $J_i$.
We remark that  the product of the OPE coefficients   $P_{\nu_1,\nu_2,J_1,J_2}^{\ell}$ in (\ref{eq:residueValue}) is a linear combination
of those in (\ref{eq:OPE_coeffs}) that appear in the conformal block expansion.

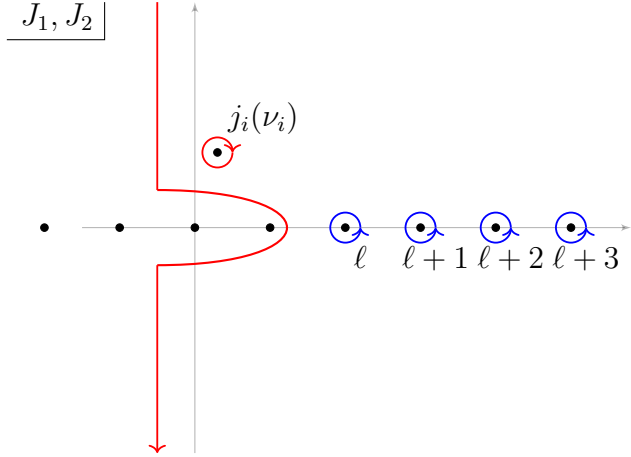
\begin{figure}[t!]
    \centering
      \begin{tikzpicture}
        \draw[] (-1.2,3) -- (-1.2,2.5) -- (-2.5,2.5);
        \node[left] at (-1.2,2.8) {$J_1,J_2$};
        \draw[-,opacity=0.3] (0,0) -- (-1.5,0);
        \draw[->,opacity=0.3,-latex'] (0,0) -- (0,3);
        \draw[-,opacity=0.3] (0,0) -- (0,-3);
        \draw[->,opacity=0.3,-latex'] (0,0) -- (5.8,0);
        \draw[fill=black] (-2,0) circle (0.05);
        \draw[fill=black] (-1,0) circle (0.05);
        \draw[fill=black] (0,0) circle (0.05);
        \draw[fill=black] (1,0) circle (0.05);
        \draw[fill=black] (2,0) circle (0.05);
        \draw[fill=black] (3,0) circle (0.05);
        \draw[fill=black] (4,0) circle (0.05);
        \draw[fill=black] (5,0) circle (0.05);
        \draw[fill=black] (0.3,1) circle (0.05);

        \draw[->,line width=0.7,blue] (2.2,0) to[out=90,in=0] (2,0.2) to[out=180,in=90] (1.8,0) to[out=-90,in=180] (2,-0.2) to[out=0,in=-90] (2.2,0);
        \draw[->,line width=0.7,blue] (3.2,0) to[out=90,in=0] (3,0.2) to[out=180,in=90] (2.8,0) to[out=-90,in=180] (3,-0.2) to[out=0,in=-90] (3.2,0);
        \draw[->,line width=0.7,blue] (4.2,0) to[out=90,in=0] (4,0.2) to[out=180,in=90] (3.8,0) to[out=-90,in=180] (4,-0.2) to[out=0,in=-90] (4.2,0);
        \draw[->,line width=0.7,blue] (5.2,0) to[out=90,in=0] (5,0.2) to[out=180,in=90] (4.8,0) to[out=-90,in=180] (5,-0.2) to[out=0,in=-90] (5.2,0);

        \node[below] at (2.2,-0.1) {$\ell$};
        \node[below] at (3.2,-0.1) {$\ell+1$};
        \node[below] at (4.2,-0.1) {$\ell+2$};
        \node[below] at (5.2,-0.1) {$\ell+3$};
        \draw[<-,line width=0.7,red] (0.5,1) to[out=90,in=0] (0.3,1.2) to[out=180,in=90] (0.1,1) to[out=-90,in=180] (0.3,0.8) to[out=0,in=-90] (0.5,1);
        \draw[<-,line width=0.7,red] (-0.5,-3) -- (-0.5,-0.5);
        \draw[-,line width=0.7,red] (-0.5,0.5) -- (-0.5,3);

        \node[above right] at (0.3,1.1) {$j_i(\nu_i)$};

        \path [draw=red,thick] (-0.5, 0.5)..controls (1.8,0.5) and (1.8,-0.5)..(-0.5,-0.5) node[blue,below,xshift=+3mm]{};
      \end{tikzpicture}
	\caption{Integration contour in spins $J_1,J_2$.
    The blue contour can be deformed to the red contour. We assume the leading Regge pole in the $J_i$-plane is located at $j_i(\nu)$ and we don't draw any further dynamical singularities that might exist to the left.
    Red contour is understood to be deformed to the right of the other infinite series of poles depending on $ \ell $ lying on the left in the $J_i$-plane.
  }
  \label{fig:contourDeformationJ1J2}
\end{figure}

We would like to provide a Sommerfeld-Watson representation of (\ref{eq:partialWave}).
First, we swap the range of summations as
\begin{align}
  \mathcal{M}(s_{ij},t_{ij} )
  = \sum_{\ell=0}^{ \infty}
  \sum_{J_1,J_2=\ell}^{\infty}
  \int \frac{d \nu_1}{2\pi i}\,\frac{d \nu_2}{2\pi i}\, b_{J_1,J_2,\ell}(\nu_1,\nu_2)
  \mathcal{M}_{J_1,J_2,\ell}(s_{ij},t_{ij})
  \,.
  \label{eq:partialWaveSwapped}
\end{align}
Next, we   analytically continue in the spin quantum numbers.
However, the $ b_{J_1,J_2,\ell} $ are not expected to have a unique
analytic  continuation in   the quantum numbers. For that reason we need to consider
their    \textit{signatured} counterparts.

Let us remind the reader the analogous construction \cite{Costa:2012cb} for the four-point correlator $ \mathcal{A}( u,v ) $ in terms of the cross ratios
\begin{align}
  u =  \frac{x_{12}^2 x_{34}^2 }{x_{13}^2 x_{24}^2 } \,, \qquad v = \frac{x_{14}^2 x_{23}^2 }{x_{13}^2 x_{24}^2}\,.
\end{align}
After expanding in Euclidean partial waves, we can write the correlation function as
\begin{align}
  \mathcal{A}( u,v)
  =
  \displaystyle\sum_{J=0}^{\infty}
  \displaystyle\int_{\frac{d}{2}+i \mathbb{R}}
  \frac{d\Delta}{2\pi i}\,
  c_{\Delta,J}\,
  F_{\Delta,J} ( u,v )
  \,,
  \label{eq:EuclideanInversion}
\end{align}
where $ c_{\Delta,J} $ denotes the OPE function and $ F_{\Delta,J} $ is the Euclidean partial wave.
It can be transformed to Mellin space as
\begin{align}
  \mathcal{M}(s,t)
  =
  \displaystyle\sum_{J=0}^{\infty}
  \int_{\frac{d}{2}+i \mathbb{R}}
  \frac{d\Delta}{2\pi i}\,
  c_{\Delta,J}\,
  \mathcal{M}_{\Delta,J} (s,t)
  \,.
  \label{eq:MellinInversion}
\end{align}
Again, the OPE function $ c_{\Delta,J} $ is not uniquely defined in the complex $ J $ plane.
Thus, we define the signatured OPE function $ c_{\Delta,J}^{\theta} $ by
\begin{align}
  \mathcal{M}_{\theta} (s,t)
  =
  \displaystyle\sum_{J=0}^{\infty}
  \displaystyle\int_{\frac{d}{2}+i \mathbb{R}}
  \frac{d\Delta}{2\pi i}\,
  c_{\Delta,J}^{\theta}\,
  \mathcal{M}_{\Delta,J}^{\theta} (s,t)
  \,,
  \label{eq:signaturedMellin}
\end{align}
where the signatured Mellin partial waves are given by
\begin{align}
  \mathcal{M}_{\Delta,J}^{\theta} (s,t)
  =
  \frac{1}{2}
  \big[
    \mathcal{M}_{\Delta,J} (s,t)
    + \theta
    \mathcal{M}_{\Delta,J}(-s,t)
    \big]
  \,,
  \label{eq:signaturedMellinAmp4pt}
\end{align}
with $\theta=\pm$.
The signatured Mellin amplitude allows for a unique analytic continuation of the
signatured OPE function $ c_{\Delta,J}^{\theta} $ \cite{Caron-Huot:2017vep}.
The problem of the non-signatured OPE function can be traced back  the factor of $(-1)^J$ that appears
in the transformation $s\to-s$,
which follows
from the large $s$
behavior $\mathcal{M}_{\Delta,J}(s,t) \approx s^J$.

A similar construction can be done for five-point functions.
We split the full correlator into eight parts depending on the \emph{signature} denoted by $ \mathbb{\theta} = (\theta_1,\theta_2,\theta_{12}) $ where each component can be $\pm$.
We define the  \textit{signatured} amplitudes as
\begin{align}
   & \mathcal{M}_{\theta}(s_{25},s_{45},s_{13})
  =
  \frac{1}{8}
  \big[
    \mathcal{M}(s_{25},s_{45},s_{13})
    +
    \theta_1
    \mathcal{M}(-s_{25},s_{45},s_{13})
    +
    \theta_{2}
    \mathcal{M}(s_{25},-s_{45},s_{13})
  \nonumber                                     \\
   &
    +
    \theta_1 \theta_{2}
    \mathcal{M}(-s_{25},-s_{45},s_{13})
    +
    \theta_{12}
    \mathcal{M}(-s_{25},-s_{45},-s_{13})
    +
    \theta_1 \theta_{12}
    \mathcal{M}(s_{25},-s_{45},-s_{13})
    \nonumber
  \\
   &
    +
    \theta_{12} \theta_{2}
    \mathcal{M}(-s_{25},s_{45},-s_{13})
    +
    \theta_1 \theta_{12} \theta_{2}
    \mathcal{M}(s_{25},s_{45},-s_{13})
    \big]
  \,.
  \label{eq:signaturedMellinAmp}
\end{align}
This equation is suitable only for $s_{ij} \gg 1$.
We also suppress the dependence on $ t_{ij} $ for brevity.
We justify it by using the properties of the Mellin partial wave (\ref{eq:ReggeizePartialWaveMellinSpace})  which,
in terms of $ J_i'=J_i-\ell $,  behaves in the  Regge limit as
\begin{align}
  \mathcal{M}_{J_1',J_2',\ell}(s_{ij},t_{ij}) & = s_{25}^{J_1'}s_{45}^{J_2'} \, s_{13}^{\ell}\, f(t_{12},t_{34})\,.
\end{align}
By analogy with the four-point case, we expect that OPE functions associated with the expansion of the signatured amplitudes in
(\ref{eq:signaturedMellinAmp}) have a unique analytic continuation in all quantum numbers $J_1',J_2',\ell$.
It would be interesting to put this on a firm footing by deriving dispersion relations along the lines of \cite{Caron-Huot:2017vep}.
The full Mellin amplitude can then be written in terms of the signatured Mellin amplitude as
\begin{align}
  \label{eq:fullMellinFromSignature}
  \mathcal{M}(s_{25},s_{45},s_{13} ) & = \sum_{\theta \in \{ -1,1 \}^{3}} \mathcal{M}_{\theta} (s_{25},s_{45},s_{13} )
  \,.
\end{align}

In terms of the signatured analogue of partial waves defined through (\ref{eq:signaturedMellinAmp}), we write
\begin{align}
  \mathcal{M}_{\theta} ( s_{ij},t_{ij} )=
  \sum_{\ell=0}^{\infty} \sum_{J_1',J_2'=0}^{\infty}
  \int \frac{d \nu_1}{2\pi i}\,
  \frac{d \nu_2}{2\pi i}\, b_{J_1',J_2',\ell}^{\theta}(\nu_1,\nu_2) \mathcal{M}_{J_1',J_2',\ell}^{\theta}(s_{ij}, t_{ij})
  \,.
  \label{eq:partialWaveSwappedOddEven}
\end{align}
Next we  perform a Sommerfeld-Watson transform on the $ \ell $ contour
\begin{align}
  \mathcal{M}_{\theta} ( s_{ij},t_{ij}) & =
  \int_{\mathbf{C}} \frac{d \ell}{2\pi i} \frac{1}{\sin ( \pi \ell)}
  \sum_{J_1',J_2'=0}^{\infty}
  \int \frac{d \nu_1}{2\pi i}\,
  \frac{d \nu_2}{2\pi i}\, b_{J_1',J_2',\ell}^{\theta}(\nu_1,\nu_2)\mathcal{M}_{J_1',J_2',\ell}^{\theta}(s_{ij},t_{ij})
  \,.
\end{align}
where ${\mathbf{C}}$ is the contour that encircles the poles at non-negative integers in $\ell$ complex plane counterclockwise.

Next we analytically continue in $ J_1'$ and $J_2'$ by means of two Sommerfeld-Watson transforms. The analytic structure in these variables is analogous to the case of four-point correlation functions. Figure \ref{fig:contourDeformationJ1J2} shows the analytic structure of the integrand.
In particular, there is a leading Regge pole in the $ J_i $ plane at $ J_i =j_i( \nu_i ) $ given by
\begin{align}
  \left[ \Delta_i\big(j_i( \nu_i )\big)- h \right]^{2}  + \nu_i^{2}  = 0\,.
\end{align}
Picking the poles in the complex spin planes at $J_1'= j_1(\nu_1)-\ell$ and $J_2'= j_2(\nu_2)-\ell$,  we obtain
the following expression for the signatured correlators
\begin{align}
  \mathcal{M}_{\theta}
  =
  \int
  \frac{d \nu_1}{2\pi i}\,
  \frac{d \nu_2}{2\pi i}
  \, s_{25}^{j_1( \nu_1)} s_{45}^{j_2( \nu_2)}
  \int_{\mathbf{C}} \frac{d \ell}{2\pi i}\,
  \,
  \frac{
    b_{j_1(\nu_1),j_2(\nu_2),\ell}^{\theta}(\nu_1,\nu_2) \,
    f_{\nu_1,\nu_2,\ell}^{\theta}(t_{ij})\, \eta^{\ell}
  }{\sin(\pi \ell) \sin\!\big(\pi (j_1(\nu_1)-\ell)\big) \sin\!\big(\pi (j_2(\nu_2)-\ell)   \big)}
  \,,\label{eq:pickingJ1J2poles}
\end{align}
where $ f_{\nu_1,\nu_2,\ell}^{\theta} $ is defined as the signatured analogue of $ f $ in (\ref{eq:HfunctionMellin}). 

The integral over $\ell$ remains to be performed as we did not consider any particular limit in $\eta$. One should however comment on the anticipated singularities in this complex plane. Indeed, we expect no dynamical poles in $\ell$, {\textit i.e.}  $b_{j_1(\nu_1),j_2(\nu_2),\ell}^{\theta}(\nu_1,\nu_2)$ should not have poles in $\ell$ and therefore all the singularities to be considered are determined by the explicit sine functions in the integrand.  This assumption is inspired by an analogous procedure for the five-particle S-matrix. Indeed, the quantum number $ \ell $ here labels a choice of tensor basis for three-point functions but it also seems to control the scaling and the asymptotic behaviour of the amplitude in multi-Regge limit. It seems unreasonable that the asymptotic scaling can be basis dependent and therefore we expect the singularities in $\ell$ to be fully determined by the singularities in spin. This is just like the discussion for the
helicity quantum number in flat space following \cite{Abarbanel:1972ayr}. 

In the Regge limit the full correlator takes the form
\begin{align}
  \mathcal{M}
  =
  \int
  \frac{d \nu_1}{2\pi i}\,
  \frac{d \nu_2}{2\pi i}
  \, s_{25}^{j_1\left( \nu_1 \right)} s_{45}^{j_2\left( \nu_2 \right)}
  \int_{\mathbf{C}} \frac{d \ell}{2\pi i}\, \eta^{\ell}  g_{\nu_1,\nu_2,\ell}(t_{ij})
  \,,\label{eq:final}
\end{align}
where the function $g_{\nu_1,\nu_2,\ell} (t_{ij})$ is defined from replacing (\ref{eq:pickingJ1J2poles}) in (\ref{eq:fullMellinFromSignature}).
This allows us to represent the Reggeized Mellin amplitude in terms of the operator content of the leading Regge trajectories and their couplings to the external states.

\section{Conclusions}
\label{sec:conclusions}
In this paper, we have discussed and analyzed the generalization of the Regge limit to five-point correlation functions in conformal field theories in general dimensions. The kinematics of this limit is similar to the four point \cite{Costa:2012cb} case with one crucial difference, the insertion of one extra point. In particular the fifth point is essential to have different rapidities between first set of four operators.  The location of the fifth point is also important for the interpretation of the five point Regge limit as an OPE limit in the second sheet. In particular, the dimension and spin of the exchanged operators in this second sheet OPE is transformed  to $(1-J,1-\Delta)$, which can be interpreted in terms of light-ray operators\cite{Kravchuk:2018htv}.

Our proposal for the five point Regge limit is also confirmed by the exploration of this kinematics in Mellin space. More concretely, we have verified that the Regge limit  is dominated by a region in Mellin space characterized by some large Mellin variables in close analogy to the Regge limit of scattering amplitudes in flat space.
This similarity leads us to focus more on Mellin amplitudes to study the Regge limit.
In particular, we have analyzed the generalization of Mack polynomials for Mellin amplitudes and discussed some of its properties. We have also derived the conformal partial wave in Mellin space in the Regge limit  when the exchanged operators have the same spin. We have also studied the behavior of the conformal block in this limit, using the recent results on lightcone blocks for higher-point functions.
To compute the monodromies of the conformal block, we also used some techniques from the study of hypergeometric functions.

Equipped with a new formula for conformal partial wave in the Regge limit we have extended Regge theory to five point correlation functions in Mellin space by borrowing the methods that were used for flat space scattering amplitudes. Our final result for the five point correlation function in the Regge limit corresponds to the exchange of two pomeron like operators.

In the process, we also discussed a novel basis of three-point functions of operators with spin $\left(  J_1,J_2,0  \right)$, respectively.
This basis is useful for study of Euclidean OPE limit as it leads to the analytic expression for the conformal block in this limit.
These expressions appear to be a natural generalization of the Wigner $ d $ function used in the S matrix.
This suggests that the basis used in the literature for three-point functions of operators with spin $\left(  J_1,J_2,J_3  \right)$ might not be the most natural one for the study of Euclidean conformal blocks.
Given the importance of conformal blocks in numerical bootstrap, it would be interesting to study the properties of this basis in more detail.
The case of $\left(  J_1,J_2,J_3  \right)$ three-point function is accessible in the Euclidean OPE limit of six-point function in the snowflake channel.

The new basis appears to be useful to arrive at a Euclidean inversion formula for the five-point functions, mainly due to simple orthogonality properties.
While the study of analytic structure of the correlator of higher-point functions is still in its infancy, we expect that it admits a drastic simplification in the multi-Regge limit.
In particular, it would be interesting to investigate if the anomalous thresholds affect the Regge limit.
It would also be interesting to derive an inversion formula and dispersion relation in the multi-Regge limit for CFTs, along the lines of \cite{NotesOnMultiRegge}.
An important ingredient is the multivariable generalization of the Cauchy formula, called the Bargman-Weil formula.

It would be interesting to generalize the Regge limit considered in this work to higher point functions.
The generalization of the partial wave expansion in the S matrix is done in \cite{White:1976qm}, for four dimensional quantum field theories.
Analogous generalization of the partial wave expansion is within reach for three dimensional CFTs for $n$-point functions.
We would benefit from the fact that there are no representations of the rotation subgroup of the conformal group in three dimensions with more than one rows in the Young's tableaux.
However, for higher dimensions, there will be proliferation of indices labelling the internal vertices.

Finally, in the S matrix case, a crucial ingredient for the absence of singularities in $ \ell $ was the use of Steinmann relations.
It would be interesting to explore the analogue of Steinmann relations in CFTs, with or without large central charge limit.

\section*{Acknowledgements}
We would like to thank
Ant{\'o}nio Antunes,
Jo\~ao Caetano,
Simon Caron-Huot,
Miguel Correia,
Petr Kravchuk,
Jo\~ao Penedones,
Pedro Ribeiro,
David Simmons-Duffin,
Pedro Vieira,
Sasha Zhiboedov for discussions.
V.G. was supported by Simons Foundation grants \#488637 (Simons collaboration on the non-perturbative bootstrap) and Fundacao para a Ciencia e Tecnologia (FCT) under the grant CEECIND/03356/2022. JVB is funded by FCT with fellowship DFA/BD/5354/2020, co-funded by
Norte Portugal Regional Operational Programme (NORTE 2020), under the PORTUGAL
2020 Partnership Agreement, through the European Social Fund (ESF).

\appendix

\section{Lightcone blocks}\label{app:Blocks}
The scalar five-point conformal blocks, mentioned in the main text, can be expressed in terms of an expansion around the lightcone (\ref{eq:scalarblockslightcone})  by acting with (\ref{eq:lightconeScalarAll}) on a three-point function. In (\ref{eq:scalarblockslightcone})  we have written it in terms of a function $\tilde{\mathcal{I}}_{n_1,n_2}$ given by
\begin{align}
   & \tilde{\mathcal{I}}_{n_1,n_2}=\tfrac{\left(\frac{a-\Delta_5}{2}\right)_{n_1}\left(\frac{2 n_1-\Delta_5+a}{2} \right)_{n_2} \left(\frac{a+4-\Delta_5-d}{2} \right)_{n_1} \left(\frac{2 n_1-\Delta_5+a+4-d}{2} \right)_{n_2} }{  \left(t_1^2 u_1 u_4-t_1 \left(t_2 \left(1-u_5\right)+t_2 u_4 \left(u_2 u_5-1\right)+u_1 u_4+u_5-1\right)+u_5 \left(t_2^2 u_3-t_2 \left(u_3-u_2 u_4+1\right)+1\right)\right)^{\frac{a+2 n_1+2 n_2-\Delta_5}{2}} } \label{eq:blockscalarposallmi} \\
   & \prod_{i=1}^2\tfrac{(-1)^{n_i} \Gamma (2 n_i+\Delta_{k_i})  \left(\frac{\Delta_{k_i}}{2}\right)_{n_i}^2  (t_i(1-t_i))^{\frac{\Delta_{k_i}+2n_i}{2}-1}    }{n_i!(\Delta_{k_i})_{2 n_i} \left(\frac{2\Delta_{k_i}+4-d }{2}\right)_{n_i} \Gamma^2 \left(\frac{2n_i+\Delta_{k_i}}{2}\right) }\nonumber
\end{align}
where $a=\Delta_{k_1}+\Delta_{k_2}$. One nice feature of this result is that it allows to to analytic continuations in $u_2,u_4,u_5$ at all orders in $u_1$ and $u_3$, this is specially useful to verify that the analytic continuation of the conformal block has a distinct behavior in the Regge limit. With this expression in our hands we can also do a Mellin transform and obtain the Mellin amplitude associated with the scalar conformal block. For  instance the function $Q_{m_1,m_2}$ in (\ref{eq:MellinamplitudepolesQ}) is given in this case by
\begin{align}
   & Q_{m_1,m_2}=\sum_{n_i=0}\prod_{i=1}^2\tfrac{2(-1)^{m_i} \Gamma \left(\Delta _{k_i}\right) }{ \left(m_i-n_i\right)! \Gamma^2 \left(\frac{\Delta _{k_i}}{2}\right) \Gamma \left(\Delta_{\phi}-m_i-\frac{\Delta _{k_i}}{2}\right)  \left(1-\frac{d }{2}+\Delta _{k_i}\right)_{m_i-n_i}  }\tfrac{\left(\frac{\bar{\Delta} -\Delta_{\phi} }{2} \right)_{m_1-n_1} \binom{n_1+n_2+\frac{\Delta_{\phi}}{2}-m_1-m_2-\frac{\bar{\Delta} }{2}  }{n_1}}{   \Gamma \left(\frac{\Delta_{\phi}+2 m_1-2 m_2+\Delta _{k_1}-\Delta _{k_2}}{2} \right)  }\nonumber \\
   & \times\tfrac{ \left(\frac{\bar{\Delta}+2-d -\Delta_{\phi}}{2} \right)_{m_1-n_1} \left(\frac{2 m_1+\bar{\Delta}-2 n_1-\Delta_{\phi}}{2}  \right)_{m_2-n_2} \left(\frac{2 m_1+\bar{\Delta}+2 -d -2 n_1-\Delta_{\phi}}{2}  \right)_{m_2-n_2} \binom{\frac{2 n_2+\Delta_{\phi}-2 m_1-2 m_2-\bar{\Delta} }{2} }{n_2}  }{   \Gamma \left(\frac{\Delta_{\phi}-2 m_1+2 m_2-\Delta _{k_1}+\Delta _{k_2} }{2} \right) \Gamma \left(\frac{2 m_1+2 m_2+\bar{\Delta} -\Delta_{\phi}}{2} \right) }\label{eq:ScalarblockMellin}
\end{align}
where $\bar{\Delta}=\Delta_{k_1}+\Delta_{k_2}$. Note that it does not depend on the variables $s_{ij}$ as expected since the exchanged operators are scalars. The apparent asymmetry in the channels $(12)$ and $(34)$ is related with the choice of which differential operator $F_k$ we decide to act first on a three-point function. Another advantage of having the Mellin amplitude for the scalar conformal block is that it can be used to generate some solutions for spinning blocks as we have shown in section \ref{sec:MellinMellin}.

We have checked that the solution (\ref{eq:ScalarblockMellin}) satisfies the Casimir recurrence equation, in the channel $(12)$,    given by
\begin{align}
   & \big[\mathbf{d}_{00000} \big(2 \textrm{c}_{\Delta_1J_1}-a_1^2+a_1 (2 a_4+a_5-a_3-2 d+3 \Delta_\phi )-2 a_2 (a_3+a_4)+2 a_3^2-2 a_3 a_4-2 a_3 a_5\nonumber           \\
   & +\Delta_\phi  (5 a_3-4 a_4-2 a_5+4 d)+2 a_4^2+a_4 a_5-2 \Delta_\phi ^2\big)\nonumber                                                                                \\
   & +\mathbf{d}_{00-200} (a_1-a_3+a_4-2 \Delta_\phi ) (a_3-2 a_2-a_5+2 \Delta_\phi )\nonumber                                                                           \\
   & -\mathbf{d}_{000-20} (a_1-2 a_2+a_4-\Delta_\phi ) (a_1-a_3+a_4-2 \Delta_\phi )-a_3 \mathbf{d}_{002-20} (a_1-2 a_2+a_4-\Delta_\phi )\nonumber                        \\
   & +a_1 \mathbf{d}_{-2000-2} (a_1-2 a_2-a_5+\Delta_\phi )+a_1 \mathbf{d}_{-2002-2} (a_1-2 a_2-a_5+\Delta_\phi )+2 a_1 a_2 \mathbf{d}_{-2-2-200}\nonumber               \\
   & +2 a_1 a_2 \mathbf{d}_{-2-2000}+a_1 a_5 \mathbf{d}_{-20-220}+a_1 a_5 \mathbf{d}_{-20000}+a_4 \mathbf{d}_{00-220} (2 a_2-a_3+a_5-2 \Delta_\phi )\nonumber            \\
   & +a_4 \mathbf{d}_{00020} (a_3-a_4-a_5+\Delta_\phi )+a_3 \mathbf{d}_{00200} (a_4+a_5-a_3-\Delta_\phi )\big]f(t_{ij},s_{ij})=0\label{eq:recurrencerelationsMellinCasi}
\end{align}
where $\mathbf{d}_{i_1i_2i_3i_4i_5}$ is defined by $\mathbf{d}_{i_1i_2i_3i_4i_5}f(t_{12},t_{34},s_{13},s_{25},s_{45})=f(t_{12}+i_1,\dots,s_{45}+i_5)$ and the coefficients $a_i$ are given by
\begin{align}
   & a_1= 2 \Delta_\phi -t_{12},\, \ \ \  a_2= \frac{2 \Delta_\phi -t_{34}}{2} ,\, \ \ \ a_3 =  \Delta_\phi -s_{13}, \\
   & a_4= s_{25}+t_{12}-t_{34}+\Delta_\phi ,\, \ \ a_5=  s_{45}-t_{12}+t_{34}+\Delta_\phi\,.\nonumber
\end{align}
This recurrence relation is also valid for spinning conformal blocks.
\subsection{Spinning recursion relations}
In \cite{Poland:2021xjs} the authors have derived identities that blocks with different values of spin satisfy. It is possible to verify part of these relations using lightcone blocks for unequal external dimensions introduced in the previous subsection. Using (\ref{eq:lightconeLeadingSpin}) we can verify that the lightcone blocks satisfy
\begin{align}
   & (2 J_1+2 \ell+\tau_1+\tau_2-2-\Delta_5) G_{\tau_1,J_1,\tau_2,J_2,\ell}^{\Delta_{1},\Delta_5,\Delta_3}
  + 2 (J_2-\ell) G_{\tau_1,J_1,\tau_2,J_2,\ell+1}^{\Delta_{1},\Delta_5,\Delta_3}
  \label{eq:RecurrencerelationJ}                                                                                                                                                                                                                                              \\
   & +\frac{4 (2 J_1+\tau_1-2) (2 J_1+\tau_1-1) }{(2 J_1+\tau_1-\Delta_{12}-2) }\bigg[\frac{\sqrt{u_5}  G_{\tau_1+1,J_1-1,\tau_2,J_2,\ell}^{\Delta_{1}+1,\Delta_5+1,\Delta_3} }{\sqrt{u_1} }-G_{\tau_1,J_1-1,\tau_2,J_2,\ell}^{\Delta_{1},\Delta_5,\Delta_3}\bigg]=0\nonumber
\end{align}
where $G_{\tau_1,J_1,\tau_2,J_2,\ell}^{\Delta_1,\Delta_5,\Delta_3}$ represents the lightcone conformal block for the exchange of a twist $\tau_i$ and spin $J_i$ in the channels $(12)$ and $(34)$ for external with dimension $\Delta_i$,\begin{align}
   & G_{\tau_1,J_1,\tau_2,J_2,\ell}^{\Delta_1,\Delta_5,\Delta_3} =  u_1^{\frac{\tau_1}{2}}  u_3^{\frac{\tau_2}{2}}u_5^{\frac{\Delta_5}{2}}(1-u_2)^{\ell} \int [dt_1dt_2] (1-u_2 u_5+t_2 (u_2-1) u_5)^{J_1-\ell }                                                                  \\
   & \tfrac{(1-u_2 u_4+t_1 (u_2-1) u_4)^{J_2-\ell } }{(1+t_1 (u_5-1))^{\frac{\Delta_5+2 J_1+\tau_1-\tau_2-2 \ell }{2} }  (1+t_2 (u_4-1))^{\frac{\Delta_5+2 J_2-\tau_1+\tau_2-2 \ell}{2}}  (1+(1-t_1) (1-t_2) (u_2-1))^{\frac{2 J_1+2 J_2+\tau_1+\tau_2-\Delta_5}{2} } } \nonumber
\end{align}
where $[dt_1dt_2] = \prod_{i=1}^2\frac{dt_i\,\Gamma(2J_i+\tau_i)t^{\frac{\tau_i+2J_i+a_i}{2}-1}(1-t_i)^{\frac{\tau_i+2J_i-a_i}{2}-1}}{\Gamma\left(\frac{\tau_i+2J_i+a_i}{2}\right)\Gamma\left(\frac{\tau_i+2J_i-a_i}{2}\right) }$ with $a_1=\Delta_{12}, a_2=\Delta_{34}$.
As before, the index $\ell$ labels a particular structure in the three-point function (\ref{eq:threepontfunction})

By considering the Mellin transform of $G_{\tau_1,J_1,\tau_2,J_2,\ell}^{\Delta_1,\Delta_5,\Delta_3}$ we can phrase the recurrence relation in spin (\ref{eq:RecurrencerelationJ})  in terms of a Mellin amplitudes
\begin{align}
   & \frac{2 (2 J_1+\tau_1-2) (2 J_1+\tau_1-1) ((\Delta_5-2 \delta_{25}+\tau_{12}) \mathcal{M}_{\tau_1+1,J_1-1,\tau_2,J_2,l}^{\Delta_1+1,\Delta_5+1,\Delta_3}-\mathcal{M}_{\tau_1,J_1-1,\tau_2,J_2,\ell}^{\Delta_1,\Delta_5,\Delta_3})}{(2 J_1+\tau_1-\Delta_{12}-2) } \nonumber \\
   & +(2 J_1+2 \ell+\tau_1+\tau_2-\Delta_5-2)\mathcal{M}_{\tau_1,J_1,\tau_2,J_2,\ell}^{\Delta_1,\Delta_5,\Delta_3}+2 (J_2-\ell) \mathcal{M}_{\tau_1,J_1,\tau_2,J_2,\ell+1}^{\Delta_1,\Delta_5,\Delta_3}=0\label{eq:MellinRecurrencespin}
\end{align}
with
\begin{align}
  G_{\tau_1,J_1,\tau_2,J_2,\ell}^{\Delta_1,\Delta_5,\Delta_3} = u_1^{\frac{\tau_1}{2}}  u_3^{\frac{\tau_2}{2}}\! \int [d\delta_{ij}] \mathcal{M}_{\tau_1,J_1,\tau_2,J_2,\ell}^{\Delta_1,\Delta_5,\Delta_3}(\delta_{ij})  u_4^{-\delta_{45}} u_5^{\delta_{ 25}+\frac{ \tau_2-\tau_1 }{2}} u_2^{\delta_{13} -\delta_{45} +\frac{\Delta_5-a_2-\tau_1}{2}} \prod_{i<j}\Gamma(\delta_{ij})
\end{align}
where we have used the constraints to eliminate the some of the $\delta_{ij}$ and $\delta_{12}$ and $\delta_{34}$ are set to $\frac{\Delta_1+\Delta_2-\tau_1}{2}$ and $\frac{\Delta_3+\Delta_4-\tau_2}{2}$ respectively. We have suppressed the dependence on Mellin variables in (\ref{eq:MellinRecurrencespin}) since there are no shifts in them.

There is an extra identity that is needed to turn (\ref{eq:RecurrencerelationJ}) into a self-consistent recurrence relation
\begin{align}
   & \frac{4 (\tau_1+2 \ell -1) (\tau_2+2 \ell -1) (\tau_1+\tau_2+4 \ell-\Delta_5 -4) }{(\tau_1+2 \ell-\Delta_{12} -2) (\Delta_{34}+\tau_2+2 \ell -2) }\bigg[\frac{G_{\tau_1+1,\ell -1,\tau_2+1,\ell -1,\ell -1}^{\Delta_1+1,\Delta_5,\Delta_3-1}}{\sqrt{u_1} \sqrt{u_3} }-G_{\tau_1,\ell -1,\tau_2,\ell -1,\ell -1}^{\Delta_1,\Delta_5,\Delta_3}\bigg]\nonumber   \\
   & +\frac{ (\Delta_5-\tau_{12}) (\tau_1+2 \ell -1)  }{(\tau_2+2 \ell -2) (\tau_1+2 \ell-\Delta_{12} -2)}    \bigg[\frac{\sqrt{u_5} (\Delta_5+\tau_{12}) G_{\tau_1+1,\ell -1,\tau_2,\ell ,\ell -1}^{\Delta_1+1,\Delta_5+1,\Delta_3}}{\sqrt{u_1} }-2G_{\tau_1,\ell -1,\tau_2,\ell ,\ell -1}^{\Delta_1,\Delta_5,\Delta_3}\bigg]                           \nonumber \\
   & -\frac{(\Delta_5+\tau_{12}) (\tau_2+2 \ell -1) (\tau_1+\tau_2+4 \ell-\Delta_5 -4) G_{\tau_1,\ell ,\tau_2,\ell -1,\ell -1}^{\Delta_1,\Delta_5,\Delta_3}}{(\tau_1+2 \ell -2) (\Delta_{34}+\tau_2+2 \ell -2)}\nonumber                                                                                                                                           \\
   & + (\tau_1+\tau_2+4 \ell -\Delta_5-2)G_{\tau_1,\ell ,\tau_2,\ell,\ell }^{\Delta_1,\Delta_5,\Delta_3} =0.
\end{align}

\section{Scalar Mellin partial-wave}\label{app:ScaMellinpar}
In this appendix, we derive the Mellin partial-wave for scalar exchange within a five-point function. We start from partial-wave definition in position space
\begin{equation}
  F_{\nu_1, \nu_2,0,0,0}(x_i)=\int dx_6 dx_7 \langle\phi_{1}\phi_2 \phi(x_6)\rangle\langle\tilde{\phi}(x_6)\phi_5 \tilde{\phi}(x_7)\rangle\langle\phi_{3}\phi_4 \phi(x_7)\rangle
\end{equation}
where the subscripts ans superscript $0$ in $F_{\nu_1, \nu_2,0,0,0}$ denote the scalar exchanges and the superscripts refer to principal series representations of the exchanged operators. The notation $\langle \phi_i \phi_j \phi_k\rangle$ denotes kinematical structure of three-point functions
\begin{align}
  \langle\phi_1 \phi_2 \phi_3\rangle=\frac{1}{\left(-2P_{1}\cdot P_{2}\right)^{\frac{1}{2}(\Delta_1+\Delta_2-\Delta_3)}\left(-2P_{1}\cdot P_{3}\right)^{\frac{1}{2}(\Delta_1+\Delta_3-\Delta_2)}\left(-2P_{2}\cdot P_{3}\right)^{\frac{1}{2}(\Delta_2+\Delta_2-\Delta_1)}}\,,
\end{align}
where we use embedding space where $-2 P_{i}\cdot P_{j}=x_{ij}^2$.
Note that as we only consider scalar exchanges there is no sum over different possible tensor structures. In general, we consider unequal scalar fields labelled by their scaling dimensions $\Delta_i$. For operators of fixed position we do the abuse of notation $\phi_i\equiv\phi(x_i)$ but we retain the dependence on integrated variables using $\phi(x_i)$. The latter notation corresponds to scalar operators of scaling dimension $h+i \nu_i$ with $h=d/2$. Moreover, shadow operators of scaling dimension $h-i \nu_i$ are denoted with an extra tilde.

In order to integrate over $x_6$ and $x_7$ we use the Schwinger parametrization
\begin{align}
  \label{eq:schwingerparam}
  \frac{1}{(-2 P_{i}\cdot P_j)^a}=\frac{1}{\Gamma\left(m+a\right)}\int_{0}^{\infty} \frac{dt_{ij}}{t_{ij}}\, t_{ij}^{m+a}(-\partial_{t_{ij}})^{m} e^{2t_{ij} P_{i}\cdot P_j}\,.
\end{align}
for any power $a$, $(-P_{i}\cdot P_j)>0$ and some integer $m$ such that $\text{Re}(m+a)>0$. For our purposes here, it is enough to take $m=0$. It will also be useful to consider the following change of variables
\begin{align}
  \label{eq:changevariablests}
   & t_{12}=2 t_1 t_2\,,\quad t_{16}=2 t_1 t \,,\quad t_{26}=2 t_2 t \,, \quad t_{34}= 2 t_3 t_4\,,\quad t_{37}=2 t_3 s\,,\nonumber \\
   & t_{47}=2 t_4 s\,,\quad t_{56}=2 t_5 \bar{t}\,,\quad t_{57}=2 t_5 \bar{s}\,,\quad t_{67}=2 \bar{t} \bar{s}\,,
\end{align}
which is introduced to reproduce the form of integral one finds from considering a tree-level Witten diagram with two scalar exchanges using the notation of~\cite{Penedones:2010ue}. Here $t_i$'s are related with bulk-to-boundary propagators of the external scalars whereas $t, \bar{t}, s, \bar{s}$ refer to split representations of the bulk-to-bulk propagators.

The integrals over $x_6$ and $x_7$ are easy to compute successively by noting~\cite{Penedones:2010ue}
\begin{equation}
  \int_{0}^{\infty} \frac{dt d\bar{t}}{t \bar{t}} t^{\Delta_t} \bar{t}^{\Delta_{\bar{t}}} \int dP e^{2 P\cdot (t X+ \bar{t} Y)} = 2 \pi^{h}\int_{0}^{\infty} \frac{dt d\bar{t}}{t \bar{t}} t^{\Delta_t}\bar{t}^{\Delta_{\bar{t}}} e^{(t X+ \bar{t} Y)^2}
\end{equation}
where $\Delta_{t}+\Delta_{\bar{t}}= 2h$ with $X$ and $Y$ two timelike vectors.
We then find (dropping constants)
\begin{align}
   & F_{\nu_1, \nu_2,0,0,0}(x_i)\sim \int \frac{dt d\bar{t} ds d\bar{s}}{t\bar{t} s \bar{s}}t^{h+i \nu_1} \bar{t}^{h-i\nu_1} s^{h+i\nu_2} \bar{s}^{h-i\nu_2}\left(\prod_{i=1}^{5}\int\frac{dt_i}{t_i} t_i^{\Delta_i}\right) \exp\left[-t_1 t_2 x_{12}^2 \left(t^2 \left(\bar{s}^2 \bar{t}^2+1\right)+1\right)\right.\nonumber \\
   & \left.-t_1 t_3 x_{13}^2 t\bar{t} s \bar{s}-t_1 t_4 x_{14}^2 t \bar{t} s \bar{s}- t_1 t_5 x_{15}^2 t \bar{t}
  \left(\bar{s}^2 \left(\bar{t}^2+1\right)+1\right)- t_2 t_3 x_{23}^2 t \bar{t} s \bar{s}-t_2 t_4 x_{24}^2 t  \bar{t} s \bar{s}\right.\nonumber                                                                                                                                                                               \\
   & \left.-t_2 t_5 x_{25}^2 t \bar{t} \left(\bar{s}^2
    \left(\bar{t}^2+1\right)+1\right)-t_3 t_4 x_{34}^2 \left(s^2+1\right)- t_3 t_5 x_{35}^2 s
    \bar{s} \left(\bar{t}^2+1\right)- t_4 t_5  x_{45}^2 s \bar{s} \left(\bar{t}^2+1\right)\right]
\end{align}
which is of the form of Symanzik's formula~\cite{Symanzik:1972wj}
\begin{equation}
  \label{eq:symanzikformula}
  2\int_{0}^{\infty}\prod_{i=1}^{n}\frac{dt_i}{t_i}t_i^{\Delta_i}e^{-\sum_{i<j}^{n} t_i t_j Q_{ij}}=\frac{1}{(2\pi i)^{(n(n-3))/2}}\int d\delta_{ij}\prod_{i<j}^{n}\Gamma(\delta_{ij})Q_{ij}^{-\delta_{ij}}\,,
\end{equation}
with $Q_{ij}>0$. The Mellin variables $\delta_{ij}$ are integrated along a contour parallel to the imaginary axis with $\text{Re} (\delta_{ij})>0$ and obey the constraints
\begin{equation}
  \sum_{j\ne i}^{n}\delta_{ij}=\Delta_i\,.
\end{equation}
This allows us to find the inverse Mellin transform of the position-space partial-wave and the Mellin partial-wave
\begin{align}
  F_{\nu_1, \nu_2,0,0,0}(x_i)= \frac{1}{(2\pi i)^5}\int d\delta_{ij}  \mathcal{M}_{\nu_1, \nu_2,0,0,0}(\delta_{ij})\prod_{i<j}^{n}\Gamma(\delta_{ij})x_{ij}^{-2\delta_{ij}}
\end{align}
The remaining integrations in $t, \bar{t}, s$ and $\bar{s}$ are straightforward to do. We then find
\begin{align}
  \mathcal{M}_{\nu_1, \nu_2,0,0,0}(\delta_{ij})= & {\textstyle\frac{\pi^{2h}\left(\left(\prod_{i=1}^{2}\Gamma\left(\frac{\Delta_{2i-1}+\Delta_{2i}-t_{2i-1\,2i}}{2}\right)\left(\prod_{\sigma=\pm}\Gamma\left(\frac{h+\sigma \left(\Delta_{2i-1}-\Delta_{2i}\right)+i \nu_i}{2}\right)\right)\right)\right)^{-1}}{\Gamma \left(\Delta _5\right) \Gamma \left(\frac{\Delta _5-i \nu _1+i \nu _2}{2}\right) \Gamma \left(\frac{t_{12}-t_{34}+\Delta _5}{2}\right) \Gamma \left(\frac{2 h-\Delta _5-i \nu _1-i \nu _2}{2}\right) \Gamma \left(\frac{h-t_{12}+\Delta _5-i \nu _2}{2}\right)}} \\
                                                 & {\textstyle\left[\left(\prod_{\sigma=\pm}\Gamma\left(\frac{h-t_{12}+\sigma\Delta_5-i \nu_2}{2}\right)\Gamma\left(\frac{\Delta_5+\sigma i \nu_1+i \nu_2}{2}\right)\right)\Gamma\left(\frac{h-t_{34}+i \nu_2}{2}\right)\Gamma\left(\frac{t_{12}-t_{34}+\Delta_5}{2}\right)\right.}\nonumber                                                                                                                                                                                                                                              \\
                                                 & {\textstyle\left. _3F_2\left(_{\Delta_5\,,\frac{2-h+t_{12}+\Delta_5+i\nu_2}{2}}^{\frac{t_{12}-t_{34}+\Delta_5}{2}\,,\frac{\Delta_5-i \nu_1+i \nu_2}{2},\frac{\Delta_5+i \nu_1+i \nu_2}{2}};1\right)+\Gamma\left(\Delta_5\right)\Gamma\left(\frac{t_{12}-h+\Delta_5+i \nu_2}{2}\right)\right.}\nonumber                                                                                                                                                                                                                                 \\
                                                 & {\textstyle \left.\left(\prod_{\sigma=\pm}\prod_{i=1}^{2}\Gamma\left(\frac{h-t_{2i-1\,2i}+\sigma i\nu_i}{2}\right)\right)\,_3F_2\left(_{\frac{2+h-t_{12}-\Delta_5-i\nu_2}{2}\,,\frac{h-t_{12}+\Delta_5-i\nu_2}{2}}^{\frac{h-t_{12}-i\nu_1}{2}\,,\frac{h-t_{12}+i\nu_1}{2}\,,\frac{h-t_{34}-i\nu_2}{2}};1\right)\right]}\nonumber
\end{align}
where we use the notation $\delta_{ij}=\frac{\Delta_i+\Delta_j-t_{ij}}{2}$.

Let us finish this appendix by noting that a similar computation can be performed for spinning exchanges using Schwinger parametrization~(\ref{eq:schwingerparam}). To do so, at each moment, we multinomially expand the integrand decomposing it into sums over integrands of similar form to the ones encountered for scalar exchanges. In the end, one finds a spinning Mellin partial wave written as several sums over scalar-type Mellin partial waves. In particular the sums are bounded by the values of spin of the exchanged operators. This is no-good for an analytic continuation in spin that we want to consider here. For that reason and due to its length we do not write that result here.

\section{Explicit examples in position space}
In this appendix, we single-out a conformal block contribution in position space and compute its Regge-limit behavior.

We start with five-point conformal block lightcone limit in its integral representation
\begin{align}
   & G_{k_1k_2\ell}(u_i)= u_1^\frac{\tau_1}{2} u_3^\frac{\tau_2}{2} (1-u_2)^{\ell} u_5^{\frac{\Delta_\phi}{2}} \int_{0}^{1} [dt_1][dt_2]
  \label{eq:5ptlightconeblockdef}                                                                                                                                                                                                                                                                       \\
   & \frac{\big(1-t_1(1-u_2)u_4-u_2u_4\big)^{J_2-\ell}\big(1-t_2(1-u_2)u_5-u_2u_5\big)^{J_1-\ell}}{\big(1-(1-u_4)t_2\big)^{\frac{h_2-\tau_1-2\ell+\Delta_\phi}{2}}\big(1-(1-u_5)t_1\big)^{\frac{h_1-\tau_2-2\ell+\Delta_\phi}{2}} \big(1-(1-t_1)(1-t_2)(1-u_2)\big)^{\frac{h_1+h_2-\Delta_\phi}{2}}}\,,
  \nonumber
\end{align}
where $\tau_i=\Delta_i-J_i$ is the twist and $h_i=\Delta_i+J_i$ the conformal spin of the $i$-th exchanged operator.
The measure is given by $[dt]=\frac{\Gamma(\Delta_i+J_i)}{\Gamma^2(\frac{\Delta_i+J_i}{2})}(t(1-t))^{\frac{\Delta_i+J_i}{2}-1}$.

Generically, we do not know how to evaluate these integrals in terms of known analytic functions.
However, when the exponents in the denominator of the integrand are integers, this is no longer the case.\footnote{The package HyperInt~\cite{Panzer:2014caa} is particularly useful to evaluate these integrals.} As a matter of example we consider the simple case of $\Delta_i=\Delta_\phi=2$ and $J_1=J_2=\ell=0$. Note that this is just a choice and spinning cases would also have a similar discussion but with longer explicit expressions. In this case, equation~\eqref{eq:5ptlightconeblockdef} can be integrated and yields (apart from an overall constant)
\begin{align}
  \label{eq:simplecase}
   & \frac{u_1 u_3 u_5}{1-u_5+u_4 \left(u_2 u_5-1\right)}
  \left[\text{Li}_2\left(u_2 u_4\right)-\text{Li}_2\left(u_4\right)+\text{Li}_2\left(u_2 u_5\right)-\text{Li}_2\left(u_5\right)-\text{Li}_2\left(u_2\right)\right.    \\
   & \left.-\log \left(1-u_2\right) \log \left(u_2\right)-\log \left(1-u_4\right) \log \left(u_4\right)-\log \left(1-u_5\right) \log \left(u_5\right)\right.\nonumber \\
   & \left.+\log \left(u_4\right)
    \log \left(u_5\right)+\log
    \left(u_2 u_5\right) \log \left(1-u_2 u_5\right)+\log
    \left(u_2 u_4\right)\log \left(1-u_2 u_4\right)+\zeta(2)\right]\nonumber\,.
\end{align}

As we perform the analytic continuation from an Euclidean to double-Reggeon exchange kinematics that we presented in~(\ref{eq:Reggelimitcoordinates}), we cross block branch-cuts and it mixes with other solutions of the Casimir equations. In particular, the discontinuities of the block contain the leading contributions in the Regge limit. Having an explicit expression to work with we can tell the full story.

\begin{figure}
  \centering
  \begin{subfigure}{.5\textwidth}
    \centering
    \includegraphics[scale=.55]{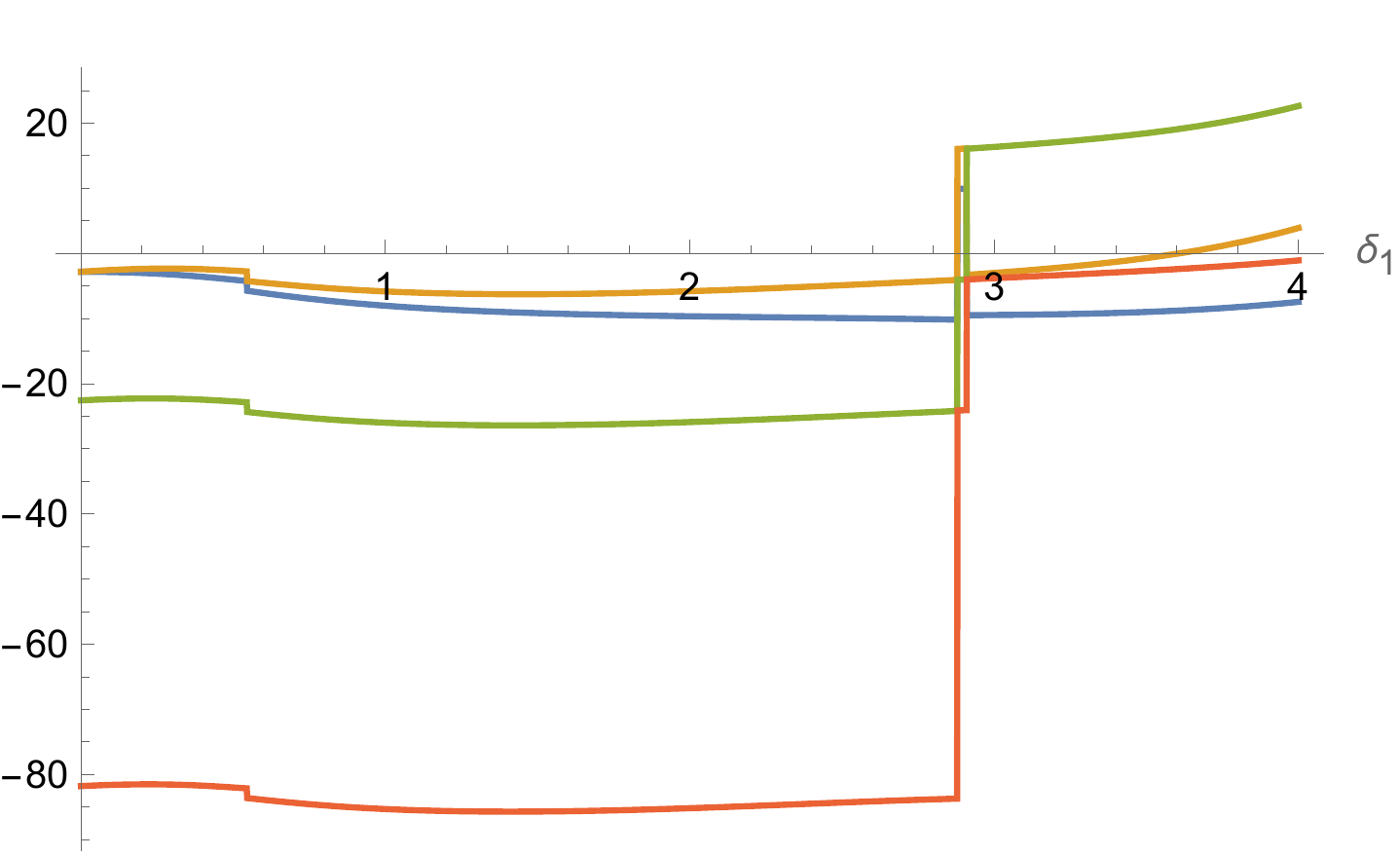}
  \end{subfigure}%
  \begin{subfigure}{.5\textwidth}
    \centering
    \includegraphics[scale=.55]{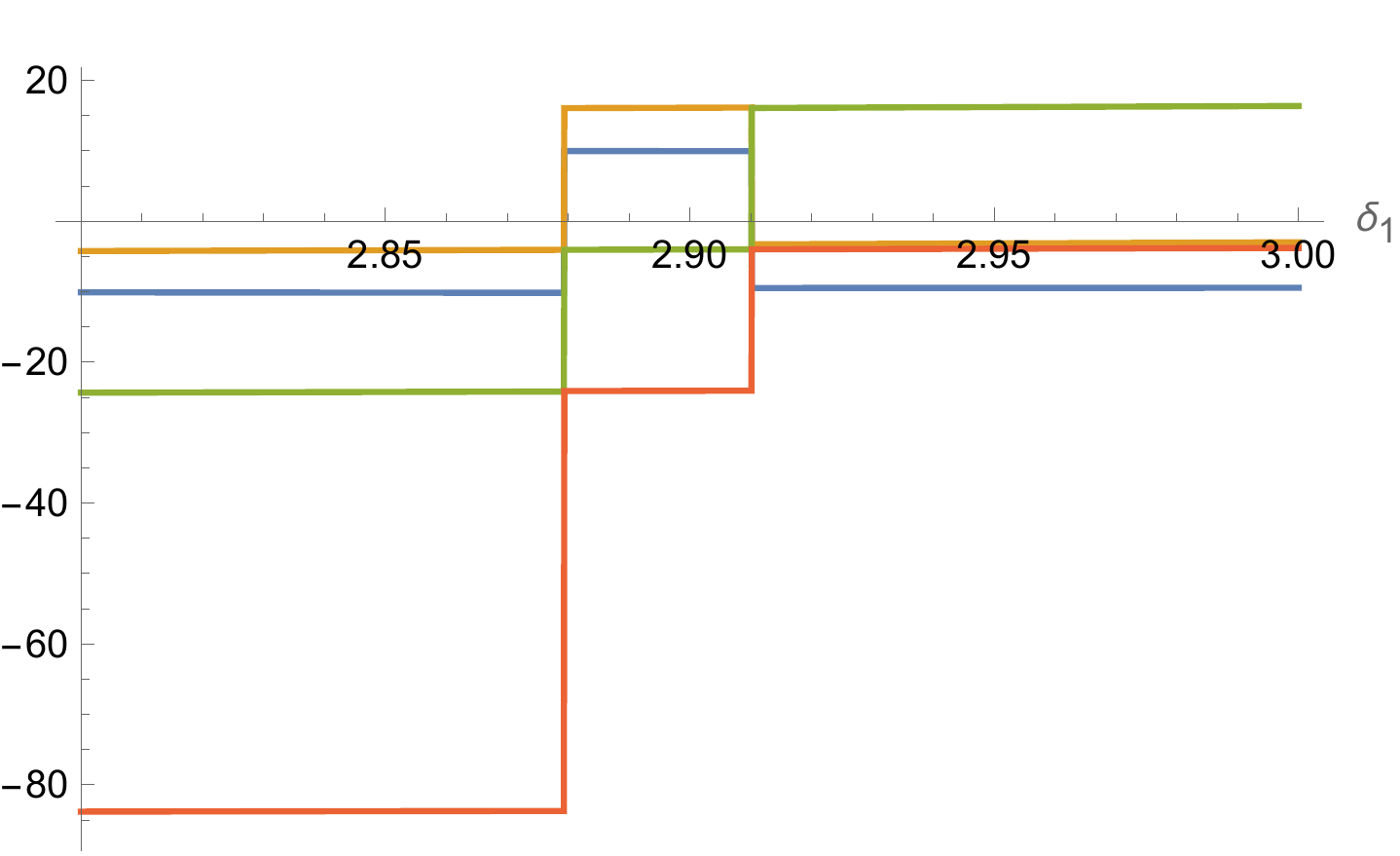}
  \end{subfigure}
  \caption{Discontinuities of lightcone block under analytic continuation~\eqref{eq:Reggelimitcoordinates}. In blue, the real part of the stripped-off lightcone block. In orange, the real part of the block with $\log(u_2)\to \log(u_2)+2\pi i$. In green, the previous with $\log(u_4)\to \log(u_4)-2\pi i$ and in red, the latter with $\log(u_5)\to \log(u_5)-2\pi i$. On the right, a zoomed-in version of the same plot. The plots are obtained with $\delta_2=0.73 \delta_1$. }
  \label{fig:discsforsimplecase}
\end{figure}

As we perform the analytic continuation and as the lightcones are crossed, pairs of operators become timelike separated and cross-ratios $u_2, u_4$ and $u_5$ go around 0. Note then that we are indeed crossing branch-cuts of the expression~\eqref{eq:simplecase}. In particular, we observe that only $\log$ terms in~\eqref{eq:simplecase} can contribute to the discontinuity as $u_i$ goes around 0 with $\log(x)\to \log(x)\pm 2\pi i$. The actual sign one picks is determined by how one moves around branch-cuts. As we reviewed in the main text, this depends on the ordering of operators of the Wightman function we consider. As before, here we take an ordering compatible with the time-ordering of Regge kinematics, i.e. $\langle \phi_4 \phi_1 \phi_2  \phi_5 \phi_3\rangle$. Taking this ordering and the associated $i\epsilon$-prescription, we can perform the path continuation to Regge kinematics in our explicit-lightcone-block contribution and observe its discontinuities concretely. This is plotted in figure~\ref{fig:discsforsimplecase}.\footnote{In this plot we only considered the terms within the brackets in~\eqref{eq:simplecase}. Note that only this part is relevant for the discontinuities we want to study.}
As we move according to the chosen path for analytic continuation, we observe that the lightcone block (blue) has discontinuities. The first one can be removed if one replaces $\log(u_2) \to \log(u_2)+ 2\pi i$ as shown by the orange line. Clearly, this shows that the discontinuity of the lightcone block is due to a logarithmic discontinuity in $u_2$. Similarly, when the orange line has a discontinuity, there is a continuation provided by the green line. The latter is defined from the former with the replacement $\log(u_4)\to \log(u_4)-2\pi i$. We conclude that a discontinuity in $u_4$ has taken place. The same is true for the red line which provides the continuation of the green line once we take $\log(u_5)\to \log(u_5)-2\pi i$ and once again a discontinuity, this time in $u_5$, has to be considered. This simple example shows in practice what we had already guessed: the lightcone block has discontinuities associated with $u_2, u_4$ and $u_5$ going around 0 and all of them are important. Let us then study the discontinuities of~\eqref{eq:simplecase} on these variables.

It is possible to use the integral representation of the lightcone block to argue that there are no sequential discontinuities involving $u_2$, i.e.
\begin{equation}
  \text{Disc}_{u_2} \text{Disc}_{u_4\, \text{or}\, u_5}G_{k_1,k_2, \ell} =\text{Disc}_{u_4\, \text{or}\, u_5}\text{Disc}_{u_2}  G_{k_1,k_2, \ell}=0\,.
\end{equation}
In the expression~\eqref{eq:simplecase} this is straightforward to see as there are no products of the type $\log(u_2)\log(u_4)$ or $\log(u_2)\log(u_5)$. As it was stated in the main text and as we will see below, it is actually the sum  $\text{Disc}_{u_2}  G_{k_1,k_2, \ell}+ \text{Disc}_{u_5}\text{Disc}_{u_4} G_{k_1,k_2,\ell}$ that dominates the Regge behavior of the correlation function.

The discontinuity of expression~\eqref{eq:simplecase} as $u_2$ goes around 0 with fixed $u_4, u_5 >0$ is given by
\begin{equation}
  \pm 2\pi i\frac{u_1 u_3 u_5}{1-u_5+u_4(u_2 u_5-1)}\log\left(\frac{1-u_2}{(1-u_2 u_4)(1-u_2 u_5)}\right)\,,
\end{equation}
which in the limit $u_4, u_5 \to 1$ with $\chi_2=\frac{1-u_2}{(1-u_4)(1-u_5)}$ fixed simplifies to
\begin{equation}
  \label{eq:discu2simple}
  \pm 2 \pi i\frac{\sqrt{u_1} \sqrt{u_3}}{\left(\chi_2-1\right) \chi_4\chi_5}\log \left(\chi_2\right)\,,
\end{equation}
where we also use $\chi_4=\frac{1-u_4}{\sqrt{u_3}}$ and $\chi_5=\frac{1-u_5}{\sqrt{u_1}}$ which approach infinity due to the order of limits considered. This order of limits does not correspond to the actual Regge limit: indeed, we will call this ordered limit a boundary condition for Regge limit. The name simply follows from the fact that we use it below as a boundary condition for a set of recursion relations where we compute the Regge limit of a conformal block starting from the lightcone. Note, moreover, that the scaling in both $u_1$ and $u_3$ in the expression above agrees with the expected $u_i^{(1-J_i)/2}$ of Regge limit. As stated above we are indeed describing a double Reggeon exchange. This clearly contrasts with the Euclidean OPE scaling, $u_i^{\Delta_i/2}$, manifesting the difference between Regge and Euclidean kinematics. Perhaps a more striking example would follow from considering a spinning case from the beginning. The story is no different in those cases but the expressions grow considerably in size.
We also note the existence of a $\log$ term in the case at hand. We point out that some other examples where the lightcone block can be integrated do not have these contributions in the above limit. Its existence in this case suggests however that a generic function for the discontinuity of the lightcone block as $u_2$ goes around 0 must contain $\log$ terms when the representation labels of the external and exchanged operators conspire in a certain way.

We now consider the discontinuity in $u_4$ with fixed and positive $u_2, u_5$. This gives
\begin{equation}
  \label{eq:discu4simplecomplete}
  \pm 2\pi i\frac{u_1 u_3 u_5}{1-u_5+u_4(u_2 u_5-1)}\log\left(\frac{1-u_4}{(1-u_2 u_4)u_5}\right)\,,
\end{equation}
which yields a boundary condition for Regge limit
\begin{equation}
  \label{eq:discu4simple}
  \pm 2 \pi i \frac{u_1 \sqrt{u_3}}{\chi_4}\,.
\end{equation}
From the symmetry of~\eqref{eq:simplecase} between $u_4$ and $u_5$ we immediately see that a similar result follows for the discontinuity in $u_5$. Note that these terms are subdominant in the limit $u_1, u_3 \to 0$ when compared to~\eqref{eq:discu2simple}. In particular, in expression~\eqref{eq:discu4simple} $u_1$ scales as $u_1^{\Delta_1/2}$ whereas $u_3$ scales as $u_3^{(1-J_2)/2}$. The converse happens in the discontinuity in $u_5$ complex plane. This behavior should correspond to single Reggeon exchanges. Notably, the sequential discontinuity in $u_4$ and $u_5$ produces a dominant contribution for the double Reggeon kinematics. To see this, consider~\eqref{eq:discu4simplecomplete} and take the sequential discontinuity in $u_5$. This gives
\begin{equation}
  \pm 4 \pi ^2 \frac{u_1 u_3 u_5}{1-u_5+u_4 \left(u_2 u_5-1\right)}\,,
\end{equation}
which fixes the boundary condition for Regge limit
\begin{equation}
  4 \pi ^2\frac{\sqrt{u_1} \sqrt{u_3}}{\left(\chi _2-1\right) \chi _4 \chi _5}\,,
\end{equation}
that is as dominant as~\eqref{eq:discu2simple}. We conclude in this simple example, the generic statement we have made in the main text that $\text{Disc}_{u_2}  G_{k_1,k_2, \ell}+ \text{Disc}_{u_5}\text{Disc}_{u_4} G_{k_1,k_2,\ell}$ provide the dominant contributions of the correlation function in Regge limit.

Even though the existence of $\log$ terms in the boundary condition for the Regge limit is not generic, we should however show how to deal with them when we compute the conformal blocks at Regge limit. If there are no $\log$ terms in your case of interest, simply set those terms to zero in the procedure below.
We consider the Casimir equations in the limit of $u_1, u_3 \to 0$ with a block that scales as
\begin{equation}
  \mathcal{G}_{k_{1}^{\prime} k_{2}^{\prime}\ell}(x_i) \propto u_1^{\frac{1-J_1}{2}} u_3^{\frac{1-J_2}{2}}\mathcal{H}(\chi_2,\chi_4,\chi_5)\,\,.
\end{equation}
In this limit, the Casimir equations for $\mathcal{H}$ simplify and read
\begin{align}
  \label{eq:casinHregge}
   & \left[\chi_4^2 \left(4 (2\chi_2-1) (\partial_{\chi_2}-\chi_5 \partial_{\chi_2}\partial_{\chi_5})-(d-1) \chi_5^3 \partial_{\chi_5}-\left(\chi_5^2-4\right) \chi_5^2 \partial_{\chi_5}^2\right)\right.\nonumber \\
   & \left.+4\left((\chi_2-1) \chi_2 \chi_4^2+1\right)\partial_{\chi_2}^2+(\Delta_1-1)(\Delta_1-d+1)\chi_4^2 \chi_5^2\right]\mathcal{H}(\chi_2,\chi_4,\chi_5)=0
\end{align}
with an entirely similar second equation obtained from the above by replacing $\Delta_1$ by $\Delta_2$ and permuting the roles of $\chi_4$ and $\chi_5$. In our particular case of study, in the limit of $u_1, u_3 \to 0$, large $\chi_4, \chi_5$ and fixed $\chi_2$, the leading Regge contribution of the block behaves as
\begin{equation}
  \label{eq:bdycondition222}
  \frac{\sqrt{u_1}\sqrt{u_3}}{(\chi_2-1)\chi_4 \chi_5}\left(a+b \log(\chi_2)\right)
\end{equation}
where $a$ and $b$ are constants. We can thus further impose in~\eqref{eq:casinHregge}
\begin{equation}
  \mathcal{H}(\chi_2,\chi_4,\chi_5)=\frac{\mathbb{H}(\chi_2,\chi_4,\chi_5)}{\chi_4  \chi_5}\,.
\end{equation}
According to~\eqref{eq:bdycondition222} and considering a small-$\chi_2$  limit, we can look for solutions of the Casimir equations of the form
\begin{equation}
  \mathbb{H}(\chi_2,\chi_4,\chi_5)=\sum_{n_1,n_2,n_3} a_{n_1,n_2,n_3} \chi_2^{n_1} \chi_4^{-n_2} \chi_5^{-n_3}+ b_{n_1,n_2,n_3} \log(\chi_2) \chi_2^{n_1} \chi_4^{-n_2} \chi_5^{-n_3}
\end{equation}
where the coefficients $a_{n_1,n_2,n_3}$ and $b_{n_1, n_2,n_3}$ reduce to $a$ and $b$, respectively, when all $n_i$ are 0. The remaining expansion coefficients $a_{n_1,n_2,n_3}$ and $b_{n_1, n_2,n_3}$ are fixed by the Casimir equations. It is easy to see that this ansatz gives rise to terms in the Casimir equations of the form
\begin{equation}
  \chi_2^{c_1+n_1}\chi_4^{c_2-n_2} \chi_5^{c_3-n_3}\times\begin{cases}a_{n_1,n_2,n_3} \\
    b_{n_1,n_2,n_3} \\
    b_{n_1,n_2,n_3}\log(\chi_2)\end{cases}\,.
\end{equation}
Clearly, the terms that depend on $\log$ should cancel among each other in order to satisfy the Casimir equation. This leads to two constraints per Casimir equation, one for the $\log$-dependent terms and one for the remaining. For the isolated $\log$ terms, we find recursion relations for the coefficients by removing the $\chi$-dependence from the equations. To do so, we shift each term accordingly, i.e. $n_1 \to n_1-c_1, n_2 \to n_2+c_2$ and $n_3 \to n_3+ c_3$. This leads to the following recursion relations
\begin{align}
  b_{n_1,n_2,n_3}= \frac{1}{n_3 (d-n_3-4)} & \left[ 4 (n_1+1) \left((n_1+n_3)
    b_{n_1+1,n_2,n_3-2}-(n_1+2)
  b_{n_1+2,n_2-2,n_3-2}\right)\right.\nonumber                                \\
                                           & \left. -4 (n_1+n_3-1) (n_1+n_3)
    b_{n_1,n_2,n_3-2}\right]
\end{align}
with a similar one where we exchange the roles of $n_3$ and $n_2$. Clearly, the above recursion relation cannot be used whenever $n_3=0$. In such case, the other recursion relation can be used instead (and vice-versa). An entirely similar argument follows for the non-$\log$-dependent terms. We find the recursion relations
\begin{align}
   & a_{n_1,n_2,n_3}= \frac{4}{n_2 (4-d+n_2)}\left[(n_1+n_2-1) (n_1+n_2)
    a_{n_1,n_2-2,n_3}-(n_1+1) (n_1+n_2)
  a_{n_1+1,n_2-2,n_3}\right.\nonumber                                    \\
   & \left.+(n_1+1) (n_1+2)
    a_{n_1+2,n_2-2,n_3-2}+(2 n_1+2 n_2-1)
    b_{n_1,n_2-2,n_3}-(2 n_1+n_2+1)
  b_{n_1+1,n_2-2,n_3}\right.\nonumber                                    \\
   & \left.+(2 n_1+3)
    b_{n_1+2,n_2-2,n_3-2}\right]
\end{align}
with another equivalent relation where the roles of $n_2$ and $n_3$ are swapped.

These recursion relations are only meaningful once one prescribes a boundary condition. We impose that
\begin{align}
   & a_{n_1,n_2,n_3}=b_{n_1,n_2,n_3}=0\qquad \text{if} \qquad n_1<0\, \lor\, n_2<0\, \lor\, n_3<0\,,\nonumber \\
   & a_{0,0,0}= a \qquad \text{and} \qquad b_{0,0,0}=b\,.
\end{align}
It is easy to check that these recursion relations fix the behavior of all the coefficients up to those of the form $a_{n_1,0,0}$ and $b_{n_1,0,0}$ but note that these can be read from~\eqref{eq:bdycondition222} by expanding it on small $\chi_2$ limit.

\section{Other Regge kinematics}
\label{sec:otherregge}
In this short appendix, we detail other possible Regge kinematics that we did not explore in detail in this paper but that might be worth studying in the future.

\subsubsection*{Single Reggeon exchange}

Within a five-point function, one can consider a single Reggeon exchange. In terms of Mandelstam invariants $s_{25}$ or $s_{45}$ of figure~\ref{fig:mandelstaminv5pt} only one of the two becomes large. In the context of CFTs, this translates to having only two operators, one in the first and one in the second Poincare patches, approaching each other in such a way that there is only one cross-ratio going to $0$ rather than two.

One possible analytic continuation that describes single Reggeon exchange is given by
\begin{align}
   & x_{1}= -r\left(\sinh(\delta_1), \cosh(\delta_1), \textbf{0}_{d-2} \right)\, &  & x_{3}= \left(0, 1, \textbf{0}_{d-2} \right)\,   & x_5=(0,h_1,h_2,\textbf{0}_{d-3})\nonumber \\
   & x_{2}= r\left(\sinh(\delta_2),  \cosh(\delta_2), \textbf{0}_{d-2} \right)\, &  & x_{4}= \left(0, -1, \textbf{0}_{d-2} \right)\,.
\end{align}
with positive rapidities $\delta_i$ and $\textbf{0}_{d}$ denoting a $d$-dimensional vector of zeros. In the large-rapidities limit, one can check that $u_1 \to 0$ and $u_2, u_5 \to 1$ with unfixed $u_3$ and $u_4$. This agrees with the Euclidean OPE limit in the (12) channel. Again, we emphasize that this limit is attained after branch-cuts are crossed and thus in an intrinsically Lorentzian Regge sheet.

\subsubsection*{Six-point snowflake}

The six-point conformal block of external scalars is known in the lightcone limit in the snowflake topology~\cite{Antunes:2021kmm}. Even though we did not attempt in this paper to analyze the cut-structure of this block, we nonetheless write down an analytic continuation prescription to achieve a Regge limit configuration that is consistent at the level of the cross-ratios with the OPE on channels (12), (34) and (56).

We use the set of 9 cyclic cross-ratios
\begin{align}
   & u_1=\frac{x_{12}^2 x_{35}^2}{x_{13}^2 x_{25}^2} \quad u_{i+1}=u_i\vert_{x_i\to x_{i+1}} \quad \text{mod 6}\,\nonumber \\
   & U_1=\frac{x_{13}^2 x_{46}^2}{x_{14}^2 x_{36}^2}\quad U_{i+1}=U_i\vert_{x_i\to x_{i+1}} \quad \text{mod 3}\,.
\end{align}
In the snowflake OPE limit $u_1, u_3, u_5 \to 0$ and the remaining all go to 1. In the Regge limit one should reobtain the same limiting values of the cross-ratios after some lightcones are crossed. We start with a totally spacelike configuration and perform the analytic continuation
\begin{align}
  \label{eq:analyticcont6pt}
   & x_{1}= -r_1\left(\sinh(\delta_1),\cosh(\delta_1), \textbf{0}_{d-2} \right)\, &  & x_{4}= \left(\sinh(\delta_1), -\cosh(\delta_1), \textbf{0}_{d-2}\right)\,\nonumber \\
   & x_{2}= r_1\left(\sinh(\delta_2),\cosh(\delta_2), \textbf{0}_{d-2}\right)\,   &  & x_5=(r_2\sinh(\delta_3), r_3, h, r_2 \cosh(\delta_3),\textbf{0}_{d-4})\nonumber\,  \\
   & x_{3}= \left(-\sinh(\delta_2), \cosh(\delta_2), \textbf{0}_{d-2} \right)\,   &  & x_6=(-r_2 \sinh(\delta_3), r_4, h,- r_2 \cosh(\delta_3),\textbf{0}_{d-4})\,.
\end{align}
where one can see that we use 9 degrees of freedom. Note as well that for six-point functions one can at most use the conformal symmetry to state that any generic correlation function is related to one that lives in some half-subspace in 4 dimensions~\cite{Kravchuk:2016qvl}. Perhaps the most notorious difference in this case is the need to boost a pair of points along some different plane. It is easy to check however that this prescription indeed leads to the expected OPE behavior for a snowflake six-point function.

\nocite{*}
\printbibliography
\end{document}